\documentclass[aps,preprint,amsmath,amssymb,amsfonts]{revtex4}
\usepackage{epsfig}
\usepackage{graphicx}
\usepackage{subfigure}
\usepackage{dcolumn}
\usepackage{bm}
\usepackage{amsthm}
\usepackage{enumitem}
\usepackage{slashed}
\usepackage{braket}
\usepackage{amsmath}

\newcommand{\del}{\partial}

\newcommand{\tr}{\operatorname{tr}}

\newcommand{\bbR}{\mathbb{R}}
\newcommand{\bbC}{\mathbb{C}}

\newcommand{\bbP}{\mathbb{P}}

\newcommand{\calC}{\mathcal{C}}
\newcommand{\calD}{\mathcal{D}}

\newcommand{\calM}{\mathcal{M}}

\newcommand{\calO}{\mathcal{O}}

\newcommand{\scri}{\mathcal{I}}

\begin{document}

\title{Higher-spin symmetry vs. boundary locality, and a rehabilitation of dS/CFT}

\author{Adrian David}
\email{adrian.david@oist.jp}
\affiliation{Okinawa Institute of Science and Technology, 1919-1 Tancha, Onna-son, Okinawa 904-0495, Japan}
\author{Yasha Neiman}
\email{yashula@icloud.com}
\affiliation{Okinawa Institute of Science and Technology, 1919-1 Tancha, Onna-son, Okinawa 904-0495, Japan}

\date{\today}

\begin{abstract}
We consider the holographic duality between 4d type-A higher-spin gravity and a 3d free vector model. It is known that the Feynman diagrams for boundary correlators can be encapsulated in an HS-algebraic twistorial expression. This expression can be evaluated not just on separate boundary insertions, but on entire finite source distributions. We do so for the first time, and find that the result $Z_{\text{HS}}$ disagrees with the usual CFT partition function. While such disagreement was expected due to contact corrections, it persists even in their absence. We ascribe it to a confusion between on-shell and off-shell boundary calculations. In Lorentzian boundary signature, this manifests via wrong relative signs for Feynman diagrams with different permutations of the source points. In Euclidean, the signs are instead ambiguous, spoiling would-be linear superpositions. Framing the situation as a conflict between boundary locality and HS symmetry, we sacrifice locality and choose to take $Z_{\text{HS}}$ seriously. We are rewarded by the dissolution of a long-standing pathology in higher-spin dS/CFT. Though we lose the connection to the local CFT, the precise form of $Z_{\text{HS}}$ can be recovered from first principles, by demanding a \emph{spin-local} boundary action. 
\end{abstract}
\maketitle
\tableofcontents
\newpage

\section{Introduction} \label{sec:intro}

\subsection{Summary and structure of the paper}

This paper arose from the study of non-local expressions for the partition function in higher-spin (HS) holography, with the long-term goal of extracting physics inside a cosmological horizon in dS/CFT. During these explorations, we came to notice that the conflict between boundary locality and global HS symmetry is sharper than was apparently expected in the literature. On the other hand, boundary \emph{spin-}locality is perfectly compatible with HS symmetry, and can even be invoked to fix those details of the partition function that HS symmetry leaves unconstrained. Choosing the spin-local, HS-symmetric partition function over the standard local one, we arrive at closed-form expressions for the partition function at finite sources, decomposed in $SO(4)$ modes. Aside from its simplicity, this partition function is free of a long-standing pathology in higher-spin dS/CFT, wherein the Hartle-Hawking wavefunction for a constant spin-0 source was not globally peaked at zero \cite{Anninos:2012ft}. 

The conflict that we find between local and HS-symmetric partition functions appears for both Euclidean and Lorentzian boundary signatures; however, the details are different. For a Lorentzian boundary, with $SO(2,3)$ symmetry, a subtle sign disagreement appears already at the level of correlators. For a Euclidean boundary, with $SO(1,4)$ symmetry, the correlators can be made to agree, but then a subtle failure of linearity occurs when integrating them into partition functions. In the interest of painting a full picture, we will discuss both cases; however, our main focus is the $SO(1,4)$ case, which is the relevant one for dS/CFT.

The paper is structured as follows. In the remainder of this section, we review some relevant background and recent developments in HS gravity and its holography (leaving the dS-specific discussion to section \ref{sec:discuss:dS}). In section \ref{sec:correlators}, after introducing the necessary geometry and algebra, we construct the Euclidean $n$-point correlators in HS-algebraic language. In the process, we present for the first time the dictionary between local CFT sources and spacetime-independent twistor functions, and fix some crucial sign mistakes/ambiguities in our previous work \cite{Neiman:2017mel,Neiman:2018ufb}. In section \ref{sec:partition}, we upgrade the correlators into an HS-algebraic formula for the Euclidean partition function, and show that it disagrees with the local CFT calculation. In section \ref{sec:signs}, we explain the disagreement in terms of sign ambiguities in the HS-algebraic correlators. In section \ref{sec:Lorentzian}, we switch to a Lorentzian boundary. There, the signs in HS-algebraic correlators are no longer ambiguous, but they \emph{disagree} with the CFT calculation. In section \ref{sec:shell}, we provide another explanation for the disagreements, in terms of a confusion between on-shell and off-shell boundary particle states. In section \ref{sec:discuss}, we discuss the implications of choosing the HS-algebraic partition function over that given by the local CFT. Specifically, in section \ref{sec:discuss:spin_locality}, we discuss how boundary spin-locality can replace the standard structure of the CFT as a guiding principle. Finally, in section \ref{sec:discuss:dS}, we discuss the consequences for dS/CFT.

\subsection{The bulk situation: HS gravity, locality and spin-locality}

Higher-spin gravity \cite{Vasiliev:1995dn,Vasiliev:1999ba} is a most fascinating and frustrating specimen of mathematical physics. It can be thought of as an exploration of infinite-dimensional, bosonic extensions of spacetime symmetry, analogous to the finite-dimensional fermionic extensions that one encounters in supergravity. While the theory can be formulated in various dimensions, we will be interested here in the 4d case.

Like string theory, HS gravity a theory of infinitely many fields, including a massless spin-2 ``graviton''. In its simplest version, there is a single field of every integer (or every even) spin, all massless. The theory does not have a known action principle, but is given instead by Vasiliev's equations of motion. In addition to spacetime coordinates $x$, these equations involve two twistor-like pairs of spinor variables. The first, known as $Y$, is analogous to the fermionic coordinates in a superspace formalism: it enumerates the component fields in the HS multiplet, along with their derivatives (due to the bosonic nature of $Y$, we actually get the \emph{full infinite tower} of independent derivatives). The second twistor-like variable, known as $Z$, is entirely auxiliary. Thus, the Vasiliev equations are not quite field equations, but rather ``equations for the field equations'': one must first solve the $Z$-dependence, to get the actual field equations in $x$ and $Y$. Moreover, to fully define the system, one must decide on boundary data in the $Z$ space, which until recently has not been done in a principled manner. Ambiguity in the $Z$ boundary data translates into a freedom of redefining the ``physical'' $(x,Y)$-dependent fields. 

In a more standard field theory, the freedom of field redefinitions is crucially curtailed by the restriction of locality \cite{Barnich:1993vg}. However, higher-spin gravity is a theory of infinitely many massless fields, which interact at all orders in derivatives. Therefore, the task of properly defining the theory becomes tightly intertwined with understanding the non-locality involved. We want a notion of locality that is loose enough to be true, but tight enough to constrain the theory \cite{Vasiliev:2015wma,Skvortsov:2015lja}. This is now a subject of intense work \cite{Gelfond:2018vmi,Didenko:2018fgx,Didenko:2019xzz,Gelfond:2019tac}, centered around the criterion of ``spin-locality''. This concept refers to constraining non-locality not in the spacetime coordinates $x$ directly, but in the spinor coordinates $Y$ (note that derivatives in $x$-space and in $Y$-space are related by the Vasiliev equations). So far, it seems that this criterion is non-trivially fulfilled by a particular choice of boundary data in the auxiliary $Z$ space, and thus succeeds in constraining the theory and completing its definition. 

These recent developments were sparked in part by holographic computations. Similarly to string theory, higher-spin gravity appears on the bulk side of AdS/CFT holographic dualities \cite{Maldacena:1997re,Witten:1998qj,Aharony:1999ti}. Generally speaking, higher-spin theory is holographically dual to vector models, with a Chern-Simons gauging of the internal symmetry \cite{Giombi:2012ms}. In the simplest case \cite{Klebanov:2002ja}, the boundary theory is just a \emph{free} vector model of spin-0 fields. This will be our case of interest, since it manifests HS symmetry in the most direct manner. The vector model's internal symmetry group can be either $U(N)$ (for all integer spins in the bulk), or $O(N)$ (for even spins only). 

If one assumes the holographic duality, one can bypass some of the difficulties with the bulk equations. In particular, one can reconstruct the interaction vertices of physical bulk fields, starting from the known $n$-point functions of the boundary theory \cite{Sleight:2016dba}. In \cite{Bekaert:2015tva,Sleight:2017pcz}, this was done for the quartic scalar vertex. The result negated earlier hopes that the theory's non-locality may be restricted to the cosmological radius. Instead, the quartic vertex remains non-local at all distances, much like a massless propagator (note that the original calculation in \cite{Bekaert:2015tva} actually reached the opposite conclusion, and was corrected in \cite{Sleight:2017pcz}; for early hints of the quartic vertex's non-locality, see \cite{Fotopoulos:2010ay,Taronna:2011kt}). The authors of \cite{Sleight:2017pcz} were thus led to claim that locality can in no way be used to constrain field redefinitions in the bulk theory, rendering the latter's interactions completely arbitrary. The most far-reaching parts of this claim have been challenged \cite{Ponomarev:2017qab}. Either way, the apparent failure of spacetime locality that was noted in \cite{Sleight:2017pcz} is one of the main motivations behind the ongoing exploration \cite{Gelfond:2018vmi,Didenko:2018fgx,Didenko:2019xzz,Gelfond:2019tac} of spin-locality as an alternative guiding principle.

In this paper, we will enact a similar drama in miniature, this time on the boundary side of higher-spin holography. While the boundary CFT is necessarily local, we will point out a contradiction between its local structure and global HS symmetry. As we will now review, \emph{some} contradiction of this general nature was long anticipated in the literature.

\subsection{A boundary puzzle: CFT correlators, partition function and HS symmetry} \label{sec:intro:boundary}

When speaking of the boundary CFT and its locality, one must distinguish between two levels of discussion. In AdS/CFT, the bulk fields are mapped to the CFT's single-trace primary operators and their sources. In the free vector model of \cite{Klebanov:2002ja}, these operators are all conserved currents $J_{k_1\dots k_s}^{(s)}$, one for each spin $s$, including an ``honorary current'' of spin 0 \cite{Craigie:1983fb,Anselmi:1999bb} (note that the definition here differs by a factor of $2^s$ from the one in \cite{Neiman:2017mel}):
\begin{align}
  J^{(s)}_{k_1\dots k_s} = \frac{1}{i^s}\,\bar\phi_I
  \left(\sum_{m=0}^s (-1)^m \binom{2s}{2m} \overset{\leftarrow}{\del}_{(k_1}\dots\overset{\leftarrow}{\del}_{k_m} \overset{\rightarrow}{\del}_{k_{m+1}}\dots\overset{\rightarrow}{\del}_{k_s)} - \text{traces}\right) \phi^I \ .
\label{eq:explicit_J}
\end{align}
Here, $(i,j,k,\dots)$ are boundary spacetime indices, and $\phi^I$ is the vector model's fundamental field with color index $I$. The sources for these operators are spin-$s$ gauge potentials $A^{(s)}_{k_1\dots k_s}$. The most basic objects of study are the $n$-point correlation functions of the currents $J_{k_1\dots k_s}^{(s)}$. These can be extracted from the CFT path integral, in which one adds to the Lagrangian a linear coupling $A^{(s)}_{k_1\dots k_s} J_{(s)}^{k_1\dots k_s}$ to the external potentials $A^{(s)}_{k_1\dots k_s}$. For example, for spin 0, we add a mass-type coupling $\sigma\bar\phi_I\phi^I$ of the scalar operator $J^{(0)} = \bar\phi_I\phi^I$ to a source $\sigma$; for spin 1, we add a coupling $A_i J^i$ of the charge current $J_i = J^{(1)}_i = -i\bar\phi_I\overset{\leftrightarrow}{\del}_i\phi^I$ to a Maxwell potential $A_i$; for spin 2, a coupling $\frac{1}{2} h_{ij}T^{ij}$ of the stress-energy tensor $T_{ij} = \frac{1}{8}J^{(2)}_{ij}$ to a metric perturbation $h_{ij}$, and so on. The $n$-point correlators are then given by derivatives at zero of the path integral with respect to sources at $n$ distinct points $(\ell_1,\dots,\ell_n)$. For the free vector model, all such correlators can be written down without calculation: they are given by 1-loop Feynman diagrams, which (in spacetime, as opposed to momentum space) don't involve any integrals.

At a more advanced level, one can ask for the partition function at some nonzero value of the sources $A^{(s)}_{k_1\dots k_s}(\ell)$. Naively, this can be written as a Taylor series, whose coefficients are just found by integrating the $n$-point correlators over the insertion points $(\ell_1,\dots,\ell_n)$. The Taylor series might miss some non-perturbative effects, but those will not bother us here. Even perturbatively, the naive integration over $(\ell_1,\dots,\ell_n)$ can fail: the integrals are typically UV-divergent! The solution is to include additional, non-linear couplings to the sources in the Lagrangian, i.e. terms in which two or more sources at the same point are multiplied together. For the special case in which the sources are gauge fields, these ``contact corrections'' can also be deduced from gauge invariance. For spin 0, such corrections are absent; for spin 1, we have the single contact correction $A_i A^i \bar\phi_I\phi^I$, familiar from scalar electrodynamics; for spin 2, there is an infinite tower of contact corrections, which comprises the non-linear coupling of $\phi^I$ to a curved background metric. Analogously, for spin $s>2$, the contact corrections should represent a higher-spin generalization of coupling to a curved background. The explicit form of these corrections remains largely unknown, though they can all be derived in principle from the construction of \cite{Segal:2002gd} (and even uplifted to form interacting bulk HS theories \cite{Bekaert:2017bpy,Grigoriev:2018wrx}). At the end of the day, even in the free vector model, in which all the correlators can be written out explicitly, the partition function remains a difficult open problem (though calculations for some low-spin backgrounds are of course possible -- see e.g. \cite{Anninos:2012ft,Bobev:2016sap,Bobev:2017asb}).

Can we do better? While individual higher spins may be difficult, we might expect some simplification once the entire HS multiplet is treated in a unified, HS-covariant manner. Specifically, the free vector model (or, equivalently, the bulk HS theory with appropriate boundary conditions) should enjoy global higher-spin symmetry, since the higher-spin currents \eqref{eq:explicit_J} are all conserved. In particular, it should be possible to package all the $n$-point correlators in an HS-covariant manner. This has been carried out several times, in slightly different versions. From a bulk point of view, the general philosophy of constructing correlators from HS invariants was proposed in \cite{Colombo:2012jx}, then streamlined in \cite{Didenko:2012tv}. There, all the correlators were packaged HS-covariantly, and fixed up to a single coefficient for each $n$ (which, assuming the existence of a boundary CFT, can in turn be fixed by matching OPE structures). In \cite{Neiman:2017mel}, the correlators \emph{with} their coefficients were derived from the boundary theory in an HS-algebraic manner, starting from a boundary-bilocal formalism \cite{Das:2003vw,Douglas:2010rc}. A similar answer was given in \cite{Gelfond:2013xt}, using a boundary HS-covariant formalism from \cite{Gelfond:2003vh}, embedded in the larger framework of so-called multiparticle algebra \cite{Vasiliev:2012tv}. In section \ref{sec:correlators} of this paper, we will again derive these correlators, working up from the spin-0 case to the bilocal language, and from there to all spins. The minimal version of all the above procedures can be distilled as:
\begin{enumerate}
	\item Encode the source $\sigma(\ell)$ for the spin-0 operator $J^{(0)}$ in an HS-covariant form. In practice, this means replacing the dependence on spacetime coordinates $\ell$ with some appropriate spinor or twistor variables $Y$.
	\item Find an HS-covariant expression that reproduces the correlators of $J^{(0)}$.
	\item By HS symmetry, this expression can now be extended to provide the correlators for all other spins.
\end{enumerate}
So far, no controversy. Confusion begins when one attempts to upgrade from correlators to the partition function at finite sources. The authors of \cite{Colombo:2012jx} were careful to warn that the correlators are just the ``trivial'', kinematic part of the theory, and that the dynamical ``meat'' is in the contact corrections. Indeed, for spin 2, in some sense the entire notion of curved geometry, and thus the gravitational nature of the bulk theory, arises from the contact corrections to the $T_{ij}$ correlators! Also, the currents $J^{(s)}$ in the free vector model are only conserved \emph{away from sources}; thus, in the presence of sources, we should expect contact corrections to restore (gauged) higher-spin symmetry. 

On the other hand, we know that for spin 0, contact corrections are absent. Thus, in that sector, we should be able to just integrate the $n$-point correlators and obtain the partition function. But then, by global HS symmetry, shouldn't this also give the full partition function for all spins? This was proposed explicitly in \cite{Neiman:2017mel}, based on two pieces of evidence. First, the HS-covariant twistorial expression that one obtains in this way does not show any sign of a UV divergence. Second, one can derive from it expectation values for the currents $J^{(s)}_{k_1\dots k_s}$ as a function of finite sources, and these are manifestly conserved. Thus, from the point of view of gauge invariance and also of UV divergences, the need for contact corrections appears to be gone! This is a strange conclusion, since, as discussed above, it seems to do away with curved backgrounds on the boundary, and with dynamical geometry in the bulk. However, even this may not be cause for panic: as pointed out in \cite{Neiman:2015wma}, the Vasiliev equations can be written equally well on a non-dynamical, pure (A)dS geometry.

This is all very confusing! Can we naively integrate the correlators to make a sensible partition function, or not? The usual local treatment of the CFT says ``no''. Global HS symmetry says ``yes''. If the two formalisms agree on the correlators, how can they disagree on their integrals? Is linearity somehow failing? 

In this paper, we claim to resolve the above confusion. In Euclidean boundary signature, there is indeed a failure of linearity hidden inside the dictionary between the local and twistor languages. For Lorentzian boundary, there is instead a subtle disagreement between the CFT path integral and HS algebra already at the level of the correlators. In either case, we have a conflict between boundary locality and global HS symmetry. Surprisingly, the disagreements begin already at spin 0, so they are \emph{not a consequence} of contact corrections. 

After carefully analyzing the disagreements and their causes, we will choose one partition function over the other. We will choose to retain HS symmetry at the cost of boundary locality, replacing the latter by spin-locality. The resulting formalism will be essentially the one already put forward in \cite{Gelfond:2003vh,Gelfond:2013xt}. Contact corrections will be absent, and the notions of curved background and a dynamical bulk geometry indeed seem to be lost. Radical though this outcome may seem, we will argue for its merits in section \ref{sec:discuss}, both in general and with regard to the project of dS/CFT.

\subsection{Embedding space and spacetime-independent twistors}

In this paper, we will use spinors and twistors after the fashion of \cite{Neiman:2017mel}. Namely, we define the 4d bulk and 3d boundary in terms of a 5d embedding space. We then define twistors as the spinors of this embedding space, with no need to reference any bulk or boundary point. Specific bulk or boundary points are associated with local spinor spaces, which can be viewed as subspaces (or quotient spaces) of twistor space. Bulk and boundary fields can be written with spinor indices, or packaged into generating functions (``master fields'') with spinor arguments. In this case, there remains a dependence on the spacetime point under consideration. There is, however, an alternative: the field can be encoded, via the Penrose transform, as a function of \emph{just} a twistor argument, which makes no reference, implicit or explicit, to any spacetime point.

This is all standard in Penrose's twistor theory \cite{Penrose:1986ca,Ward:1990vs}; the main difference is that Penrose would usually consider conformal symmetry in 4d, whereas we fix a nonzero cosmological constant, which only leaves isometries of the 4d bulk (or, equivalently, conformal symmetry on the 3d boundary). Within HS theory, Penrose-style twistors and the Penrose transform were introduced by one of the authors \cite{Neiman:2014npa,Neiman:2015wma,Neiman:2017mel}. This is in contrast to most of the HS literature, where the word ``twistor'' refers not to a global geometric object in spacetime, but to a pair of local spinors at a particular spacetime point. 

The spacetime-independent twistor language of \cite{Neiman:2014npa,Neiman:2015wma,Neiman:2017mel} will provide us with a convenient, covariant middle ground between the ``reference frames'' of different spacetime points. These ``reference frames'' themselves, which we will also use extensively, are just the standard (bulk and boundary) master-field formalisms from the HS literature.

\section{Euclidean boundary correlators from HS algebra} \label{sec:correlators}

\subsection{Geometry} \label{sec:correlators:geometry}

In this subsection, we introduce our geometric formalism, in a condensed version of Section 3 of \cite{Neiman:2017mel}.

\subsubsection{Spacetime}

We consider first Euclidean signature, i.e. a Euclidean anti-de Sitter ($EAdS_4$) bulk with an $S_3$ boundary. We will represent these as embedded in a flat $\bbR^{1,4}$ Minkowski space, with a (mostly-plus) flat metric $\eta_{\mu\nu}$, which is used to raise and lower spacetime indices $(\mu,\nu,\dots)$. The 4d bulk is defined as the hyperboloid of unit future-pointing timelike vectors in $\bbR^{1,4}$:
\begin{align}
 EAdS_4 = \left\{x^\mu\in\bbR^{1,4}\, |\, x_\mu x^\mu = -1, \ x^0 > 0 \right\} \ . \label{eq:EAdS}
\end{align}
Vectors in $EAdS_4$ are simply vectors $v^\mu\in \bbR^{1,4}$ that are tangential to the hyperboloid \eqref{eq:EAdS}, i.e. that satisfy $v\cdot x \equiv v_\mu x^\mu = 0$. The covariant derivative $\nabla_\mu$ on $EAdS_4$ is just the flat derivative $\del_\mu$ in $\bbR^{1,4}$, followed by projecting all tensor indices back into the $EAdS_4$ tangent space with the projector $q_\mu^\nu(x) = \delta_\mu^\nu + x_\mu x^\nu$.

The conformal boundary of $EAdS_4$ is the 3-sphere of lightlike directions in $\bbR^{1,4}$, i.e. the projective lightcone. We represent boundary points by lightlike vectors $\ell^\mu$ with $\ell\cdot\ell = 0$, modulo rescalings $\ell^\mu \rightarrow \rho\ell^\mu$. Such rescalings induce Weyl transformations $d\ell_\mu d\ell^\mu\rightarrow \rho^2 d\ell_\mu d\ell^\mu$ on the boundary's conformally flat metric. A boundary quantity that depends on the scaling as $f(\rho\ell^\mu) = \rho^{-\Delta}f(\ell^\mu)$ is said to have conformal weight $\Delta$. Boundary \emph{vectors} at a point $\ell$ can be described as embedding-space vectors $v^\mu\in \bbR^{1,4}$ that are tangential to the lightcone, i.e. $v\cdot\ell = 0$, modulo the equivalence relation $v^\mu\cong v^\mu + \alpha\ell^\mu$. 

A conformal frame on the boundary corresponds to choosing a particular vector $\ell^\mu$ along each null direction, i.e. choosing a section of the $\bbR^{1,4}$ lightcone. An especially useful conformal frame is defined by the lightcone section $\ell\cdot x_0 = -1$, where $x_0^\mu\in \bbR^{1,4}$ is a unit future-pointing timelike vector, i.e. a point in the $EAdS_4$ bulk. In this frame, the boundary is a unit 3-sphere $S_3$. Another useful frame is defined by $\ell\cdot\ell_\infty = -\frac{1}{2}$, where $\ell_\infty^\mu\in \bbR^{1,4}$ is a future-pointing \emph{lightlike} vector, i.e. a boundary point. In this frame, the boundary becomes flat space $\bbR^3$, with $\ell_\infty$ acting as the ``point at infinity''. The distance between two points $\ell_1,\ell_2$ in this flat $\bbR^3$ is $r = \sqrt{-2\ell_1\cdot\ell_2}$.

Finally, a bulk point $x$ can be said to ``approach the boundary point $\ell$'' via the limiting procedure $x^\mu\rightarrow\ell^\mu/z$, where $z$ goes to zero. Sometimes, it is important to also specify the direction of the approach. In that case, we choose a second boundary point $l^\mu$, normalized with respect to $\ell^\mu$ as $\ell\cdot l = -\frac{1}{2}$, and locate $x$ along the geodesic connecting $\ell$ with $l$:
\begin{align}
 x^\mu = \frac{1}{z}\ell^\mu + zl^\mu \ . \label{eq:approach}
\end{align}

\subsubsection{Twistor space} \label{sec:correlators:geometry:twistors}

The isometry group of $EAdS_4$, or the conformal group of its boundary, is just the rotation group $SO(1,4)$ in the embedding space $\bbR^{1,4}$. The double cover of this group is $USp(2,2)$. Its spin-$\frac{1}{2}$ representation, composed of 4-component Dirac spinors, is what we will refer to as \emph{twistor space}. We will denote twistors by the indices $(a,b,\dots)$. Twistor space has a symplectic metric $I_{ab}$ with inverse $I^{ab}I_{ac} = \delta^b_c$, which we use to raise and lower indices as $U_a = I_{ab}U^b$, $U^a = U_b I^{ba}$. From $I_{ab}$, we can construct a Levi-Civita symbol $\epsilon_{abcd} = 3I_{[ab}I_{cd]}$, along with a measure $d^4U$ (in which we include $2\pi$ factors for convenience):
\begin{align}
 d^4U \equiv \frac{\epsilon_{abcd}}{4!(2\pi)^2}\,dU^a dU^b dU^c dU^d \ . 
\end{align} 
Twistor space and spacetime are related by the gamma matrices $(\gamma_\mu)^a{}_b$, which satisfy the Clifford algebra $\{\gamma_\mu,\gamma_\nu\} = -2\eta_{\mu\nu}$. We also introduce the antisymmetrized products $\gamma_{\mu\nu} \equiv \frac{1}{2}[\gamma_\mu,\gamma_\nu]$, which generate the $SO(1,4)$ spacetime symmetry. As twistor matrices, the $\gamma_\mu$ are traceless and antisymmetric, while the $\gamma_{\mu\nu}$ are symmetric. We use $\gamma_\mu$ to map between twistor matrices and $\bbR^{1,4}$ vectors, and similarly for $\gamma_{\mu\nu}$ and $\bbR^{1,4}$ bivectors:
\begin{align}
 \xi^{ab} = \gamma_\mu^{ab}\xi^\mu \ ; \quad \xi^\mu = -\frac{1}{4}\gamma^\mu_{ab}\xi^{ab} \ ; \quad 
 f^{ab} = \frac{1}{2}\gamma_{\mu\nu}^{ab}f^{\mu\nu} \ ; \quad f^{\mu\nu} = \frac{1}{4}\gamma^{\mu\nu}_{ab} f^{ab} \ .
\end{align}
In our $SO(1,4)$ signature, the properties of twistors under complex conjugation follow those of $SO(3)$ spinors. The complex conjugation operation $U^a\rightarrow \bar U^a$ that is consistent with $SO(1,4)$ symmetry squares to $-1$. Therefore, it can't be represented as \emph{component-wise} complex conjugation $(U^a)^*$, and there is no invariant notion of a real twistor. The symplectic metric $I_{ab}$ can be defined as real, both component-wise $(I_{ab})^* = I_{ab}$ and under the twistor complex conjugation, $\bar I_{ab} = I_{ab}$. The gamma matrices $(\gamma_\mu)^a{}_b$ are real under the twistor complex conjugation $\bar\gamma_\mu = \gamma_\mu$, but cannot all be real component-wise. Instead, they are quaternionic, i.e. they can be written in $2\times 2$ blocks made up of $(1,-i\boldsymbol{\sigma})$, where $1$ is the $2\times 2$ identity matrix, and $\boldsymbol{\sigma}$ are the Pauli matrices.

We will often use condensed index-free notation, in which twistor indices are implicitly contracted bottom-to-top. Thus, for example, $Ux\ell V \equiv U_a(x^\mu\gamma_\mu)^a{}_b(\ell^\nu\gamma_\nu)^b{}_c V^c$. We define the delta function $\delta(U)$ on twistor space via the formal integral:
\begin{align}
 \delta(U) = \int d^4V e^{iVU} \ ; \quad \int d^4U\,\delta(U) f(U) = f(0) \ , \label{eq:delta_U}
\end{align}
and note the formula for Gaussian integrals:
\begin{align}
 \int d^4U\, e^{(UAU)/2} = \frac{\pm 1}{\sqrt{\det A}} \ ; \quad \det A = \frac{1}{8}\left(\tr A^2\right)^2 - \frac{1}{4}\tr A^4 \ , \label{eq:Gaussian}
\end{align} 
where $A_{ab}$ is a symmetric twistor matrix. Sign ambiguities of complex Gaussian integrals such as the one in \eqref{eq:Gaussian} will be important below.

\subsubsection{Local spinor spaces} \label{sec:correlators:geometry:spinors}

Choices of (bulk or boundary) spacetime points pick out various $\bbC^2$ subspaces out of the $\bbC^4$ twistor space. These are the local spinor spaces at each spacetime point. In fact, Penrose essentially \emph{defines} spacetime points as $\bbC^2$ subspaces of twistor space (though these are usually described as $\bbC\bbP^1$, since he prefers to work projectively). To see how this works, consider a spacetime point represented by a vector $\xi^\mu \in \bbR^{1,4}$. The $\bbC^2$ subspace associated with the point $\xi^\mu$ is spanned by the rank-2 twistor matrix:
\begin{align}
 P^a{}_b(\xi) = \frac{1}{2}\left(\sqrt{-\xi\cdot\xi}\,\delta^a_b + \xi^a{}_b\right) \ ,
\end{align}
or $P(\xi) = \frac{1}{2}\left(\sqrt{-\xi\cdot\xi} + \xi \right)$ in index-free notation. We will also use the notation $P(\xi)$ for the $\bbC^2$ subspace itself. When we want to emphasize that a twistor belongs to the subspace $P(\xi)$, we will denote it as e.g. $u^a_{(\xi)}$ rather than simply $U^a$. The subspaces $P(\xi)$ and $P(-\xi)$ are totally orthogonal under the twistor metric. 

A measure $d^2u_{(\xi)}$ on $P(\xi)$ can be defined as:
\begin{align}
 \frac{du_{(\xi)}^a du_{(\xi)}^b}{2\pi} \equiv P^{ab}(\xi)\, d^2u_{(\xi)} \ . \label{eq:measure}
\end{align}
With this measure, we define a delta function, whose support is on $P(-\xi)$, i.e. on the subspace orthogonal to $P(\xi)$:
\begin{align}
 \delta_\xi(U) = \int_{P(\xi)} d^2v_{(\xi)}\,e^{iv_{(\xi)}U} \ . \label{eq:delta_xi}
\end{align}
A useful integral is that of a delta function associated with one spacetime point $\xi$ over the spinor space associated with another point $\xi'$:
\begin{align}
 \int_{P(\xi')} d^2u_{(\xi')}\,\delta_{\xi}(u_{(\xi')}) f(u_{(\xi')}) = \frac{2}{\sqrt{(\xi\cdot\xi)(\xi'\cdot\xi')} - \xi\cdot\xi'}\,f(0) \ . \label{eq:spinor_delta_integral}
\end{align}
The Gaussian integral on $P(\xi)$ reads:
\begin{align}
\int_{P(\xi)} d^2u_{(\xi)}\, e^{u_{(\xi)}Au_{(\xi)}/2} = \frac{\pm 1}{\sqrt{\det_\xi(A)}} \ ; \quad \det\nolimits_\xi(A) = -\frac{1}{2}\tr\left(P(\xi)A\right)^2 \ , \label{eq:Gaussian_spinor}
\end{align}
where $A_{ab}$ is again a symmetric twistor matrix. Finally, a twistor $U$ can be decomposed along a pair of spinor spaces $P(\xi),P(\xi')$ as $U = u_{(\xi)}+u_{(\xi')}$, via:
\begin{gather}
  u_{(\xi)} = \frac{2P(\xi)P(-\xi')U}{\sqrt{(\xi\cdot\xi)(\xi'\cdot\xi')} + \xi\cdot\xi'} \ ; \quad u_{(\xi')} = \frac{2P(\xi')P(-\xi)U}{\sqrt{(\xi\cdot\xi)(\xi'\cdot\xi')} + \xi\cdot\xi'} \ ; \label{eq:decomposition} \\
  d^4U = \frac{1}{2}\left(\sqrt{(\xi\cdot\xi)(\xi'\cdot\xi')} + \xi\cdot\xi'\right) d^2u_{(\xi)}\,d^2u_{(\xi')} \ . \label{eq:measure_decomposition}
\end{gather}

Let's now consider more concretely the local spinor spaces at bulk vs. boundary points. At a bulk point, we take $\xi^\mu = x^\mu$ with $x\cdot x = -1$. The choice of point $x$ breaks the spacetime symmetry $SO(1,4)$ down to the bulk rotation group $SO(4)$ at $x$. Accordingly, twistor space decomposes into the right-handed and left-handed Weyl spinor spaces at $x$. The rank-2 twistor matrix $P(x) = (1+x)/2$ is a projector, which spans (or projects onto) the right-handed spinor space. The restriction of the twistor metric onto this spinor space is simply $P_{ab}(x)$. To describe the left-handed spinor space at $x$, we must use $\xi^\mu = -x^\mu$, which gives the projector $P(-x) = (1-x)/2$. Note that $-x^\mu$ itself is a point on the antipodal $EAdS_4$, i.e. on the other branch of the double-sheeted hyperboloid $x_\mu x^\mu = -1$. In Lorentzian bulk signature, $x^\mu$ and $-x^\mu$ both lie on the same connected spacetime, and have the same Weyl spinor spaces $P(\pm x)$, but with opposite handedness roles \cite{Neiman:2013hca}. The measure \eqref{eq:measure} in the case of a bulk point can be defined equivalently as: 
\begin{align}
 d^2u_{(\pm x)} = \frac{P_{ab}(\pm x)}{2(2\pi)}\,du_{(\pm x)}^a du_{(\pm x)}^b \ . \label{eq:measure_bulk}
\end{align}
The decomposition $U = P(x)U + P(-x)U \equiv u_{(x)} + u_{(-x)}$ of a twistor into right-handed and left-handed spinors at $x$ is a special case of \eqref{eq:decomposition}, with the measure decomposition \eqref{eq:measure_decomposition} being simply $d^4U = d^2u_{(x)} d^2u_{(-x)}$.

The $EAdS_4$ covariant derivative $\nabla_\mu$ can now be extended to objects with Weyl spinor indices: it is again the flat derivative $\del_\mu$, followed by projecting every tensor index back into the tangent space at $x$ using $q_\mu^\nu(x) = \delta_\mu^\nu + x_\mu x^\nu$, and every spinor index back into the appropriate spinor space using $P(\pm x) = \frac{1}{2}(1\pm x)$. It will be convenient to define a covariant derivative with spinor indices as:
\begin{align}
 \hat\nabla_{ab} \equiv P^c{}_a(-x) P^d{}_b(x) \gamma^\mu_{cd} \nabla_\mu \ ,
\end{align}
so that the left-handed spinor index is forced into the first position, and right-handed one into the second position.

Now, consider a choice of \emph{boundary} point, i.e. $\xi^\mu = \ell^\mu$ with $\ell\cdot\ell = 0$. This breaks $SO(1,4)$ down to $ISO(3)$. Twistor space now acquires only \emph{one} preferred $\bbC^2$ subspace -- the space spanned by the matrix $P(\ell) = \frac{1}{2}\ell$, which is totally null under the twistor metric. From the point of view of the 3d boundary, $P(\ell)$ is the space of \emph{cospinors} at $\ell$; in particular, the measure \eqref{eq:measure} on it scales inversely with $\ell^\mu$. \emph{Spinors} at $\ell$ are represented by the quotient space $P^*(\ell)$ of twistors modulo terms in $P(\ell)$: 
\begin{align}
  (u_{(\ell)}^*)^a \cong (u^*)_{(\ell)}^a + u_{(\ell)}^a \ , \quad u_{(\ell)}^a \in P(\ell) \ . \label{eq:dual_measure}
\end{align}
The spinor metric on $P^*(\ell)$ is simply $P_{ab}(\ell) = \frac{1}{2}\ell_{ab}$, and the integration measure can be defined analogously to \eqref{eq:measure_bulk}:
\begin{align}
 d^2u_{(\ell)}^* &\equiv \frac{P_{ab}(\ell)}{2(2\pi)}\,(du_{(\ell)}^*)^a (du_{(\ell)}^*)^b \ ; \\
 d^4U &= -d^2u_{(\ell)}^* d^2u_{(\ell)} \ .
\end{align}
Multiplication by $P^a{}_b(\ell) = \frac{1}{2}\ell^a{}_b$ defines a natural isomorphism $u^*_{(\ell)}\rightarrow u_{(\ell)}$ between $P^*(\ell)$ and $P(\ell)$, which is consistent with their corresponding metrics/measures. In intrinsic boundary terms, this operation is just index-lowering from spinors into cospinors, using the spinor metric $P_{ab}(\ell)$.

Finally, if we fix \emph{two} boundary points $\ell$ and $l$, there's no longer a need to introduce quotient spaces: elements of $P(l)$ become canonical representatives for the equivalence classes in $P^*(\ell)$, and vice versa. The corresponding measures are related via:
\begin{align}
 d^2u_{(\ell)} = -\frac{2}{\ell\cdot l}\,d^2u^*_{(l)} = -\frac{l_{ab}\,du^a_{(\ell)} du^b_{(\ell)}}{4\pi(\ell\cdot l)} \ , \label{eq:measure_relationship}
\end{align}
and vice versa. If we fix the relative scaling of the null vectors $\ell^\mu,l^\mu$ as $\ell\cdot l = -\frac{1}{2}$, the decomposition \eqref{eq:decomposition}-\eqref{eq:measure_decomposition} and measure relationships \eqref{eq:measure_relationship} simplify into:
\begin{align}
 \begin{split}
   U &= u_{(\ell)} + u_{(l)} \ ; \quad u_{(\ell)} = \ell l U \in P(\ell) \ ; \quad u_{(l)} = l\ell U \in P(l) \ ; \\
   d^4U &= -\frac{1}{4}d^2u_{(\ell)}d^2u_{(l)} \ ; \quad d^2u_{(\ell)} = 4d^2u^*_{(l)} = \frac{l_{ab}\,du^a_{(\ell)} du^b_{(\ell)}}{2\pi} \ . 
 \end{split} \label{eq:twistor_boundary_decomposition}
\end{align}

\subsection{Higher-spin algebra and the Penrose transform} \label{sec:correlators:HS}

In this subsection, we introduce some fundamentals of HS algebra and the Penrose transform, in a condensed version of Sections 4, 5 and 8B of \cite{Neiman:2017mel}.

\subsubsection{HS algebra} \label{sec:correlators:HS:algebra}

Consider again the Clifford algebra $\{\gamma_\mu,\gamma_\nu\} = -2\eta_{\mu\nu}$. Higher-spin algebra is just an application of the same concept, but to a twistor $Y^a$ instead of a vector $\gamma^\mu$. Since the twistor metric is antisymmetric, we get a commutator $[Y_a,Y_b]_\star = 2iI_{ab}$ instead of an anti-commutator. Imposing a symmetric ordering convention, we arrive at the following non-commutative product:
\begin{align}
 Y^a\star Y^b = Y^a Y^b + iI^{ab} \ . \label{eq:star}
\end{align}
This can be extended to arbitrary functions (and distributions) via an integral formula:
\begin{align}
  f(Y)\star g(Y) = \int d^4U d^4V f(Y+U)\, g(Y+V)\, e^{-iUV} \ . \label{eq:star_int}
\end{align}
The HS symmetry algebra is the infinite-dimensional Lie algebra of even (i.e. integer-spin) functions $f(Y)$ with the associative product \eqref{eq:star_int}. The quadratic elements $Y_a Y_b$ generate the spacetime symmetry group $SO(1,4)$, in analogy to the role of $\gamma_{[\mu}\gamma_{\nu]}$ in Clifford algebra.

The product \eqref{eq:star_int} respects a trace operation, defined simply by:
\begin{align}
\tr_\star f(Y) = f(0) \ , \label{eq:str}
\end{align}
which satisfies $\tr_\star(f\star g) = \tr_\star(g\star f)$ for even functions $f(Y),g(Y)$. Another important object is the delta function $\delta(Y)$, as defined in \eqref{eq:delta_U}. A star product with $\delta(Y)$ implements a Fourier transform in twistor space:
\begin{align}
 f(Y)\star\delta(Y) = \int d^4U f(U)\,e^{iUY} \ ; \quad \delta(Y)\star f(Y) = \int d^4U f(U)\,e^{-iUY} \ . \label{eq:delta_Fourier}
\end{align}
$\delta(Y)$ squares to unity $\delta(Y)\star\delta(Y) = 1$, and commutes with even functions $f(Y)$. More generally, eqs. \eqref{eq:delta_Fourier} imply:
\begin{align}
 \delta(Y)\star f(Y)\star\delta(Y) = f(-Y) \ . \label{eq:delta_flip}
\end{align}

\subsubsection{Star products with spinor delta functions}

Recall that choices of spacetime points, i.e. of embedding-space vectors $\xi^\mu\in\bbR^{1,4}$, pick out some subspaces or quotient spaces of twistor space, which describe local spinors. The delta functions $\delta_\xi(Y)$ associated with these spinor spaces have the following properties under the HS star product:
\begin{align}
&f(Y)\star\delta_\xi(Y) = \int_{P(\xi)} d^2u_{(\xi)}\,f(Y+u_{(\xi)})\,e^{iu_{(\xi)}Y} \ ; \label{eq:delta_Fourier_xi_first} \\ 
&\delta_\xi(Y)\star f(Y) = \int_{P(\xi)} d^2u_{(\xi)}\,f(Y+u_{(\xi)})\,e^{-iu_{(\xi)}Y} \ ; \\
&\delta_\xi(Y)\star\delta(Y) = \delta(Y)\star\delta_\xi(Y) = \delta_{-\xi}(Y) \ . \label{eq:spinor_twistor_delta}
\end{align}
Using the techniques from section \ref{sec:correlators:geometry:spinors} for manipulating spinor spaces, one can work out the star product of a pair of spinor delta functions:
\begin{align}
 \delta_\xi(Y)\star\delta_{\xi'}(Y) = \frac{2}{\sqrt{(\xi\cdot\xi)(\xi'\cdot\xi')} - \xi\cdot\xi'}\,\exp\left(\frac{-iY\xi\xi'Y/2}{\sqrt{(\xi\cdot\xi)(\xi'\cdot\xi')} - \xi\cdot\xi'}\right) \ . \label{eq:general_two_point}
\end{align}
The most efficient way to derive this result is to first express the product as a single spinor integral using eq. \eqref{eq:delta_Fourier_xi_first}, and then evaluate that integral using eq. \eqref{eq:spinor_delta_integral}.

Star products with any additional delta functions will continue to result in Gaussians. A particularly simple case is when all points are on the boundary, i.e. when the vectors $(\xi^\mu,\xi'^\mu,\dots)$ are null $(\ell^\mu,\ell'^\mu,\dots)$. Here, the 3-point product reduces to the 2-point product:
\begin{align}
 \delta_\ell(Y)\star\delta_{\ell'}(Y) &= -\frac{2}{\ell\cdot\ell'}\exp\frac{iY\ell\ell' Y}{2\ell\cdot\ell'} \ ; \label{eq:2_pt_product} \\
 \delta_\ell(Y)\star\delta_{\ell'}(Y)\star\delta_{\ell''}(Y) &= \pm i \sqrt{-\frac{\ell\cdot\ell''}{2(\ell\cdot\ell')(\ell'\cdot\ell'')}} \, \delta_\ell(Y)\star\delta_{\ell''}(Y) \ , \label{eq:3_pt_product}
\end{align}
where we note that the square root as written is real, i.e. the quantity inside it is positive. Again, the best way to evaluate the product \eqref{eq:3_pt_product} is by first expressing it as a single spinor integral using eq. \eqref{eq:delta_Fourier_xi_first}. The resulting integral this time is a Gaussian one, which can be evaluated using eq. \eqref{eq:Gaussian_spinor}. The Gaussian integral \eqref{eq:Gaussian_spinor} in this case is over the spinor space $P(\ell'')$, with the quadratic form $A$ taken directly from the 2-point product \eqref{eq:2_pt_product}:
\begin{align}
 A_{ab} = \frac{i(\ell\ell' - \ell'\ell)_{ab}}{2\ell\cdot\ell'} \ . \label{eq:A}
\end{align} 
The 3-point product \eqref{eq:3_pt_product} is thus the result of an imaginary Gaussian integral over a complex spinor space. This is the reason for the sign ambiguity in \eqref{eq:3_pt_product}, which will be crucial in section \ref{sec:signs} below.

Applying \eqref{eq:3_pt_product} recursively, we arrive at a closed-form expression for the $n$-point product:
\begin{align}
 \delta_{\ell_1}(Y)\star\delta_{\ell_2}(Y)\star\ldots\star\delta_{\ell_n}(Y) = \frac{4(\pm i)^{n-2}}{\sqrt{\prod_{p=1}^n (-2\ell_p\cdot\ell_{p+1})}}\, \exp\frac{iY\ell_1\ell_n Y}{2\ell_1\cdot\ell_n} \ , \label{eq:n_pt_product}
\end{align}
where the product inside the square root is understood cyclically, i.e. $\ell_{n+1}\equiv \ell_1$. 

Also useful are the generalizations of \eqref{eq:general_two_point} where one of the delta functions is either shifted, or multiplied by an exponent:
\begin{align}
 \begin{split}
  \delta_\xi(Y-M)\star\delta_{\xi'}(Y) &= \frac{2}{\sqrt{(\xi\cdot\xi)(\xi'\cdot\xi')} - \xi\cdot\xi'}\,\exp\left(\frac{2iMP(\xi)P(\xi')Y - iY\xi\xi'Y/2}{\sqrt{(\xi\cdot\xi)(\xi'\cdot\xi')} - \xi\cdot\xi'}\right) \ ; \\
  \left(\delta_\xi(Y)e^{iMY}\right)\star\delta_{\xi'}(Y) &= \frac{2}{\sqrt{(\xi\cdot\xi)(\xi'\cdot\xi')} - \xi\cdot\xi'}\,\exp\left(\frac{2iMP(-\xi)P(-\xi')Y - iY\xi\xi'Y/2}{\sqrt{(\xi\cdot\xi)(\xi'\cdot\xi')} - \xi\cdot\xi'}\right) \ .
  \end{split} \label{eq:shifted_2pt_product} 
\end{align}

\subsubsection{The Penrose transform}

The Penrose transform relates twistor functions $F(Y)$ to solutions of the free massless field equations (of all spins) in the $EAdS_4$ bulk. We again restrict to integer spins, i.e. to even twistor functions $F(-Y) = F(Y)$. In HS language, the Penrose transform is just a star product of the form \eqref{eq:delta_Fourier_xi_first}:
\begin{align}
 C(x;Y) = iF(Y)\star\delta_x(Y) \ , \label{eq:Penrose}
\end{align}
where we recall that $\delta_x(Y)$ is the delta function on the right-handed Weyl spinor space at $x$. More explicitly, using \eqref{eq:delta_Fourier_xi_first} and shifting the integration variable, the star product in \eqref{eq:Penrose} decomposes $Y$ into a right-handed spinor $y_{(x)}\in P(x)$ and a left-handed one $y_{(-x)}\in P(-x)$, and Fourier-transforms $y_{(x)}$ while leaving $y_{(-x)}$ untouched:
\begin{align}
 C(x;y_{(x)}+y_{(-x)}) = i\int_{P(x)} d^2u_{(x)}\,F(u_{(x)}+y_{(-x)})\,e^{iu_{(x)} y_{(x)}}  \ . \label{eq:Penrose_explicit}
\end{align}
 The twistor function $F(Y)$ in \eqref{eq:Penrose} has no spacetime dependence whatsoever. Through the transform \eqref{eq:Penrose}, it is converted into an $x$-dependent \emph{master field} $C(x;Y)$, which encodes the higher-spin field strengths at $x$, as well as their derivatives. The factor of $i$ follows the conventions of \cite{Neiman:2017mel}. The transform \eqref{eq:Penrose} is easy to invert, since, as a special case of \eqref{eq:general_two_point}, we have:
\begin{align}
 \delta_x(Y)\star\delta_x(Y) = 1 \ .
\end{align}
Let's now describe how the various bulk master fields are encoded inside $C(x;Y)$. The right-handed spin-$s$ field strength at $x$ is an object $C^{(s)}_{a_1\dots a_{2s}}(x)$ with $2s$ totally symmetrized spinor indices, all of which lie in the right-handed spinor space $P(x)$. Similarly, the left-handed field strength $C^{(-s)}_{a_1\dots a_{2s}}(x)$ has $2s$ totally symmetrized indices in the left-handed spinor space $P(-x)$. For spin 0, there is no right-handed/left-handed distinction, and we have just the scalar field $C^{(0)}(x)$. To extract these field strengths from the master field $C(x;Y)$, we first decompose the twistor argument $Y=y_{(x)}+y_{(-x)}$ into its Weyl-spinor pieces $y_{(x)}\in P(x)$ and $y_{(-x)}\in P(-x)$. The field strengths are now encoded as the $2s$'th powers of $y_{(x)}$ and $y_{(-x)}$, respectively:
\begin{align}
 \begin{split}
   C(x;y_{(x)}) &= \sum_{s=0}^\infty \frac{1}{(2s)!}\,C^{(s)}_{a_1\dots a_{2s}}(x)\, y_{(x)}^{a_1}\dots y_{(x)}^{a_{2s}} \ ; \\
   C(x;y_{(-x)}) &= \sum_{s=0}^\infty \frac{1}{(2s)!}\,C^{(-s)}_{a_1\dots a_{2s}}(x)\, y_{(-x)}^{a_1}\dots y_{(-x)}^{a_{2s}} \ .
 \end{split} \label{eq:packaging}
\end{align}
The right-handed and left-handed gauge field strengths \eqref{eq:packaging} can be combined into a field-strength tensor for each spin, via:
\begin{align}
  C^{(s)}_{\mu_1\nu_1\cdots\mu_s\nu_s}(x) = \frac{1}{4^s}\,\gamma_{\mu_1\nu_1}^{a_1 b_1}\dots \gamma_{\mu_s\nu_s}^{a_s b_s}\left( C^{(s)}_{a_1 b_1\dots a_s b_s}(x) + C^{(-s)}_{a_1 b_1\dots a_s b_s}(x) \right) \ ,
  \label{eq:Weyl_tensor}
\end{align}
while $C^{(0)}(x) = C(x;0)$ is the spin-0 field. Eq. \eqref{eq:Weyl_tensor} describes a Maxwell field strength for $s=1$, a linearized Weyl tensor for $s=2$, and their generalizations for higher spins. The Penrose transform \eqref{eq:Penrose} automatically ensures that these fields satisfy the free massless field equations in $EAdS_4$:
\begin{align}
 \begin{split}
   s=0: \quad &\nabla_\mu\nabla^\mu C^{(0)}(x) = -2C^{(0)}(x) \ ; \\
   s=1: \quad &\nabla^\mu C^{(1)}_{\mu\nu}(x) = \nabla_{[\mu} C^{(1)}_{\nu\rho]}(x) = 0 \ ; \\
   s\geq 2: \quad &\nabla^{\mu_1} C^{(s)}_{\mu_1\nu_1\cdots\mu_s\nu_s}(x) = 0 \ .
 \end{split}
\end{align}
Note that the field equation for $C^{(0)}(x)$ is that of a \emph{conformally} massless scalar. 

Finally, those Taylor coefficients of $C(x;Y)$ which don't appear in \eqref{eq:packaging} are identified via the Penrose transform as derivatives of the fields \eqref{eq:packaging}:
\begin{align}
\begin{split}
  \left.\frac{\del^{2(s+k)} C(x;Y)}{\del y_{(x)}^{a_1}\dots\del y_{(x)}^{a_{2s+k}}\del y^{(-x)}_{b_1}\dots\del y^{(-x)}_{b_k}}\right|_{Y=0} 
     &= i^k\, \hat\nabla^{(b_1}{}_{(a_1}\dots\hat\nabla^{b_k)}{}_{a_k} C^{(s)}_{a_{k+1}\dots a_{2s+k})}(x) \ ; \\
  \left.\frac{\del^{2(s+k)} C(x;Y)}{\del y_{(x)}^{a_1}\dots\del y_{(x)}^{a_k}\del y^{(-x)}_{b_1}\dots\del y^{(-x)}_{b_{2s+k}}}\right|_{Y=0} 
     &= i^k\, \hat\nabla^{(b_1}{}_{(a_1}\dots\hat\nabla^{b_k}{}_{a_k)} C_{(-s)}^{\,b_{k+1}\dots b_{2s+k})}(x) \ .
\end{split} \label{eq:unfolding}
\end{align}
Knowing the master field at one point, we can calculate it at any other, using either the field equations or the Penrose transform \eqref{eq:Penrose}. A particularly simple case is the master field at the \emph{antipodal} point $-x^\mu$, which takes the form:
\begin{align}
 C(-x;Y) = C(x;Y)\star\delta(Y) \ , \label{eq:C_antipodal}
\end{align}
thanks to the star-product identity \eqref{eq:spinor_twistor_delta}. Recalling eq. \eqref{eq:delta_Fourier}, we conclude that the antipodal map $x\rightarrow -x$ acts on a master field $C(x;Y)$ as a Fourier transform in $Y^a$.

\subsection{Boundary correlators} \label{sec:correlators:correlators}

In this section, we construct the HS-algebraic expression for the $n$-point correlators of the free boundary CFT.

\subsubsection{The CFT picture}

Consider the free vector model on the 3d boundary, with the currents $J^{(s)}_{k_1\dots k_s}$ from \eqref{eq:explicit_J} coupled linearly to external sources $A^{(s)}_{k_1\dots k_s}$. We can upgrade the indices on $J^{(s)}$ and $A^{(s)}$ from 3d indices $k_1\dots k_s$ into $\bbR^{1,4}$ indices $\mu_1\dots\mu_s$, keeping in mind the orthogonality to $\ell^\mu$ and the equivalence under $\ell^\mu$-proportional shifts. The action of the vector model can now be written as:
\begin{align}
 S_{\text{CFT}} = -\int d^3\ell\,\bar\phi_I\Box\phi^I - \int d^3\ell \sum_{s=0}^\infty A^{(s)}_{\mu_1\dots\mu_s}(\ell)\, J_{(s)}^{\mu_1\dots\mu_s}(\ell) \ . \label{eq:S_local}
\end{align}
The fields $\phi^I,\bar\phi_I$ have conformal weight $\frac{1}{2}$, and $\Box$ is the conformal Laplacian. The internal index $I$ takes values from $1\dots N$. The fields $\phi^I$ and $\bar\phi_I$ have Bose statistics, but switch to Fermi statistics if we analytically continue to a $dS_4$ bulk \cite{Anninos:2011ui}. In \eqref{eq:S_local}, we are assuming a $U(N)$ vector model, with $\phi^I$ and $\bar\phi_I$ linearly independent; in this case, the sum is over all integer spins $s$. If $\phi^I$ and $\bar\phi_I$ are not independent but linearly related, then only the even-spin currents are non-vanishing. The internal symmetry is then $O(2N)$, or, in the $dS_4$ case, $Sp(2N)$, where we redefined the range of $I$ as $1\dots 2N$ for the sake of uniformity of the results below.

For notational efficiency, we will package the (symmetric, traceless) currents $J^{(s)}_{\mu_1\dots\mu_s}(\ell)$ into their scalar contractions with a null boundary vector $\lambda^\mu$:
\begin{align}
 J^{(s)}(\ell,\lambda) = \lambda^{\mu_1}\dots\lambda^{\mu_s} J^{(s)}_{\mu_1\dots\mu_s}(\ell) \ , \label{eq:J_contracted}
\end{align}
where $\lambda\cdot\lambda = \lambda\cdot\ell = 0$. 

In a flat conformal frame, the propagator of the fundamental field $\phi^I$ is $G = \Box^{-1} = -1/(4\pi r)$. More covariantly, we can write:
\begin{align}
 G(\ell,\ell') = -\frac{1}{4\pi\sqrt{-2\ell\cdot\ell'}} \ . \label{eq:G}
\end{align}
The connected $n$-point correlators of the free vector model consist of 1-loop Feynman diagrams. For the spin-0 operator $J^{(0)} = \bar\phi_I\phi^I$, the correlators read:
\begin{align}
 \left<J^{(0)}(\ell_1)\dots J^{(0)}(\ell_n)\right>_{\text{connected}} = N(-1)^n \left(\prod_{p=1}^n G(\ell_p,\ell_{p+1}) + \text{permutations}\right) \ . \label{eq:correlators_scalar}
\end{align}
Here, the product is cyclic, i.e. $\ell_{n+1}\equiv\ell_1$, and ``+ permutations'' denotes a sum over the $(n-1)!$ cyclically inequivalent permutations of $(\ell_1,\dots,\ell_n)$.

The correlators of currents $J^{(s)}$ with $s>0$ are morally similar, with the addition of some derivatives acting on the propagators, after the pattern of the derivatives in \eqref{eq:explicit_J}. We can encapsulate them all at once by switching temporarily to a bilocal formalism \cite{Das:2003vw,Douglas:2010rc}, where the $J^{(s)}$ are replaced by bilocal scalar operators $\bar\phi_I(\ell')\phi^I(\ell)\equiv\calO(\ell,\ell')$, coupled to sources $\Pi(\ell',\ell)$:
\begin{align}
 S_{\text{CFT}}[\Pi(\ell',\ell)] = -\int d^3\ell\,\bar\phi_I\Box\phi^I - \int d^3\ell' d^3\ell\,\bar\phi_I(\ell')\Pi(\ell',\ell)\phi^I(\ell) \ . \label{eq:S_bilocal}
\end{align}
The correlators now read:
\begin{align}
 \left<\calO(\ell_1,\ell_1')\dots \calO(\ell_n,\ell_n')\right>_{\text{connected}} = N(-1)^n \left(\prod_{p=1}^n G(\ell_p',\ell_{p+1}) + \text{permutations}\right) \ , \label{eq:correlators_bilocal}
\end{align}
where the product is again cyclic, and the sum is again over cyclically inequivalent permutations of $(1,\dots,n)$. The correlators for the local currents can be extracted from \eqref{eq:correlators_bilocal} by plugging in eqs. \eqref{eq:explicit_J},\eqref{eq:J_contracted}, which we can express covariantly as a differential operator $\calD^{(s)}$ acting on $\calO(\ell,\ell')$:
\begin{align}
 \begin{split}
  J^{(s)}(\ell,\lambda) &= \calD^{(s)}\!\left[\calO(\ell,\ell')\right] \\
   &\equiv i^s\lambda^{\mu_1}\dots\lambda^{\mu_s}\left.
        \sum_{m=0}^s (-1)^m \binom{2s}{2m} \frac{\del}{\del\ell^{(\mu_1}}\dots\frac{\del}{\del\ell^{\mu_m}} \frac{\del}{\del\ell'^{\mu_{m+1}}}\dots\frac{\del}{\del\ell'^{\mu_s)}} \calO(\ell,\ell') \right|_{\ell'=\ell} \ . 
 \end{split} \label{eq:J_embedding}
\end{align}
Note that the switch from 3d flat derivatives in \eqref{eq:explicit_J} to $\bbR^{1,4}$ flat derivatives in \eqref{eq:J_embedding} is legitimate. In particular, derivatives $\lambda^\mu\del_\mu$ never take $\ell,\ell'$ off the $\bbR^{1,4}$ lightcone, while the conformal weight of $\calO(\ell,\ell')$ and the combinatorial factors in \eqref{eq:J_embedding} conspire to make the result invariant under shifts $\lambda^\mu\rightarrow \lambda^\mu + \alpha\ell^\mu$.

\subsubsection{The twistor picture: spin 0} \label{sec:correlators:correlators:scalar}

Let us now translate the CFT correlators into HS-algebraic language. The approach is first to replace the CFT sources with HS algebra elements, i.e. twistor functions, and then to combine them into HS invariants. In the literature, this has been done from both boundary \cite{Gelfond:2008ur,Gelfond:2013xt} and bulk \cite{Giombi:2009wh,Colombo:2012jx,Didenko:2012tv} starting points. In our present framework of embedding space and spacetime-independent twistors, this procedure has been carried out \cite{Neiman:2017mel} for \emph{bilocal} sources, as in eq. \eqref{eq:S_bilocal}. Here, we will complete the picture by presenting the dictionary from \emph{local} boundary sources to the spacetime-independent twistor language. We will begin with spin 0. The lessons of the spin-0 case will enable us to clarify the bilocal results of \cite{Neiman:2017mel}, which we will then use to construct the dictionary for nonzero spins.

Traditionally, twistor functions are more closely associated with bulk fields (via the Penrose transform) than with boundary quantities. Thus, it's helpful to think in the following way: the twistor functions we're looking for are the \emph{Penrose transform of the boundary-to-bulk propagators} which correspond to the boundary sources (here, we are basically following the logic in \cite{Didenko:2012tv}, though the authors there stopped at bulk master fields, not taking the final step of the Penrose transform \eqref{eq:Penrose}). 

It's easy to guess the twistor function that should correspond to a spin-0 operator at a boundary point $\ell$. It must be a multiple of $\delta_\ell(Y)$ -- the unique twistor function that depends just on $\ell$. To apply the Penrose transform, we will need the star product formula:
\begin{align}
 \delta_\ell(Y)\star\delta_x(Y) = -\frac{2}{\ell\cdot x}\exp\frac{iY\ell x Y}{2\ell\cdot x} \ , \label{eq:ell_x}
\end{align}
which is a special case of \eqref{eq:general_two_point}. Thus, the spin-0 bulk field $C^{(0)}(x)$ corresponding to $\delta_\ell(Y)$ is $-2i/(\ell\cdot x)$, which is indeed (up to an imaginary prefactor) the boundary-to-bulk propagator for the conformally massless scalar $C^{(0)}(x)$. Now, following \cite{Colombo:2012jx,Didenko:2012tv}, let us combine a sequence of these twistor functions into the unique HS-invariant trace:
\begin{align}
 \tr_\star\big(\delta_{\ell_1}(Y)\star\ldots\star \delta_{\ell_n}(Y) \big) = \frac{4(\pm i)^{n-2}}{\sqrt{\prod_{p=1}^n (-2\ell_p\cdot\ell_{p+1})}} = 4(\pm i)^{n-2}(-4\pi)^n\prod_{p=1}^n G(\ell_p,\ell_{p+1}) \ , \label{eq:trace_n_point_raw}
\end{align}
which follows directly from \eqref{eq:n_pt_product} and \eqref{eq:G}. Crucially, this matches (up to normalization) with the Feynman diagrams encoded in the correlator \eqref{eq:correlators_scalar}. To remove the $n$-dependent normalization factors in \eqref{eq:trace_n_point_raw}, we define the twistor function corresponding to a spin-0 insertion at $\ell$ as:
\begin{align}
 \kappa^{(0)}(\ell;Y) = \pm\frac{i}{4\pi}\delta_\ell(Y) \ . \label{eq:kappa_0}
\end{align}
The corresponding scalar bulk field and $n$-point invariant then read:
\begin{align}
 \tr_\star(i\kappa^{(0)}(\ell;Y)\star\delta_x(Y)) &= \pm\frac{1}{2\pi(\ell\cdot x)} \ ; \label{eq:scalar_boundary_bulk} \\
 \tr_\star \big(\kappa^{(0)}(\ell_1;Y)\star\ldots\star\kappa^{(0)}(\ell_n;Y) \big) &= -4\prod_{p=1}^n G(\ell_p,\ell_{p+1}) \ . \label{eq:trace_n_point}
\end{align}
We can think of the star products in \eqref{eq:trace_n_point} as drawing the propagators $G(\ell_p,\ell_{p+1})$ in the Feynman diagram \emph{one at a time}, via the identity \eqref{eq:3_pt_product}. Plugging \eqref{eq:trace_n_point} into \eqref{eq:correlators_scalar}, we obtain the spin-0 correlators in HS-algebraic form:
\begin{align}
 \begin{split}
   \left<J^{(0)}(\ell_1)\dots J^{(0)}(\ell_n)\right>_{\text{connected}} = \frac{N}{4}(-1)^{n+1} \Big(&\tr_\star\!\big(\kappa^{(0)}(\ell_1;Y)\star\ldots\star\kappa^{(0)}(\ell_n;Y) \big) \\
     &+ \text{permutations}\Big) \ . 
 \end{split} \label{eq:HS_correlators_scalar}
\end{align} 

\subsubsection{From spin 0 to bilocals to all spins} \label{sec:correlators:correlators:spin}

Starting from \eqref{eq:trace_n_point}, we can quickly recover one of the main results of the \emph{bilocal} approach in \cite{Neiman:2017mel}. The key observation is that the Feynman diagrams \eqref{eq:correlators_bilocal} with $n$ bilocal insertions are just like diagrams with $2n$ local spin-0 insertions, as in \eqref{eq:correlators_scalar}, but with the $n$ propagators $G(\ell_p,\ell'_p)$ removed. Thus, if we define the twistor function corresponding to a bilocal source as:
\begin{align}
 K(\ell,\ell';Y) = \frac{\kappa^{(0)}(\ell;Y)\star\kappa^{(0)}(\ell';Y)}{G(\ell,\ell')} = \frac{1}{\pi\sqrt{-2\ell\cdot\ell'}}\exp\frac{iY\ell\ell' Y}{2\ell\cdot\ell'} \ , \label{eq:K_bilocal}
\end{align}
then, by virtue of \eqref{eq:trace_n_point}, the HS invariants constructed from these functions immediately evaluate to the Feynman diagrams from \eqref{eq:correlators_bilocal}, up to the same factor of $-4$ as in \eqref{eq:trace_n_point}:
\begin{align}
 \tr_\star\left(K(\ell_1,\ell'_1;Y)\star\dots\star K(\ell_n,\ell'_n;Y) \right) = -4\prod_{p=1}^n G(\ell'_p,\ell_{p+1}) \ . \label{eq:trace_bilocal}
\end{align}
Here again, each individual propagator can be attributed to a star product, according to:
\begin{align}
 K(\ell_1,\ell'_1;Y)\star K(\ell_2,\ell'_2;Y) &= G(\ell_2,\ell_1')\,K(\ell_1,\ell'_2;Y) \ ; \label{eq:K_K} \\
 \tr_\star K(\ell,\ell';Y) &= -4G(\ell,\ell') \ . \label{eq:tr_K}
\end{align}
Thus, the correlators \eqref{eq:correlators_bilocal} of bilocal operators take the same form as the spin-0 local correlators \eqref{eq:HS_correlators_scalar}:
\begin{align}
 \begin{split}
   \left<\calO(\ell_1,\ell'_1)\dots\calO(\ell_n,\ell'_n)\right>_{\text{connected}} = \frac{N}{4}(-1)^{n+1} \Big(&\tr_\star\!\big(K(\ell_1,\ell'_1;Y)\star\ldots\star K(\ell_n,\ell'_n;Y) \big) \\
     &+ \text{permutations}\Big) \ . 
 \end{split} \label{eq:HS_correlators_bilocal}
\end{align} 
Now, recall that correlators for the local currents of all spins can be extracted from the bilocal correlators, via eq. \eqref{eq:J_embedding}. Therefore, the correlators for all spins can be expressed in HS-algebraic form as:
\begin{align}
 \begin{split}
   &\left<J^{(s_1)}(\ell_1,\lambda_1)\dots J^{(s_n)}(\ell_n,\lambda_n)\right>_{\text{connected}} \\
   &\quad = \frac{N}{4}(-1)^{n+1} \Big(\tr_\star\!\big(\kappa^{(s_1)}(\ell_1,\lambda_1;Y)\star\ldots\star\kappa^{(s_n)}(\ell_n,\lambda_n;Y) \big) + \text{permutations}\Big) \ ,
 \end{split} \label{eq:HS_correlators_spins}
\end{align}
where the twistor function $\kappa^{(s)}(\ell,\lambda;Y)$, describing an insertion of $J^{(s)}(\ell,\lambda)$, is given by:
\begin{align}
 \kappa^{(s)}(\ell,\lambda;Y) = D^{(s)}\!\left[K(\ell,\ell';Y)\right] \ , \label{eq:kappa_from_K}
\end{align}
with $D^{(s)}$ the differential operator from \eqref{eq:J_embedding}. It remains to evaluate eq. \eqref{eq:kappa_from_K}, and obtain the explicit form of $\kappa^{(s)}(\ell,\lambda;Y)$. It is clear that the answer should be the Penrose transform of a spin-$s$ boundary-to-bulk propagator. Unfortunately, it's difficult to evaluate \eqref{eq:kappa_from_K} directly, because the $\ell=\ell'$ limit of $K(\ell,\ell';Y)$ is very singular. Instead, we will first take the Penrose transform of $K(\ell,\ell';Y)$, apply $D^{(s)}$ to the resulting bulk fields, and finally Penrose-transform back into twistor space. The Penrose transform of $K(\ell,\ell';Y)$ was evaluated in \cite{Neiman:2017mel}, as the master field:
\begin{align}
  C_{\ell,\ell'}(x;Y) \equiv iK(\ell,\ell';Y)\star\delta_x(Y) = \frac{\mp 1}{\pi\sqrt{2[\ell\cdot\ell' + 2(\ell\cdot x)(\ell'\cdot x)]}}\exp\frac{iY[\ell\ell' + 2(\ell'\cdot x)\ell x] Y}{2[\ell\cdot\ell' + 2(\ell\cdot x)(\ell'\cdot x)]} \ ,
 \label{eq:bilocal_master_raw}
\end{align}
where we again have a sign ambiguity due to a Gaussian spinor integral. Now, let us notice that the derivatives in \eqref{eq:J_embedding} only probe pairs of points $(\ell,\ell')$ that are null-separated, which translates in embedding space into $\ell\cdot\ell' = 0$. Specializing to this case, the master field \eqref{eq:bilocal_master_raw} becomes:
\begin{align}
 C_{\ell,\ell'}(x;Y) = \frac{\mp 1}{2\pi\sqrt{(\ell\cdot x)(\ell'\cdot x)}}\exp\frac{iY\ell\ell'Y}{4(\ell\cdot x)(\ell'\cdot x)}\exp\frac{iY\ell xY}{2(\ell\cdot x)} \ . \label{eq:bilocal_master}
\end{align}
The bulk field strengths \eqref{eq:packaging} are contained in the evaluation of \eqref{eq:bilocal_master} on a purely right-handed or purely left-handed argument $Y = y_{\pm x} \in P(\pm x)$:
\begin{align}
 \begin{split}
   C_{\ell,\ell'}(x;y_{(x)}) &= \frac{\mp 1}{2\pi\sqrt{(\ell\cdot x)(\ell'\cdot x)}}\exp\frac{iy_{(x)} \ell\ell' y_{(x)}}{4(\ell\cdot x)(\ell'\cdot x)} \ ; \\
   C_{\ell,\ell'}(x;y_{(-x)}) &= \frac{\mp 1}{2\pi\sqrt{(\ell\cdot x)(\ell'\cdot x)}}\exp\frac{iy_{(-x)} \ell\ell' y_{(-x)}}{4(\ell\cdot x)(\ell'\cdot x)} \ .
 \end{split} \label{eq:bilocal_field_strengths}
\end{align}
Note that the last exponent in \eqref{eq:bilocal_master} disappeared: it contributes only to the field strengths' derivatives \eqref{eq:unfolding}, not to the field strengths themselves. We are now ready to apply the differential operators $D^{(s)}$. It is convenient to Taylor-expand the exponent in \eqref{eq:bilocal_field_strengths}, which directly corresponds to an expansion in spins:
\begin{align}
  C_{\ell,\ell'}(x;y_{(x)}) = \frac{\mp 1}{2\pi\sqrt{(\ell\cdot x)(\ell'\cdot x)}} \sum_{j=0}^\infty \frac{1}{j!} \left(\frac{iy_{(x)} \ell\ell' y_{(x)}}{4(\ell\cdot x)(\ell'\cdot x)}\right)^j \ , \label{eq:bilocal_field_strengths_Taylor}
\end{align}
and similarly for the left-handed part $C_{\ell,\ell'}(x;y_{(-x)})$. We expect that only the terms with $j=s$ will contribute to $D^{(s)}\!\left[C_{\ell,\ell'}(x;y_{(\pm x)})\right]$, and this is easy to check explicitly. First, note that at $\ell=\ell'$, the matrix products $\ell\ell'$ vanish. Thus, they can only survive if first acted on by the derivatives in $D^{(s)}$. Moreover, each factor of $\ell\ell'$ must be acted on by \emph{exactly one} derivative, since two derivatives (when contracted with $\lambda^\mu$) will turn it into $\lambda^2$, which again vanishes. This means that only terms with $j\leq s$ can contribute, and in those, $j$ of the $s$ derivatives will act on $\ell\ell'$ factors, and the other $s-j$ will act on $(\ell\cdot x)$ or $(\ell'\cdot x)$ factors. Therefore, without keeping track of the combinatorial details in \eqref{eq:J_embedding}, we can write $D^{(s)}C_{\ell,\ell'}(x;y_{(x)})$ as:
\begin{align}
 D^{(s)}\!\left[C_{\ell,\ell'}(x;y_{(x)})\right] = \sum_{j=0}^s c_{s,j}\, \frac{(y_{(x)} \ell\lambda\, y_{(x)})^j(\lambda\cdot x)^{s-j}}{(\ell\cdot x)^{s+j+1}} \ , \label{eq:D_C_sum}
\end{align}
for some numerical coefficients $c_{s,j}$ (and similarly for the left-handed field strengths). Now, recall that the combinatorics in \eqref{eq:J_embedding} is arranged such that the result is invariant under $\lambda^\mu \rightarrow \lambda^\mu + \alpha\ell^\mu$. The RHS of \eqref{eq:D_C_sum} can satisfy this only if, as expected, the only term with a nonzero coefficient is the one with $j=s$. Finally taking into account all numerical factors, the result reads:
\begin{align}
 s=0: \qquad &D^{(0)}\!\left[C_{\ell,\ell'}(x;y_{(x)})\right] = D^{(0)}\!\left[C_{\ell,\ell'}(x;y_{(-x)})\right] = \pm \frac{1}{2\pi(\ell\cdot x)} \ ; \\
 \begin{split}
   s>0: \qquad &D^{(s)}\!\left[C_{\ell,\ell'}(x;y_{(x)})\right] = \pm \frac{(-1)^s}{4\pi} \frac{(y_{(x)} \ell\lambda\, y_{(x)})^s}{(\ell\cdot x)^{2s+1}} \ ; \\
   &D^{(s)}\!\left[C_{\ell,\ell'}(x;y_{(-x)})\right] = \pm \frac{(-1)^s}{4\pi} \frac{(y_{(-x)} \ell\lambda\, y_{(-x)})^s}{(\ell\cdot x)^{2s+1}} \ .
 \end{split} \label{eq:D_C_raw}
\end{align}
We can simplify further by noticing that $\ell^\mu$ and $\lambda^\mu$ span a totally null bivector, which can be re-expressed as the square of a twistor $m^a\in P(\ell)$, i.e. of a boundary cospinor at $\ell$:
\begin{align}
 m^a m^b = \gamma_{\mu\nu}^{ab}\ell^\mu\lambda^\nu = (\ell\lambda)^{ab} \ . \label{eq:m}
\end{align}
The field strengths \eqref{eq:D_C_raw} then become:
\begin{align}
 \begin{split}
   s>0: \qquad &D^{(s)}\!\left[C_{\ell,\ell'}(x;y_{(x)})\right] = \pm \frac{1}{4\pi} \frac{(m y_{(x)})^{2s}}{(\ell\cdot x)^{2s+1}} \ ; \\
   &D^{(s)}\!\left[C_{\ell,\ell'}(x;y_{(-x)})\right] = \pm \frac{1}{4\pi} \frac{(m y_{(-x)})^{2s}}{(\ell\cdot x)^{2s+1}} \ . 
 \end{split}
\end{align}
As expected, these are the field strengths of a spin-$s$ boundary-to-bulk propagator. Together with their derivatives, they combine into the master field:
\begin{align}
 D^{(s)}\!\left[C_{\ell,\ell'}(x;Y)\right] = \pm \frac{1}{4\pi} \frac{(m P_x Y)^{2s} + (mP_{-x} Y)^{2s}}{(\ell\cdot x)^{2s+1}}\exp\frac{iY\ell xY}{2(\ell\cdot x)} \ , \label{eq:D_C}
\end{align}
where the $s=0$ case is now included. Note the reappearance of the last exponent from \eqref{eq:bilocal_master}, to encode the fields' derivatives. This master field for boundary-to-bulk propagators was first written down in an intrinsic bulk formalism in \cite{Giombi:2009wh}, and in our present embedding-space formalism in \cite{Neiman:2014npa}. 

To find the Penrose transform of \eqref{eq:D_C}, we must rewrite the polarization twistor $m^a\in P(\ell)$ as:
\begin{align}
 m^a = \ell^a{}_b M^b \ . \label{eq:M}
\end{align}
Here, $M^a$ is an arbitrary twistor; however, since it enters only through its product with $\ell^a{}_b$, it is really an element of the quotient space $P^*(\ell)$, i.e. a spinor at $\ell$. In intrinsic boundary terms, $M^a$ is just the raised-index version of the cospinor $m^a$, with an extra factor of 2. In terms of $M^a$, the master field \eqref{eq:D_C} becomes:
\begin{align}
  D^{(s)}\!\left[C_{\ell,\ell'}(x;Y)\right] = \pm \frac{1}{4\pi} \frac{(M\ell P_x Y)^{2s} + (M\ell P_{-x} Y)^{2s}}{(\ell\cdot x)^{2s+1}}\exp\frac{iY\ell xY}{2(\ell\cdot x)} \ . \label{eq:D_C_M}
\end{align}
We can now find the inverse Penrose transform of this master field, by using a special case of the star-product formulas \eqref{eq:shifted_2pt_product}:
\begin{align}
 \delta_\ell(Y+M)\star\delta_x(Y) &= -\frac{2}{\ell\cdot x} \exp\left(\frac{iM\ell P(x)Y}{\ell\cdot x}\right) \exp\frac{iY\ell x Y}{2\ell\cdot x} \ ; \label{eq:shifted_ell} \\
 \left(e^{iMY}\delta_\ell(Y)\right)\star\delta_x(Y) &= -\frac{2}{\ell\cdot x} \exp\left(\frac{iM\ell P(-x)Y}{\ell\cdot x}\right) \exp\frac{iY\ell x Y}{2\ell\cdot x} \ . \label{eq:multiplied_ell}
\end{align}
Taylor-expanding in $M^a$ and comparing with \eqref{eq:D_C_M}, we conclude that the master field \eqref{eq:D_C_M} is the Penrose transform of the following twistor function:
\begin{align}
 \kappa^{(s)}(\ell,\lambda;Y) = \pm \frac{iM^{a_1}\dots M^{a_{2s}}}{8\pi} \left(Y_{a_1}\dots Y_{a_{2s}} + (-1)^s\frac{\del^{2s}}{\del Y^{a_1}\dots\del Y^{a_{2s}}}\right) \delta_\ell(Y) \ , \label{eq:kappa_s}
\end{align}
where we recall that the polarization twistor $M^a$ is a square root of the null polarization vector $\lambda^\mu$, in the sense defined by \eqref{eq:m},\eqref{eq:M}:
\begin{align}
 \gamma_{\mu\nu}^{ab}\ell^\mu\lambda^\nu = (\ell M)^a(\ell M^b) \ .
\end{align}
This completes the formulation \eqref{eq:HS_correlators_spins} of all the $n$-point current correlators in terms of HS algebra. Note that substituting $s=0$ in \eqref{eq:kappa_s} recovers the spin-0 twistor function \eqref{eq:kappa_0} from which we started. The star products of $\kappa^{(s)}$ in \eqref{eq:HS_correlators_spins} are similar to those of $\kappa^{(0)}$ in \eqref{eq:trace_n_point}, but with additional polynomial factors. These star products are guaranteed to correctly reproduce the spin-$s$ correlators, thanks to our construction \eqref{eq:kappa_from_K} of $\kappa^{(s)}(\ell;Y)$ from the bilocal $K(\ell,\ell';Y)$.

\subsubsection{Fixing the sign factors}

Let's now address the sign ambiguities that we've been carrying along throughout this section. As we pointed out, these arise from imaginary Gaussian integrals over complex spinor spaces, which come up in the evaluation of certain star products. As discussed in \cite{Neiman:2017mel}, these sign ambiguities can never be resolved without some inconsistency. In particular, they can't be resolved consistently with the topology of the spacetime symmetry group $SO(1,4)$. In other words, once we allow a sufficiently broad class of twistor functions (in particular, one that includes all possible Gaussians), then the star product \eqref{eq:star_int} fails to define a consistent algebra. This obstruction will play a key role in section \ref{sec:signs}. 

On the other hand, the star product certainly \emph{is} unambiguously defined for polynomials, which span the space of all functions. From this point of view, ambiguities and inconsistencies can only arise when taking superpositions of \emph{infinitely many} polynomial basis functions. Similarly, as long as our only interest is in the boundary correlators \eqref{eq:HS_correlators_scalar} and
\eqref{eq:HS_correlators_bilocal}-\eqref{eq:HS_correlators_spins}, then a consistent choice of signs \emph{is} possible. In other words, we can have a consistent star-product algebra for the functional class of the local insertions $\kappa^{(s)}(\ell,\lambda;Y)$, the bilocal insertions $K(\ell,\ell';Y)$, the particular Gaussians that arise from their star products, and finite superpositions thereof, as long as we exclude singular products that arise from coincident (or, more generally, null-separated) boundary points.

To have such a consistent algebra, we need to make a choice of sign in the 3-point product \eqref{eq:3_pt_product}, and then stick to that choice throughout. This is what we have done implicitly throughout this section: every $\pm$ sign is meant to be equal to that in \eqref{eq:3_pt_product}, and every $\mp$ sign is meant as its opposite. The sign choice in \eqref{eq:3_pt_product} propagates to the $n$-point trace \eqref{eq:trace_n_point_raw}, and from there to the sign choice \eqref{eq:kappa_0} for $\kappa^{(0)}(\ell;Y)$ that allows us to express the spin-0 correlators as in \eqref{eq:HS_correlators_scalar}. In \eqref{eq:K_bilocal}, this sign choice gets squared, producing a definite sign for $K(\ell,\ell';Y)$. In the course of our derivation of $\kappa^{(s)}(\ell,\lambda;Y)$ from $K(\ell,\ell';Y)$ via a bulk detour, we encountered a seemingly separate sign ambiguity, in the star product \eqref{eq:bilocal_master_raw} which defined the Penrose transform of $K(\ell,\ell';Y)$. However, that sign choice is actually of a piece with the one we already made in \eqref{eq:3_pt_product}: in the limit where the bulk point $x$ approaches a boundary point, the star product \eqref{eq:bilocal_master_raw} becomes a boundary 3-point product of the form \eqref{eq:3_pt_product}, and the signs as we wrote them are then consistent with one another. The sign choice in \eqref{eq:bilocal_master_raw} ends up translating into the sign choice \eqref{eq:kappa_s} for $\kappa^{(s)}$, which is in turn consistent with the one we had for $\kappa^{(0)}$ in \eqref{eq:kappa_0}. Finally, the same sign ambiguities that appeared in star products of $\kappa^{(0)}$ will also appear in star products of $\kappa^{(s)}$, and it's consistent to resolve them by following the pattern of the $s=0$ case.

[NOTE: In our previous work \cite{Neiman:2017mel}, where we considered only bilocal correlators, we made a sign choice that is inconsistent with the ones described here. It involved a different sign for the star product $K(\ell_1,\ell'_1;Y)\star K(\ell_2,\ell'_2;Y)$, which amounts to choosing the sign in \eqref{eq:3_pt_product} differently every second time. This becomes inconsistent once local insertions such as $\kappa^{(0)}(\ell;Y)\sim\delta_\ell(Y)$ are included in the algebra. The reasoning that was given for the sign choice in \cite{Neiman:2017mel} involved a bulk$\rightarrow$boundary limit combined with a limit of coincident points. It appears that the inconsistency arose from the ordering of those limits. We've now updated \cite{Neiman:2017mel} to agree with the signs in the present paper.]

\section{Conflict: Euclidean partition functions} \label{sec:partition}

\subsection{From correlators to partition function}

The correlator formulas \eqref{eq:HS_correlators_scalar},\eqref{eq:HS_correlators_spins} are tempting us to integrate them into an HS-algebraic partition function. In standard CFT language, the partition function as a functional of sources $A^{(s)}_{\mu_1\dots\mu_s}(\ell)$ would be constructed from the correlators as:
\begin{align}
 \begin{split}
   Z_{\text{local}}[A^{(s)}(\ell)] &= \exp \bigg( \sum_{n=2}^\infty \frac{1}{n!}\int d^3\ell_1\sum_{s_1=0}^\infty A^{(s_1)}_{\mu_1\dots\mu_{s_1}}\ldots \int d^3\ell_n\sum_{s_n=0}^\infty A^{(s_n)}_{\mu_1\dots\mu_{s_n}} \\
      &\qquad\qquad\qquad\qquad \times \left<J_{(s_1)}^{\mu_1\dots\mu_{s_1}}(\ell_1)\dots J_{(s_n)}^{\mu_1\dots\mu_{s_n}}(\ell_n)\right>_{\text{connected}} \bigg) \ . 
 \end{split} \label{eq:Z_local}
\end{align}
Here, we ignore for simplicity the 0-point function, and renormalize away the divergent 1-point function $\langle J^{(0)}(\ell)\rangle$. For the free vector model, we can write out eq. \eqref{eq:Z_local} more explicitly: since the path integral is Gaussian, the correlators will arrange themselves into a functional determinant over the space of boundary fields $\phi^I(\ell)$. In particular, for a spin-0 source we have:
\begin{align}
 Z_{\text{local}}[\sigma(\ell)] = \exp\left(-\frac{1}{N}\tr\!\big(\ln[1 + \sigma G] - \sigma G\big)\right) = \left(\frac{e^{\tr(\sigma G)}}{\det[1 + \sigma G]}\right)^N \ , \label{eq:Z_local_scalar}
\end{align}
where $G = \Box^{-1}$ is the propagator \eqref{eq:G}. To encode compactly the sources for all spins, it's best to embed them again in the bilocal formalism \eqref{eq:S_bilocal}-\eqref{eq:J_embedding}. We then have an expression formally identical to \eqref{eq:Z_local_scalar}:
\begin{align}
 Z_{\text{bilocal}}[\Pi(\ell',\ell)] = \exp\left(-\frac{1}{N}\tr\!\big(\ln[1 + \Pi G] - \Pi G\big)\right) = \left(\frac{e^{\tr(\Pi G)}}{\det[1 + \Pi G]}\right)^N \ , \label{eq:Z_bilocal}
\end{align}
from which the partition function \eqref{eq:Z_local} with \emph{local} sources can be obtained as a Taylor expansion around $\ell'=\ell$.

On the other hand, we can perform the analogous construction within HS algebra. We begin by unpacking the polarization indices from the twistor functions $\kappa^{(s)}$ that represented spin-$s$ insertions:
\begin{align}
 \kappa^{(s)}(\ell,\lambda;Y) \equiv \lambda_{\mu_1}\dots\lambda_{\mu_s}\,\kappa_{(s)}^{\mu_1\dots\mu_s}(\ell;Y) \ .
\end{align}
We can now assign a twistor function to any finite source distribution $A^{(s)}_{\mu_1\dots\mu_s}(\ell)$ via:
\begin{align}
 F(Y) = \int d^3\ell\,\sum_{s=0}^\infty A^{(s)}_{\mu_1\dots\mu_s}(\ell)\,\kappa_{(s)}^{\mu_1\dots\mu_s}(\ell;Y) \ . \label{eq:F}
\end{align}
Now, integrating the HS-algebraic correlators \eqref{eq:HS_correlators_spins} over source points and summing the Taylor series over $n$ as in \eqref{eq:Z_local}, we get the following partition function:
\begin{align}
 \begin{split}
  Z_{\text{HS}}[F(Y)] &= \exp\left(\frac{N}{4}\sum_{n=1}^\infty \frac{(-1)^{n+1}}{n}\,\tr_\star\Big(\underbrace{F(Y)\star\ldots\star F(Y)}_{n\text{ factors}}\Big) \right) \\
      &= \exp\left(\frac{N}{4}\tr_\star\ln_\star[1+F(Y)]\right) = \left(\textstyle\det_\star[1+F(Y)]\right)^{N/4} \ ,
 \end{split} \label{eq:Z}
\end{align}
where $\ln_\star(1+F)$ is defined in terms of a star-product Taylor series, and the determinant $\textstyle\det_\star(\dots)$ is defined as the exponent of $\tr_\star\ln_\star(\dots)$. For simplicity of the final result, we included in \eqref{eq:Z} an $n=1$ term, corresponding to the 1-point function, which strictly wasn't part of the correlators \eqref{eq:HS_correlators_spins}. As we will see, this piece will end up vanishing for the relevant class of functions $F(Y)$. Other than this, eq. \eqref{eq:Z} is formally very similar to the local and bilocal expressions \eqref{eq:Z_local_scalar}-\eqref{eq:Z_bilocal}, apart from a factor of $-\frac{1}{4}$ in the exponent, which we've been carrying from eq. \eqref{eq:trace_n_point}.

At this point, we can forget altogether about the local sources $A^{(s)}_{\mu_1\dots\mu_s}(\ell)$, and instead consider source distributions directly in terms of the twistor function $F(Y)$. At the very least, this has the advantage of removing gauge redundancy. Indeed, the local sources are gauge potentials, and the sum of their degrees of freedom is that of a function of 5 variables (3 for the point $\ell$, plus $2$ for the null polarization vector $\lambda^\mu$). In contrast, the twistor function $F(Y)$ is a function of just 4 variables, and corresponds via the Penrose transform to gauge-invariant bulk field strengths. If we ever wish to go back from the gauge-invariant $F(Y)$ to the local potentials $A^{(s)}_{\mu_1\dots\mu_s}(\ell)$, we can simply extract them (or, rather, their gauge-invariant field strengths) from the boundary asymptotics of the bulk fields corresponding to $F(Y)$.

The central questions for this section are now the following:
\begin{enumerate}
 \item What does the partition function $Z_{\text{HS}}$ actually look like, when we expand the twistor function $F(Y)$ in terms of some convenient basis of boundary modes? 
 \item Does $Z_{\text{HS}}$ coincide with $Z_{\text{local}}$? 
\end{enumerate}
For the full context of these questions, we refer back to section \ref{sec:intro:boundary}. Now, let's get to work. 

\subsection{Boundary master field, $\bbR^3$ modes} \label{sec:partition:ISO3}

\subsubsection{Bulk spinor-helicity basis and its interpretation as a boundary master field} \label{sec:partition:ISO3:spinor_helicity}

In this subsection, we describe the first of two useful bases for the argument $F(Y)$ of the HS-algebraic partition function \eqref{eq:Z}. Let's begin with the same twistor function that was the foundation of our discussion of correlators: the spin-0 boundary insertion $\pm i\delta_\ell(Y)$, whose Penrose transform \eqref{eq:scalar_boundary_bulk} is the spin-0 boundary-to-bulk propagator. Let's make the concrete sign choice $-i\delta_\ell(Y)$, for which the bulk field \eqref{eq:scalar_boundary_bulk} comes out positive. Also, rather than consider $\sim \delta_\ell(Y)$ insertions at multiple boundary points as before, let's focus instead on a \emph{single} point $\ell$. Recall that such a choice of special boundary point defines a flat $\bbR^3$ conformal frame on the boundary, for which $\ell$ itself is the ``point at infinity''. In the bulk, the choice of $\ell$ defines a Poincare patch. If we consider $dS_4$ bulk signature rather than $EAdS_4$, then $\ell$ also defines a cosmological horizon, on which the boundary-to-bulk propagator \eqref{eq:scalar_boundary_bulk} will have a pole. 

Now, of course, $\delta_\ell(Y)$ is just a single twistor function. To form a basis, we need a 4-parameter family. We already encountered \emph{2-parameter} families of twistor functions based at a boundary point; these are the functions $\delta_\ell(Y-M)$ and $e^{iMY}\delta_\ell(Y)$, with $M^a\in P^*(\ell)$, which generate respectively the right-handed and left-handed parts of boundary-to-bulk propagators for all spins -- see eqs. \eqref{eq:D_C_M}-\eqref{eq:kappa_s}. Let us now make a 4-parameter hybrid of these functions, of the form $e^{iu_L Y}\delta_\ell(Y-u_R)$. Here, the labels of the spinor parameters $u_R$ and $u_L$ refer to the handedness of the boundary-to-bulk propagators above.

There is, however, a slight problem with this construction. While $u_R$ still makes sense as an element of $P^*(\ell)$, $u_L$ does not: shifting it along $P(\ell)$ will change our function $e^{iu_L Y}\delta_\ell(Y-u_R)$ by a $u_R$-dependent factor. This can be addressed by fixing a \emph{second} boundary point $l^\mu$, normalized via $\ell\cdot l = -\frac{1}{2}$, which will serve as a choice of \emph{origin} in the flat conformal frame defined by $\ell^\mu$. Recall that elements of $P(l)$ can act as representatives of the equivalence classes in $P^*(\ell)$. Thus, we can fix $u_R$ and $u_L$ to be elements of $P(l)$. In intrinsic boundary terms, this makes them cospinors at the origin $l$, or simply cospinors in the $\bbR^3$ conformal frame defined by $\ell$. 

Putting all of the above together, we are decomposing the twistor function $F(Y)$ into basis functions $-ie^{iu_L Y}\delta_\ell(Y-u_R)$ with coefficients $\calC_{\ell,l}(u_R,u_L)$, where $\ell$ and $l$ are a fixed pair of boundary points, and $u_R,u_L$ vary over the cospinor space $P(l)$:
\begin{align}
 \begin{split}
   F(Y) &= -i\int_{P(l)} d^2u_R\,d^2u_L\,\calC_{\ell,l}(u_R,u_L)\,e^{iu_L Y}\,\delta_\ell(Y-u_R) \\
     &= -i\int_{P(l)} d^2u_L\,\calC_{\ell,l}(y_{(l)},u_L)\,e^{iu_L y_{(\ell)}} \ .
 \end{split} \label{eq:boundary_RL_basis}
\end{align}
As we can see, this transform decomposes the twistor $Y$ into a cospinor $y_{(l)}\in P(l)$ and a spinor $y_{(\ell)}\in P(\ell)$ in the $\bbR^3$ conformal frame, identifies $y_{(l)}$ with $u_R$, and Fourier-transforms $y_{(\ell)}$ into $u_L$. The inverse transform reads:
\begin{align}
 \calC_{\ell,l}(u_R,u_L) = i\int_{P(\ell)} d^2y_{(\ell)}\,F(y_{(\ell)} + u_R)\,e^{iy_{(\ell)}u_L} \ . \label{eq:C_boundary}
\end{align}
The basis functions $\sim e^{iu_L Y}\delta_\ell(Y-u_R)$ and their coefficients $\calC_{\ell,l}(u_R,u_L)$ have a number of useful meanings. In $dS_4$ bulk signature, they serve as a spinor-helicity formalism for field modes on the cosmological horizon defined by $\ell$, or in the corresponding Poincare patch \cite{David:2019mos}. In particular, when $u_R$ and $u_L$ are related by complex conjugation, their product $p_\mu \sim u_R\ell\gamma_\mu u_L$ defines the 3d spatial momentum of a mode in Poincare coordinates, while the dependence of $\calC_{\ell,l}(u_R,u_L)$ on their phase defines the mode's helicity. 

Another point of view on $\calC_{\ell,l}(u_R,u_L)$ is that it constitutes a boundary limit of the free bulk master field \eqref{eq:Penrose}-\eqref{eq:Penrose_explicit} constructed from $F(Y)$. In the more standard formalism of HS theory, without embedding space or spacetime-independent twistors, this limiting procedure was introduced in \cite{Vasiliev:2012vf}. Let us now demonstrate how the bulk$\rightarrow$boundary limit works in our present formalism. We begin with the bulk Penrose transform \eqref{eq:Penrose_explicit}. In a slight change to the notation there, we rename the twistor argument of $C(x;Y)$ into $U^a$, decompose it into a pair of Weyl spinors $u_{(\pm x)}\in P(\pm x)$ at $x$, and then treat these Weyl spinors as two separate arguments:
\begin{align}
 C(x;u_{(-x)},u_{(x)}) = i\int_{P(x)} d^2y_{(x)}\,F(y_{(x)}+u_{(-x)})\,e^{iy_{(x)} u_{(x)}} \ . \label{eq:Penrose_u}
\end{align}
Now, consider the limit in which $x$ approaches the boundary point $\ell$ along the geodesic that connects it with $l$, as in \eqref{eq:approach}. In this limit, the bulk spinor spaces $P(\pm x)$ both degenerate into $P(\ell)$, in the following detailed way:
\begin{align}
 P^a{}_b(\pm x) = \frac{1}{2}\left(\pm\frac{1}{z}\ell^a{}_b + \delta^a_b \pm z l^a{}_b \right) \ .
\end{align}
We can now replace the integration variable $y_{(x)}$ in \eqref{eq:Penrose_u} by a nearly identical variable $y_{(\ell)}\in P(\ell)$:
\begin{align}
 y_{(x)} &= 2P(x)y_{(\ell)} = (1 + zl)\,y_{(\ell)} \ . \label{eq:y_replacement}
\end{align}
The integration measure $d^2y_{(x)}$ becomes:
\begin{align}
 d^2y_{(x)} = \frac{P_{ab}(x)dy^a_{(x)} dy^b_{(x)}}{4\pi} =\frac{P_{ab}(x) dy^a_{(\ell)} dy^b_{(\ell)}}{\pi} = \frac{zl_{ab}\,dy^a_{(\ell)} dy^b_{(\ell)}}{2\pi} = zd^2y_{(\ell)} \ ,
\end{align}
where we used the measure definitions \eqref{eq:measure_bulk} and \eqref{eq:measure_relationship}. Thus, the bulk field \eqref{eq:Penrose_u} becomes just $z$ times our spinor function $\calC_{\ell,l}(u_R,u_L)$ from \eqref{eq:C_boundary}, if we replace the spinor arguments $u_{(\pm x)}\in P(\pm x)$ with $u_{L/R}\in P(l)$, and neglect the second term in \eqref{eq:y_replacement}. Note that the handedness labels on $u_{(\pm x)}$ and $u_{L/R}$ end up being opposite. The reason for this will become clearer in section \ref{sec:partition:SO4}.

Alternatively, instead of replacing the spinor arguments by brute force, we can perform an honest change of variables:
\begin{align}
 u_{(x)} = P(x)u_L = \frac{1}{2}\left(1 + \frac{1}{z}\ell \right) u_L \ ; \quad u_{(-x)} = P(-x)u_R = \frac{1}{2}\left(1 -\frac{1}{z}\ell \right) u_R \ . \label{eq:u_redefinition}
\end{align}
The RHS of \eqref{eq:Penrose_u} then becomes:
\begin{align}
 C(x;u_{(-x)},u_{(x)})\rightarrow iz\int_{P(\ell)} d^2y_{(\ell)}\,F\!\left(y_{(\ell)} + zly_{(\ell)} - \frac{1}{2z}\ell u_R + \frac{1}{2}u_R \right) e^{iy_{(\ell)} u_L} \ . \label{eq:Penrose_u_limit}
\end{align}
Shifting the integration variable from $y_{(\ell)}$ to $y_{(\ell)} + \frac{1}{2z}\ell u_R$, this becomes:
\begin{align}
 \begin{split}
   C(x;u_{(-x)},u_{(x)}) &\rightarrow iz \exp\left(\frac{iu_R\ell u_L}{2z}\right) \int_{P(\ell)} d^2y_{(\ell)}\,F(y_{(\ell)} + u_R)\, e^{iy_{(\ell)} u_L} \\
     &= z \exp\left(\frac{iu_R\ell u_L}{2z}\right) \calC_{\ell,l}(u_R,u_L) \ .
 \end{split}
\end{align}
This establishes $\calC_{\ell,l}(u_R,u_L)$ as a boundary limit of $C(x;u_{(-x)},u_{(x)})$, up to the same factor of $z$ as before, but now with by an extra exponential factor. It was important here that we did not neglect the $z$-proportional second term in \eqref{eq:y_replacement}, since it gets multiplied by $z^{-1}$, and ends up contributing non-trivially. 

\subsubsection{Boundary momentum basis and its star products} \label{sec:partition:ISO3:momentum_spinors}

An alternative choice of variables, which is sometimes more useful, is $u = u_R + u_L$ and $u' = u_R - u_L$. This defines a relabeling of the basis decomposition \eqref{eq:boundary_RL_basis}:
\begin{align}
 F(Y) &= \int_{P(l)} d^2u\,d^2u'\,\tilde\calC_{\ell,l}(u,u')\,k_{\ell,l}(u,u';Y) \ ; \label{eq:boundary_uu'_basis} \\
 \tilde\calC_{\ell,l}(u,u') &= -\frac{i}{4}\,\calC_{\ell,l}\!\left(\frac{u+u'}{2},\frac{u-u'}{2}\right) \ ; \label{eq:C_tilde_C} \\
 k_{\ell,l}(u,u';Y) &= e^{i(u - u')Y/2}\,\delta_\ell\!\left(Y - \frac{u + u'}{2}\right) \ . \label{eq:k}
\end{align}
Here, we reshuffled the factor of $-i$ for convenience, while the factor of $\frac{1}{4}$ results from the change of integration variables. The basis functions \eqref{eq:k} have an especially nice behavior under the star product:
\begin{align}
  k_{\ell,l}(u,u';Y)\star k_{\ell,l}(v,v';Y) &= \delta_\ell(u'-v)\,k_{\ell,l}(u,v';Y) \ ; \label{eq:k_k} \\
  \tr_\star k_{\ell,l}(u,u';Y) &= 4\delta_\ell(u+u') \ , \label{eq:tr_k}
\end{align}
which is reminiscent of \eqref{eq:K_K}-\eqref{eq:tr_K}, but with the propagators replaced by delta functions. Eq. \eqref{eq:k_k} implies that, in the $(u,u')$ basis, the star product becomes just a matrix product:
\begin{align}
 F(Y)\star F(Y) \quad \Longleftrightarrow \quad \int_{P(l)} d^2u''\,\tilde\calC_{\ell,l}(u,u'')\,\tilde\calC_{\ell,l}(u'',u') \ , \label{eq:star_matrix}
\end{align}
while eq. \eqref{eq:tr_k} shows that the HS trace \eqref{eq:str} takes the form:
\begin{align}
 \tr_\star F(Y) = 4\int_{P(l)} d^2u\,\tilde\calC_{\ell,l}(u,-u) \ . \label{eq:trace_matrix}
\end{align}
Thus, the HS-algebraic partition function \eqref{eq:Z} in this basis reads:
\begin{align}
 \begin{split}
   Z_{\text{HS}}[F(Y)] = \exp\Big(N\sum_{n=1}^\infty &\frac{(-1)^{n+1}}{n} \int_{P(l)} d^2u_1\dots d^2u_n\, \\
   &\ \times \tilde\calC_{\ell,l}(u_1,u_2)\,\tilde\calC_{\ell_,l}(u_2,u_3)\dots\tilde\calC_{\ell,l}(u_{n-1},u_n)\,\tilde\calC_{\ell,l}(u_n,-u_1) \Big) \ .
 \end{split} \label{eq:Z_tilde_C}
\end{align} 
Note that the factor of 4 from \eqref{eq:n_pt_product},\eqref{eq:trace_n_point},\eqref{eq:Z} gets canceled here by the one from \eqref{eq:tr_k},\eqref{eq:trace_matrix}. The slightly strange flipped sign on the final spinor argument in \eqref{eq:trace_matrix}-\eqref{eq:Z_tilde_C} has its own origin in HS algebra, being related to the star product with the twistor delta function $\delta(Y)$:
\begin{align}
 \begin{split}
   k_{\ell,l}(u,u';Y)\star\delta(Y) &= -k_{\ell,l}(u,-u';Y) \ ; \\
   \delta(Y)\star k_{\ell,l}(u,u';Y) &= -k_{\ell,l}(-u,u';Y) \ ,
 \end{split}
\end{align}
so that:
\begin{align}
 \begin{split}
   F(Y)\star\delta(Y) \quad &\Longleftrightarrow \quad -\tilde\calC_{\ell,l}(u,-u') \ ; \\
   \delta(Y)\star F(Y) \quad &\Longleftrightarrow \quad -\tilde\calC_{\ell,l}(-u,u') \ .
 \end{split} \label{eq:delta_matrix}
\end{align}
As we'll review in section \ref{sec:shell:QM}, the matrix-algebra-like formulas \eqref{eq:star_matrix}-\eqref{eq:trace_matrix} can be interpreted in terms of particle quantum mechanics on the boundary, with $u,u'$ as the spinor square roots of on-shell boundary momenta. Similar formulas have been developed from an intrinsic boundary point of view in e.g. \cite{Gelfond:2008ur,Gelfond:2013xt}, as well as in \cite{Koch:2010cy,Das:2012dt,Koch:2014aqa}, where null boundary momenta are used directly, without recourse to spinors.

\subsubsection{Importance and disadvantages of the $\bbR^3$ bases} \label{sec:partition:ISO3:pros_cons}

The decompositions \eqref{eq:boundary_RL_basis},\eqref{eq:boundary_uu'_basis} into boundary $\bbR^3$ modes have a number of advantages. First, they're more directly related than the original twistor function $F(Y)$ to actual bulk and boundary field modes. Second, they'll play an important conceptual role in the discussion of spin-locality in section \ref{sec:discuss:spin_locality}. Third, they provide the simplest possible explicit formula \eqref{eq:Z_tilde_C} for the HS-algebraic partition function $Z_{\text{HS}}$. 

Despite such nice properties, this basis isn't quite satisfactory for actually computing the partition function. Just as twistor space in $SO(1,4)$ signature is complex, so are the spinor spaces $P(\ell)$ and $P(l)$ on the Euclidean 3d boundary. As a result, the delta functions in \eqref{eq:k_k}-\eqref{eq:tr_k} are very formal constructs. As delta functions of a complex variable \emph{without its complex conjugate}, they do not have a well-defined support: we can never say for which values of its argument such a delta function vanishes! As a result, if we attempt to evaluate $Z_{\text{HS}}$ on one of the $(u_R,u_L)$ or $(u,u')$ basis functions themselves, the answer is ill-defined. When, instead of a single basis function, we consider a superposition with coefficients $\calC_{\ell,l}(u_R,u_L)$ or $\tilde\calC_{\ell,l}(u,u')$, the problem merely reappears in a new guise: just like the integrals in the star product \eqref{eq:star_int} or in the Penrose transform \eqref{eq:Penrose_explicit}, the spinor integrals in \eqref{eq:Z_tilde_C} are \emph{contour integrals} over complex variables. Therefore, they aren't fully well-defined until we specify the integration contour, and ensure that it leads to convergent integrals for the desired classes of functions $\calC_{\ell,l}(u_R,u_L)$ or $\tilde\calC_{\ell,l}(u,u')$. This is easy to do for the $n=2$ term, where the two spinor variables can be related by complex conjugation  \cite{Neiman:2018ufb}, but not so for general $n$.

Nevertheless, this problem is not a deep one. The simplest solution is to just switch signature to Lorentzian $AdS_4$, where both twistors and boundary spinors are real, and the contour and convergence issues have been studied systematically \cite{Gelfond:2008ur}. However, we'll insist for now on remaining in $EAdS_4$, with its Euclidean boundary and complex spinors. One reason is our eventual interest in a $dS_4$ bulk. Another is that a Euclidean boundary admits the most symmetric global conformal frame, namely $S_3$, and it is there that we'd like to evaluate $Z_{\text{HS}}$.

In the next subsection, we will describe a basis that doesn't suffer from contour ambiguities, and is in fact adapted to an $S_3$ conformal frame. Since it can be related back to the $\bbR^3$ mode decompositions \eqref{eq:boundary_RL_basis},\eqref{eq:boundary_uu'_basis}, one may consider it not as a separate construction, but as a particular way of addressing the contour and convergence issues.

\subsection{Bulk master field, $S_3$ modes} \label{sec:partition:SO4}

\subsubsection{Basis, star products and partition function} \label{sec:partition:SO4:basis}

In this subsection, we at last obtain some actual values for the HS-algebraic partition function \eqref{eq:Z}. This will require a switch from modes based on an $\bbR^3$ conformal frame to an $S_3$ frame. An $S_3$ conformal frame on the boundary, with residual symmetry $SO(4)$, is obtained by fixing a timelike direction in the $\bbR^{1,4}$ embedding space, i.e. a bulk point $x^\mu\in EAdS_4$. The natural modes in this frame are spin-weighted spherical harmonics on $S_3$, arranged into the integer-spin irreps of $SO(4)$. Conveniently, this is precisely the structure of a linearized master field $C(x;Y)$ at the fixed bulk point $x^\mu$. Consider, for example, modes of the spin-0 source, which can be organized in spherical harmonics on $S_3$ with angular momentum $j$. What are the corresponding values of the master field $C(x;Y)$ at the fixed point $x$? One can see from eqs. \eqref{eq:packaging},\eqref{eq:unfolding} that these will be spanned by the balanced monomials $C(x;Y) = (m_{(x)}y_{(x)})^j (m_{(-x)}y_{(-x)})^j$, with polarization defined by a pair of Weyl spinors $m_{(x)}\in P(x)$ and $m_{(-x)}\in P(-x)$. Similarly, spin-$s$ modes with $s>0$ are spanned by the unbalanced monomials $C(x;Y) = (m_{(x)}y_{(x)})^{2s+j} (m_{(-x)}y_{(-x)})^j$ and $C(x;Y) = (m_{(x)}y_{(x)})^j (m_{(-x)}y_{(-x)})^{2s+j}$. We can arrange all these modes as the Taylor expansion, in either $m_{(\pm x)}$ or $y_{(\pm x)}$, of a single master field $C(x;Y) = e^{iMY}$, where we combined the two polarization spinors $m_{(\pm x)}$ into a single Dirac spinor (or twistor) $M^a = m_{(x)}^a + m_{(-x)}^a$. The restriction to integer spins can be realized by considering only even orders in the Taylor expansion; equivalently, we can replace $e^{iMY}$ by $\cos(MY)$. 

The twistor function corresponding to the master field $C(x;Y) = e^{iMY}$ at $x$ is given by the inverse Penrose transform:
\begin{align}
 k_x(M;Y) = -ie^{iMY}\star\delta_x(Y) = -ie^{iMY}\delta_x(Y - M) = -ie^{im_{(-x)}y_{(-x)}}\delta_x(y_{(x)} - m_{(x)}) \ . \label{eq:k_S3}
\end{align}
This will be our basis of twistor functions for the $S_3$ modes, parameterized by the polarization twistor $M^a$, with $x$ held fixed. The expansion of a twistor function $F(Y)$ in this basis takes the form:
\begin{align}
 F(Y) = \int d^4M\,C(-x;M)\,k_x(M;Y) = -i\int d^2m_{(-x)}\,C(-x;m_{(-x)} + y_{(x)})\,e^{im_{(-x)}y_{(-x)}} \ , \label{eq:expansion_S3}
\end{align}
where the coefficient function $C(-x;M)$ is, as the notation suggests, just the Penrose transform \eqref{eq:Penrose_explicit} of $F(Y)$, evaluated not at $x^\mu$, but at the antipodal point $-x^\mu$, which lives on the second branch of the $EAdS_4$ hyperboloid. The appearance of $C(-x;M)$ here is not surprising, due to its Fourier-transform relationship \eqref{eq:C_antipodal} with $C(x;Y)$. Recalling that the handedness of spinors at $-x$ is the opposite of that at $x$, we now also understand the reversed handedness labels in \eqref{eq:u_redefinition}. 

Let's now evaluate the traced products of $k_x(M;Y)$ that will compose the HS-algebraic partition function \eqref{eq:Z}:
\begin{align}
 \tr_\star\!\big(k_x(M_1;Y)\star\ldots\star k_x(M_n;Y)\big) \ . \label{eq:trace_S3_raw}
\end{align}
The task is simple, since the right-handed and left-handed spinors at $x$ decouple from each other in the star product \eqref{eq:star}-\eqref{eq:star_int}. For the left-handed factors, we will only need the star product of two exponents:
\begin{align}
 e^{imy}\star e^{im'y} = e^{imm'}e^{i(m + m')y} \ , \label{eq:exp_exp}
\end{align}
where we suppressed the $(-x)$ subscript on all the spinors. For the right-handed factors, the star product alternates between delta functions and exponents:
\begin{align}
 \delta(y-m)\star\delta(y-m') &= e^{-imm'}e^{i(m-m')y} \ ; \label{eq:delta_delta} \\
 e^{imy}\star\delta(y-m') &= e^{imm'}\delta(y - m -m') \ , \label{eq:exp_delta}
\end{align}
where this time the $(x)$ subscripts were suppressed. The alternating form of the star products \eqref{eq:delta_delta}-\eqref{eq:exp_delta} is directly analogous to the alternation between ``particle-like'' and ``black hole-like'' states in HS cosmological perturbation theory, discussed in \cite{Aros:2017ror}. Applying the star products \eqref{eq:exp_exp}-\eqref{eq:exp_delta} recursively, we arrive at:
\begin{align}
   &k_x(M_1;Y)\star\ldots\star k_x(M_n;Y) = \frac{1}{i^n}\exp\left(i\sum_{p=1}^n\sum_{q=p+1}^n \left( m^{(-x)}_p m^{(-x)}_q + (-1)^{q-p}\,m^{(x)}_p m^{(x)}_q \right) \right) \nonumber \\
   &\quad \times \exp\left(i\sum_{p=1}^n m^{(-x)}_p y_{(-x)}\right) \times \left\{
     \begin{array}{ll}
        \displaystyle \exp\left(i\sum_{p=1}^n (-1)^{p+1}\,m^{(x)}_p y_{(x)}\right) & \qquad n \text{ even} \\
        \displaystyle \delta_x\!\left(y_{(x)} + \sum_{p=1}^n (-1)^p\,m^{(x)}_p \right) & \qquad n\text{ odd}
     \end{array} \right. \ . \label{eq:S3_n_pt_product}
\end{align}
Let us now take the HS trace \eqref{eq:trace_S3_raw} of this product. For even $n$, this is trivial: the trace is just the $Y$-independent exponent on the first line. On the other hand, for odd $n$, we are forced to take the trace of a delta function:
\begin{align}
 \tilde m_{(x)} \equiv \sum_{p=1}^n (-1)^p\,m^{(x)}_p \ ; \quad \tr_\star\delta_x(y_{(x)} + \tilde m_{(x)}) ={} ???
\end{align}
While this trace isn't obviously well-defined, we can see that it must vanish by a symmetry argument. First, let us Taylor-expand it as:
\begin{align}
 \tr_\star\delta_x(y_{(x)} + \tilde m_{(x)}) = \sum_{k=0}^\infty \frac{1}{k!}\,\tilde m_{(x)}^{a_1}\dots\tilde m_{(x)}^{a_k}\tr_\star\left(\frac{\del^k}{\del Y^{a_1}\dots\del Y^{a_k}}\,\delta_x(Y)\right) \ . \label{eq:Taylor_delta}
\end{align}
The traces on the RHS of order $k>0$ must clearly vanish by rotational symmetry (note that the traces are symmetric in their spinor indices, while the spinor metric $P_{ab}(x)$ is antisymmetric). This leaves the zeroth-order term $\tr_\star\delta_x(Y)$. As we will argue in detail in section \ref{sec:signs}, this term should also be regarded as zero, due to either of two \emph{discrete symmetries}: complex conjugation and spin-parity. In brief, the complex-conjugation argument is that $\tr_\star\delta_x(Y)$ should be real, but then having it nonzero would lead to a complex partition function, due to $1/i^n$ prefactor in \eqref{eq:S3_n_pt_product}. The spin-parity argument is that HS algebra distinguishes even spins, corresponding to twistor functions $f(Y)$ with homogeneity 2 mod 4, from odd spins, corresponding to $f(Y)$ with homogeneity 0 mod 4. Now, $\delta_x(Y)$ has homogeneity $-2$ (corresponding, as expected, to spin 0), while the trace operation $\tr_\star f(Y) = f(0)$ singles out the component of $f(Y)$ with homogeneity 0. Thus, we conclude that $\tr_\star\delta_x(Y)$, and with it the entire series \eqref{eq:Taylor_delta}, vanishes. Putting everything together, the traces \eqref{eq:trace_S3_raw} evaluate to:
\begin{align}
 \begin{split}
   &\tr_\star(k_x(M_1;Y)\star\ldots\star k_x(M_n;Y)) \\
   &\qquad = \left\{ 
    \begin{array}{cl}
      \displaystyle (-1)^{n/2}\exp\sum_{p=1}^n\sum_{q=p+1}^n i\left( m^{(-x)}_p m^{(-x)}_q + (-1)^{q-p}\,m^{(x)}_p m^{(x)}_q \right) & \qquad n\text{ even} \\
      \displaystyle 0 & \qquad n\text{ odd}
   \end{array} \right. \ .
 \end{split} \label{eq:trace_S3}
\end{align}
The partition function $Z_{\text{HS}}$ now follows from plugging the traces \eqref{eq:trace_S3} and the basis expansion \eqref{eq:expansion_S3} into the general HS-algebraic formula \eqref{eq:Z}:
\begin{align}
 \begin{split}
  Z_{\text{HS}}[F(Y)] = \exp\bigg(\frac{N}{8}\sum_{n=1}^\infty &\frac{(-1)^{n+1}}{n} \int d^4M_1\,C(-x;M_1)\ldots \int d^4M_{2n}\,C(-x;M_{2n}) \\
     &{}\times \exp\sum_{p=1}^{2n}\sum_{q=p+1}^{2n} i\left( m^{(-x)}_p m^{(-x)}_q + (-1)^{q-p}\,m^{(x)}_p m^{(x)}_q \right) \bigg) \ .
 \end{split} \label{eq:Z_S3}
\end{align}
Note that only even orders, labeled in \eqref{eq:Z_S3} by $2n$, are present. In section \ref{sec:signs}, we will return to the surprising absence of the odd orders, and argue from first principles that $Z_{\text{HS}}$ must in fact be even. For now, it is our first hint that $Z_{\text{HS}}$ is very different from $Z_{\text{local}}$.

One crucial feature of the $S_3$ basis is that the traces \eqref{eq:trace_S3} are perfectly regular if we set all the arguments $M_n$ equal to each other, $M_n\equiv M$. In fact, they become trivial, with no $M$-dependence remaining:  
\begin{align}
  \tr_\star(\underbrace{k_x(M;Y)\star\ldots\star k_x(M;Y)}_{n\text{ factors}}) = \left\{ 
    \begin{array}{cl}
       \displaystyle (-1)^{n/2} & \quad n\text{ even} \\
       \displaystyle 0 & \quad n\text{ odd}
    \end{array}\right. \ .
\end{align}
This allows us to evaluate $Z_{\text{HS}}$ on a \emph{single} mode of the form \eqref{eq:k_S3}:
\begin{align}
 Z_{\text{HS}}[c\,k_x(M;Y)] = (1 + c^2)^{N/8} \ , \label{eq:Z_single_mode}
\end{align}
where the scalar coefficient $c$ sets the mode's magnitude. We thus have a simple, explicit example of $Z_{\text{HS}}$ evaluated on a finite source!

\subsubsection{Real contour and relation to the $\bbR^3$ basis}

There is one final piece in our treatment of the $S_3$ basis that remains ambiguous: the basis expansion \eqref{eq:expansion_S3} of $F(Y)$ is an integral over the complex polarization spinors $M^a = m_{(x)}^a + m_{(-x)}^a$, without a specified integration contour. There are two ways to handle this issue. First, since the partition function is perfectly well-defined on individual basis elements $k_x(M;Y)$, the integral is not really necessary: we can instead consider discrete sums of the basis elements. Alternatively, we can fixed a preferred choice of ``real'' integration contour. Since the left-handed and right-handed spinors of $SO(4)$ are neither real nor related by complex conjugation, this requires some further breaking of spacetime symmetry. One natural way to proceed is to choose a spacelike unit vector $v^\mu\in\bbR^{1,4}$ orthogonal to $x^\mu$, i.e. a tangent direction at the bulk point $x$. This breaks the spacetime symmetry down to $SO(3)$, and enables us to fix a contour by imposing a ``reality condition'' that sets $m_{(-x)}$ proportional to $v\bar m_{(x)}$. 

Now, let us notice that a choice of $v^\mu$ is equivalent to choosing a geodesic that goes through $x$. The endpoints of this geodesic are the two boundary points:
\begin{align}
 \ell^\mu = \frac{1}{2}(x^\mu + v^\mu) \ ; \quad l^\mu = \frac{1}{2}(x^\mu - v^\mu) \ ,
\end{align}
from which we can construct a basis of $\bbR^3$ modes, as in section \ref{sec:partition:ISO3}. As promised in section \ref{sec:partition:ISO3:pros_cons}, it is easy to relate the $\bbR^3$ and $S_3$ bases to each other. To do this, let's consider an element \eqref{eq:k_S3} of the $S_3$ basis, and apply to it the transforms \eqref{eq:C_boundary},\eqref{eq:C_tilde_C}. Using our already familiar techniques for manipulating spinor spaces, we obtain:
\begin{align}
 C_{\ell,l}(u_R,u_L) &= 4\exp\left(iu_L\ell u_R + iM(u_L + u_R) + iM\ell(u_L - u_R) + iM\ell lM\right) \ ; \label{eq:S3_R3_raw} \\
 \tilde C_{\ell,l}(u,u') &= -i\exp\left(\frac{i}{2}u\ell u' + iMu - iM\ell u' + iM\ell lM\right) \ , \label{eq:S3_R3_tilde_raw}
\end{align}
where we recall that the spinor arguments $u_R,u_L$ and $u,u'$ are all elements of $P(l)$. As a final tweak, we can decompose the parameter $M^a$ of the $S_3$ basis function along the boundary spinor spaces $P(\ell),P(l)$, as $M = m_{(\ell)} + m_{(l)}$. The reality condition that sets $m_{(-x)}$ proportional to $v\bar m_{(x)}$ will now make $m_{(\ell)}$ proportional to $\ell\bar m_{(l)}$. In these variables, the basis transformations \eqref{eq:S3_R3_raw}-\eqref{eq:S3_R3_tilde_raw} reads:
\begin{align}
 C_{\ell,l}(u_R,u_L) &= 4\exp\left(iu_L\ell u_R + im_{(\ell)}(u_L + u_R) + im_{(l)}\ell(u_L - u_R) + im_{(l)}m_{(\ell)}\right) \ ; \label{eq:S3_R3} \\
 \tilde C_{\ell,l}(u,u') &= -i\exp\left(\frac{i}{2}u\ell u' + im_{(\ell)}u - im_{(l)}\ell u' + im_{(l)}m_{(\ell)}\right) \ , \label{eq:S3_R3_tilde} 
\end{align} 
The star products \eqref{eq:S3_n_pt_product} of the $S_3$ basis can now be recovered using the $\bbR^3$ modes' matrix-product formula \eqref{eq:star_matrix}. This demonstrates that HS algebra in the $S_3$ and $\bbR^3$ bases is mutually consistent. In fact, the $S_3$ formalism can be viewed as a regularization of the $\bbR^3$ one. The role of regularizer is played by the $u_L\ell u_R$ or $u\ell u'$ term in \eqref{eq:S3_R3_raw}-\eqref{eq:S3_R3_tilde}. Without it, the modes' star products all result in complex delta functions, as in eqs. \eqref{eq:k_k}-\eqref{eq:tr_k}; with it, the star products produce proper functions, as in eq. \eqref{eq:S3_n_pt_product} (though still only at even orders).  

\subsection{Disagreement with the CFT result} \label{sec:partition:local}

Let us now compare our result \eqref{eq:Z_single_mode} for the HS-algebraic partition function with the one obtained by a standard local calculation in the boundary CFT. As discussed in section \ref{sec:intro:boundary}, we will focus on the simplest case of a spin-0 source, where the local correlators don't require contact corrections. In particular, we will choose a scalar source $\sigma$ that is \emph{constant} in an $S_3$ conformal frame. From the discussion of $SO(4)$ representation theory in section \ref{sec:partition:SO4:basis}, it follows that the corresponding twistor function must be proportional to $-i\delta_x(Y)$, i.e. to the $M^a=0$ element of our $S_3$ basis \eqref{eq:k_S3}. We recall that $x^\mu$ here refers to the particular bulk point that defines the $S_3$ boundary frame. The $S_3$ boundary frame itself is given by the $\bbR^{1,4}$ lightcone section $\ell\cdot x = -1$.

Let us now fix our twistor function's normalization. Recall from eq. \eqref{eq:kappa_0} that a local insertion of the spin-0 operator $J^{(0)}(\ell)$ is described by the twistor function $\kappa^{(0)}(\ell;Y) = \pm\frac{i}{4\pi}\delta_\ell(Y)$ (the sign ambiguity won't matter here). Thus, the twistor function for a constant source $\sigma$ should be given by the integral:
\begin{align}
 F(Y) = \sigma\int_{S_3}\kappa^{(0)}(\ell;Y)\,d^3\ell = \pm \frac{i\sigma}{4\pi}\int_{S_3}\delta_\ell(Y)\,d^3\ell \ . \label{eq:scalar_F_raw}
\end{align}
 Unfortunately, it's unclear how to integrate $\delta_\ell(Y)$ over the boundary point $\ell$. However, we can employ a trick from section \ref{sec:correlators:correlators:spin}, and perform the integral not on the twistor functions directly, but on the corresponding free bulk master fields at $x$. The Penrose transform of each local boundary insertion $\kappa^{(0)}(\ell;Y)$ is given by eq. \eqref{eq:ell_x} as:
\begin{align}
 i\kappa^{(0)}(\ell;Y)\star\delta_x(Y) =  \mp\frac{1}{4\pi}\,\delta_\ell(Y)\star\delta_x(Y) = \pm \frac{1}{2\pi(\ell\cdot x)}\exp\frac{iY\ell xY}{2(\ell\cdot x)} = \mp \frac{1}{2\pi}\exp\frac{Y\ell xY}{2i} \ .
\end{align}
It now becomes easy to integrate $\ell$ over $S_3$. Performing a Taylor expansion in $Y^a$, it is clear from $SO(4)$ symmetry that all terms with nonzero powers must vanish upon integration. Thus, we are left with the $Y$-independent term $\mp \frac{1}{2\pi}$, for which the integral is just multiplication by the 3-sphere volume $2\pi^2$. Altogether, the Penrose transform of our desired twistor function \eqref{eq:scalar_F_raw} reads:
\begin{align}
 iF(Y)\star\delta_x(Y) = \mp \pi\sigma \ ,
\end{align}
from which we extract $F(Y)$ as:
\begin{align}
 F(Y) = \pm \pi i\sigma\delta_x(Y) = \mp \pi\sigma k_x(0;Y) \ . \label{eq:0_mode}
\end{align}
The HS-algebraic partition function \eqref{eq:Z_single_mode} for this argument reads:
\begin{align}
 Z_{\text{HS}} = (1 + \pi^2\sigma^2)^{N/8} \ . \label{eq:Z_HS_sigma}
\end{align}
Let's now perform the corresponding CFT calculation, following \cite{Anninos:2012ft}. The partition function $Z_{\text{local}}$ is given by the functional determinant \eqref{eq:Z_local_scalar}, where $\sigma$ is now a constant. This is just the partition function on $S_3$ of $N$ free scalar fields $\phi^I$ with mass $m^2 = -\sigma$:
\begin{align}
 Z_{\text{local}} = \exp\left(-N\tr\left[ \ln\left(1 + \frac{\sigma}{\Box}\right) - \frac{\sigma}{\Box} \right]\right) \ . \label{eq:Z_local_scalar_again}
\end{align}
To evaluate this, we decompose the scalar fields into $S_3$ spherical harmonics with angular momentum $j$. These are the $(\frac{j}{2},\frac{j}{2})$ representations of $SO(4)$, whose dimension is $(j+1)^2$. The conformal Laplacian on these harmonics has the eigenvalues:
\begin{align}
 \Box = \nabla^2 - \frac{3}{4} = -j(j+2) - \frac{3}{4} = -(j+1)^2 + \frac{1}{4} \ . \label{eq:Box_S3}
\end{align} 
The partition function \eqref{eq:Z_local_scalar} becomes:
\begin{align}
 \begin{split}
   Z_{\text{local}} &= \exp\left(-N\sum_{j=0}^\infty (j+1)^2\left[ \ln\left(1 - \frac{\sigma}{(j+1)^2 - 1/4}\right) + \frac{\sigma}{(j+1)^2 - 1/4} \right]\right) \\ 
    &= \exp\left(-N\sum_{k=1}^\infty k^2\left[ \ln\left(1 - \frac{\sigma}{k^2 - 1/4}\right) + \frac{\sigma}{k^2 - 1/4} \right]\right) \ , 
 \end{split}
\end{align}
where we renamed $k\equiv j+1$. With a modicum of help from Mathematica, the sum can be converted into an integral:
\begin{align}
 Z_{\text{local}} = \exp\left(-\frac{N\pi}{8}\int_1^{\sqrt{1+4\sigma}} t^2 \cot\frac{\pi t}{2}\,dt \right) \ , \label{eq:Z_local_sigma}
\end{align}
which in turn can be evaluated in terms of polylogarithms \cite{Anninos:2012ft}. The central observation of this paper is that the partition functions \eqref{eq:Z_HS_sigma} and \eqref{eq:Z_local_sigma} are not the same. To be completely concrete, let's compare the Taylor series of $\ln Z$ with respect to the source $\sigma$:
\begin{align}
 \ln Z_{\text{HS}} &= \frac{N\pi^2}{8}\left(\sigma^2 - \frac{\pi^2}{2}\sigma^4 + O(\sigma^6) \right) \ ; \label{eq:lnZ_HS_sigma_Taylor} \\
 \ln Z_{\text{local}} &= \frac{N\pi^2}{8}\left(\sigma^2 - \frac{2}{3}\sigma^3 + \left(\frac{\pi^2}{6} - 1\right)\sigma^4 + O(\sigma^5) \right) \ . \label{eq:lnZ_local_sigma_Taylor}
\end{align}
The coefficients of $\sigma^2$ match, but the higher-order coefficients do not. The most obvious difference is that $Z_{\text{HS}}$ is an even function of $\sigma$, while $Z_{\text{local}}$ is neither even nor odd. One may be tempted to pin the blame on our argument in section \ref{sec:partition:SO4:basis} that set all the odd orders in $Z_{\text{HS}}$ to zero. However, the subleading \emph{even}-order coefficients in  \eqref{eq:lnZ_HS_sigma_Taylor}-\eqref{eq:lnZ_local_sigma_Taylor} are also different! We have thus established that the local and HS-algebraic partition functions differ, despite having been constructed from the same $n$-point correlators. Moreover, they differ in the spin-0 sector, where $Z_{\text{local}}$ is indeed fully captured by the correlators, with no contact corrections. As far as we can tell, this disagreement is genuine: $Z_{\text{local}}$ and $Z_{\text{HS}}$ are computing two different things. In the following sections, we will try to understand where the disagreement is coming from, and how to think of $Z_{\text{HS}}$ in light of it.

\section{Explanation: Sign ambiguity and discrete symmetries} \label{sec:signs}

The disagreement between $Z_{\text{local}}$ and $Z_{\text{HS}}$ despite the identical correlators constitutes a \emph{failure of linearity}: somehow, the dictionary between local correlators and HS algebra fails to commute with linear superpositions. We will now point out the technical core of this failure. 

\subsection{Spontaneously symmetry breaking as the culprit} \label{sec:signs:general}

In our construction of the HS-algebraic correlators in section \ref{sec:correlators}, there was one delicate step: the sign ambiguity in the product of three local insertions \eqref{eq:3_pt_product}. As we have seen, \emph{if one is only interested in $n$-point correlators}, this ambiguity can be consistently resolved in either direction, each leading to a different sign choice in the dictionary \eqref{eq:kappa_s} between local sources and twistor functions. However, if we allow for arbitrary superpositions, leading to arbitrary twistor functions, it may become impossible to resolve the ambiguity consistently. Could this be the reason behind the apparent failure of linearity?

The answer is, at first sight, not obvious. Recall that we did not construct the basis functions of section \ref{sec:partition} as explicit linear superpositions of local boundary insertions $\kappa^{(s)}(\ell,\lambda;Y)$. The best we could do in section \ref{sec:partition:local} was to carry out the superposition indirectly, via the Penrose transform into bulk fields. How can we follow the fate of the sign ambiguity \eqref{eq:3_pt_product} through this opaque process? By keeping track of symmetries! In particular, we will be interested in two discrete symmetries: \emph{complex conjugation} and \emph{spin-parity}. We already invoked these in section \ref{sec:partition:SO4:basis}, to justify the vanishing of $\tr_\star\delta_x(Y)$, which implied the vanishing of all odd orders of $Z_{\text{HS}}$ in the $S_3$ basis. Here, we will use these same symmetries to make a more general argument.

Complex conjugation and spin-parity are well-known symmetries of HS algebra. They are respected unambiguously by HS algebra acting on \emph{polynomial} twistor functions, and formally extend to more general functions and distributions. They can be verified explicitly for the star products of the various basis functions from section \ref{sec:partition} (though not always for their HS traces: again, for the odd orders in \eqref{eq:trace_S3}, we needed to \emph{invoke} the symmetries in order to arrive at a result). 

In contrast, the star products in the local correlators \eqref{eq:HS_correlators_scalar},\eqref{eq:HS_correlators_spins} \emph{break} the discrete symmetries. To our knowledge, this basic observation has not been made before in the literature. Moreover, we can identify the symmetry breaking as \emph{spontaneous}, and precisely associate it with the sign ambiguity in the 3-point product \eqref{eq:3_pt_product}: the discrete symmetry is violated once we make a sign choice in \eqref{eq:3_pt_product}, but is maintained in the sense of relating the two choices to each other. This suggests that the sign ambiguity is indeed the mechanism behind the failure of the superposition principle: as we attempt to integrate local insertions into the finite-source modes of section \ref{sec:partition}, the sign ambiguity in the star product \emph{fails} to be resolved consistently, and this allows the discrete symmetry between the two sign choices to be recovered.

Having laid out the general logic, we will now separately discuss each of the two discrete symmetries. As we focus on their relationship with the local HS-algebraic correlators \eqref{eq:HS_correlators_scalar},\eqref{eq:HS_correlators_spins}, the reader may notice that we ignore their \emph{even more} problematic relationship with the \emph{bilocal} correlators \eqref{eq:HS_correlators_bilocal}. That issue deserves separate treatment, which will be given elsewhere.

\subsection{Complex conjugation}

The first of the two discrete symmetries is \emph{complex conjugation}. We briefly mentioned the anti-idempotent complex conjugation $Y^a\rightarrow\bar Y^a$ of twistors in section \ref{sec:correlators:geometry:twistors}. We can define the complex conjugation of twistor functions $f(Y)\rightarrow\bar f(Y)$ via $\bar f(\bar Y) = \overline{f(Y)}$. Since the gamma matrices $(\gamma_\mu)^a{}_b$ are real under the twistor complex conjugation, the same is true for the projectors $P(\ell),P(\pm x)$ onto the bulk and boundary spinor spaces, as well as for the corresponding delta functions $\delta_\ell(Y),\delta_{\pm x}(Y)$. This means that the local boundary spin-0 insertions $\sim\pm i\delta_\ell(Y)$ from \eqref{eq:kappa_0} and the zero-angular-momentum $S_3$ mode function $\sim\pm i\delta_x(Y)$ from \eqref{eq:0_mode} are \emph{imaginary}. Their analogues \eqref{eq:kappa_s},\eqref{eq:k_S3} for nonzero spins and angular momenta are also imaginary, once we impose real polarization tensors by averaging over $\lambda^\mu\leftrightarrow\bar\lambda^\mu$ or $M^a\leftrightarrow\bar M^a$ (the anti-idempotence of $M^a\rightarrow\bar M^a$ doesn't matter here, because of the restriction to integer spins, i.e. even powers).

Now, the star product \eqref{eq:star} is preserved by complex conjugation, except for a sign flip in the non-commutative term. As a result, the complex conjugation $f(Y)\rightarrow \bar f(Y)$ of twistor function acts as a Hermitian conjugation with respect to the star product: $\overline{f\star g} = \bar g\star\bar f$. If we symmetrize over the order of factors, as in the correlators \eqref{eq:HS_correlators_scalar},\eqref{eq:HS_correlators_spins}, or simply set all the factors equal, as in the partition function \eqref{eq:Z}, then the reversed multiplication order is of no consequence. The trace operation \eqref{eq:str} also obviously commutes with complex conjugation. 

Putting everything together, the complex conjugation symmetry of HS algebra implies that the correlators \eqref{eq:HS_correlators_scalar},\eqref{eq:HS_correlators_spins} should be \emph{imaginary} at odd $n$, and likewise for the odd-order pieces of $Z_{\text{HS}}$ in \eqref{eq:Z}. This is a strange conclusion indeed. The only way to satisfy it without accepting complex correlators and partition functions is for the odd-$n$ correlators and the odd orders in $Z_{\text{HS}}$ to \emph{vanish}. In section \ref{sec:partition:SO4:basis}, this was indeed what we concluded for $Z_{\text{HS}}$ in the $S_3$ basis, as expressed in \eqref{eq:trace_S3}-\eqref{eq:Z_S3}. In particular, for a constant spin-0 source, we evaluated $Z_{\text{HS}}$ as the even function \eqref{eq:Z_HS_sigma}.

On the other hand, this symmetry argument is clearly \emph{not} obeyed by the $n$-point correlators, or by the partition function $Z_{\text{local}}$ obtained by integrating them. In particular, the 3-point correlators are real and nonzero; similarly, the partition function \eqref{eq:Z_local_sigma} for constant spin-0 source is real, but is neither even nor odd. The symmetry breaking is clearly visible in the formulas \eqref{eq:HS_correlators_scalar},\eqref{eq:HS_correlators_spins} for the HS-algebraic correlators: at odd $n$, after symmetrizing over the order of factors and imposing real polarizations, the LHS should be imaginary, but the RHS is real. This feature is clearly descended from the 3-point star product \eqref{eq:3_pt_product}, where the LHS (when symmetrized over factor ordering) is real, but the RHS is imaginary. What enables this symmetry breaking is the sign ambiguity: the RHS of \eqref{eq:3_pt_product} is imaginary \emph{with an ambiguous sign}, giving a \emph{real} average, namely zero. The same is true for the HS-algebraic correlators \eqref{eq:HS_correlators_scalar},\eqref{eq:HS_correlators_spins}: at odd $n$, the RHS is real, but with an ambiguous sign, which makes for a zero average.

To sum up, by complex conjugation symmetry, the HS-algebraic correlators at odd $n$, as well as the odd part of $Z_{\text{HS}}$, should be imaginary. $Z_{\text{HS}}$ in e.g. the $S_3$ basis satisfies this symmetry, by vanishing at odd orders. The local $n$-point correlators \eqref{eq:HS_correlators_scalar},\eqref{eq:HS_correlators_spins} and the partition function $Z_{\text{local}}$ obtained by integrating them \emph{don't} satisfy the symmetry, due to spontaneous symmetry breaking via the sign ambiguity in the 3-point star product \eqref{eq:3_pt_product}. As discussed in section \ref{sec:signs:general}, this helps establish the sign ambiguity as the driving force behind the disagreement between $Z_{\text{HS}}$ and $Z_{\text{local}}$.

\subsection{Spin parity} \label{sec:signs:spin_parity}

At the holomorphic level, i.e. without invoking complex conjugation, we find another discrete symmetry with the same fate. This is the symmetry of \emph{spin parity}, i.e. the distinction between even and odd spins. In twistor space, it acts by substituting $Y^a\rightarrow iY^a$. Since spin $s$ is encoded in twistor functions of homogeneity $-2\pm 2s$, even spins correspond to functions that satisfy $F(iY) = -F(Y)$, and odd spins -- to functions that satisfy $F(iY) = +F(Y)$. In HS algebra, the commutator of two even-spin functions is again an even-spin function; this is why the restriction to even spins is a consistent truncation of HS gravity (and one that is essential for unitarity in $dS_4$). However, in the correlators \eqref{eq:HS_correlators_scalar},\eqref{eq:HS_correlators_spins} and the partition function \eqref{eq:Z}, we are dealing not with commutators, but \emph{anti-commutators}, and the anti-commutator of two even-spin twistor functions is an odd-spin function! This leads to a surprising situation, quite analogous to what we saw for complex conjugation.

Again, the problem manifests at odd orders. The symmetrized star product of an odd number of even-spin functions is again an even-spin function. On the other hand, the HS trace $\tr_\star f(Y)$ picks out the piece of $f(Y)$ with homogeneity 0, which corresponds to spin 1, and vanishes on even spins. Thus, the spin-parity symmetry of HS algebra predicts that the correlators \eqref{eq:HS_correlators_scalar},\eqref{eq:HS_correlators_spins} should vanish for even spins and odd $n$, and that $Z_{\text{HS}}$ with even-spin sources should be even. Again, for $Z_{\text{HS}}$ in the $S_3$ basis, we already reached this conclusion, in the form of the vanishing of all odd orders in \eqref{eq:trace_S3}-\eqref{eq:Z_S3}. However, the conclusion is \emph{not} obeyed by the $n$-point correlators. In particular, both the scalar $J^{(0)}$ and the stress tensor $T_{\mu\nu}\sim J^{(2)}_{\mu\nu}$ have nonzero 3-point functions despite their even spin. Again, the mismatch can be traced back to the 3-point star product \eqref{eq:3_pt_product}, where the LHS (upon symmetrizing over the order of factors) is even-spin, but the RHS is odd-spin. Again, both in \eqref{eq:3_pt_product} and in the correlators \eqref{eq:HS_correlators_scalar},\eqref{eq:HS_correlators_spins}, the culprit is spontaneous symmetry breaking via the sign ambiguity: in every case, the symmetry-violating RHS has an ambiguous sign, which makes for a zero average.

\section{Conflict 2: Lorentzian boundary correlators} \label{sec:Lorentzian}

At this point, it's worth recalling where the sign ambiguity in \eqref{eq:3_pt_product} originates: it is the result of a Gaussian integral over a complex spinor space. In much of the HS literature, complex spinors are avoided altogether by working in Lorentzian $AdS_4$. There, the twistors (i.e. spinors of $SO(2,3)$) and boundary spinors (i.e. spinors of $SO(1,2)$) have a real structure. This provides natural contours for the twistor and spinor integrals inside star products. As we will see, the sign ambiguity in this case indeed disappears, along with the associated breaking of discrete symmetries. There are no longer any hints of linearity violation in HS algebra. However, the disagreement between $Z_{\text{HS}}$ and $Z_{\text{local}}$ remains. Instead of lurking in the transition from correlators to finite sources, the disagreement now appears already in the correlators! As we will see, the products that make up the HS-algebraic correlators in this signature have unambiguous signs, \emph{but not the right ones} to reproduce the CFT result.

\subsection{Boundary topology and propagators}

Let us describe the Lorentzian $AdS_4$ setup in more detail. The embedding space is now $\bbR^{2,3}$, whose metric we again denote by $\eta_{\mu\nu}$, with mostly-plus signature. The $AdS_4$ bulk is given by the points $x^\mu\in\bbR^{2,3}$ with $x\cdot x = -1$. The boundary 2+1d spacetime is given by points $\ell^\mu\in\bbR^{2,3}$ with $\ell\cdot\ell = 0$, along with the equivalence relation $\ell^\mu\cong\rho\ell^\mu$ for $\rho>0$. A pair of boundary points $\ell,\ell'$ are timelike-separated if $\ell\cdot\ell'>0$ and spacelike-separated if $\ell\cdot\ell'<0$. The boundary has the topology and conformal metric of $S_1\times S_2$, where the $S_1$ is a timelike circle. It can be viewed as a union of two 2+1d Minkowski spaces $\calM_A\cup\calM_B$, glued along their null infinity as $\scri^+(\calM_A)=\scri^-(\calM_B)$ and vice versa, and with their timelike and spacelike infinities identified as $i^+(\calM_A)=i^-(\calM_A)=i^0(\calM_B)$ and vice versa. 

Note that our use of the embedding-space formalism is predicated on not unwrapping the $S_1$ timelike circle. In fact, we would argue that this ``default'', non-unwrapped topology is the only one that is consistent with global HS symmetry. In any case, the quantities that we'll compute here will be insensitive to this issue.

The Lorentzian CFT has the action:
\begin{align}
 S_{\text{CFT}} = \int d^3\ell\,\bar\phi_I\Box\phi^I + \int d^3\ell \sum_{s=0}^\infty A^{(s)}_{\mu_1\dots\mu_s}(\ell)\, J_{(s)}^{\mu_1\dots\mu_s}(\ell) \ . \label{eq:Lorentzian_S_local}
\end{align}
It will suffice for us to consider correlators \eqref{eq:correlators_scalar} of the scalar operator $J^{(0)}$, which now read:
\begin{align}
 \left<J^{(0)}(\ell_1)\dots J^{(0)}(\ell_n)\right>_{\text{connected}} = N i^n \left(\prod_{p=1}^n G(\ell_p,\ell_{p+1}) + \text{permutations}\right) \ , \label{eq:Lorentzian_correlators_scalar}
\end{align}
where $G=\Box^{-1}$ is a Lorentzian propagator for the fundamental fields $\phi^I$. In a standard Lorentzian QFT (with a non-compact time axis), the correlators derived directly from the path integral would be time-ordered, making the propagator on the RHS of \eqref{eq:Lorentzian_correlators_scalar} the Feynman propagator. In a causally consistent patch around $\ell'=\ell$, where the time periodicity can be ignored, the Feynman propagator reads:
 \begin{align}
   G_F(\ell,\ell') = \left\{ \def\arraystretch{1.5}
    \begin{array}{cl}
      \displaystyle \frac{1}{4\pi\sqrt{2\ell\cdot\ell'}} & \quad \ell\cdot\ell' > 0 \text{ (timelike separation)} \\
      \displaystyle \frac{-i}{4\pi\sqrt{-2\ell\cdot\ell'}} & \quad \ell\cdot\ell' < 0 \text{ (spacelike separation)}
    \end{array} \right. \ . \label{eq:G_F}
\end{align}
Globally on the $S_1\times S_2$ topology, this is not a valid inverse of $\Box$, since $\Box G_F$ has an unwanted (and imaginary) extra singularity at the antipodal point $\ell'=-\ell$. Instead, the unique propagator $G=\Box^{-1}$ that makes global sense on $S_1\times S_2$ reads:
\begin{align}
 G_{S_1\times S_2}(\ell,\ell') = \left\{ \def\arraystretch{1.2}
 \begin{array}{cl}
   \displaystyle \frac{1}{4\pi\sqrt{2\ell\cdot\ell'}} & \quad \ell\cdot\ell' > 0 \text{ (timelike separation)} \\
   0 & \quad \ell\cdot\ell' < 0 \text{ (spacelike separation)}
 \end{array} \right. \ . \label{eq:G_global}
\end{align}
Here, we'll avoid this subtlety by only considering correlators at \emph{timelike-separated points}, where $G_{S_1\times S_2}$ and $G_F$ agree, and we can simply write:
\begin{align}
 G(\ell,\ell') = \frac{1}{4\pi\sqrt{2\ell\cdot\ell'}}  \ . \label{eq:Lorentzian_G}
\end{align}

\subsection{Twistors, spinors and star products}

We now turn to twistors, spinors and HS algebra in the $SO(2,3)$ signature. Twistors $U^a$ are now the spinors of $SO(2,3)$. Their complex conjugation properties follow those of $SO(1,2)$ spinors. The complex conjugation $U^a\rightarrow \bar U^a$ now squares to $+1$, and can be represented simply as component-wise conjugation $\bar U^a = (U^a)^*$. There is an invariant notion of real twistors, and the twistor metric $I_{ab}$ is real. We take the gamma matrices $(\gamma_\mu)^a{}_b$ to be real as well. However, their anticommutator is then $\{\gamma_\mu,\gamma_\nu\} = +\eta_{\mu\nu}$, with the opposite sign from the one that led to ``real'' gamma matrices in $SO(1,4)$ signature. As a result, we'll need to reverse signs in every formula from section \ref{sec:correlators} where a star product leads to a scalar product of two vectors. The chiral spinor projectors at a bulk point $x$ are now $P(\pm x) = \frac{1}{2}(1\pm ix)$. For boundary points, we will continue to use real ``projectors'' $P(\ell) = \frac{1}{2}\ell$, associated with real delta functions $\delta_\ell(U)$. These are now delta functions with a \emph{real argument}, i.e. well-defined distributions on the real twistor space, as opposed to the more formal delta functions with complex argument from $SO(1,4)$ signature. The HS star product is again given by \eqref{eq:star}-\eqref{eq:star_int}, with the Hermitian property $\overline{f\star g} = \bar g\star\bar f$. Since twistor space is real, the integral star-product formula \eqref{eq:star_int} now comes with a natural choice of contour.

Since the delta functions $\delta_\ell(U)$ are now real, their integrals \eqref{eq:spinor_delta_integral} should be positive, enforced by an absolute value:
\begin{align}
 \int_{P(\ell')} d^2u_{(\ell')}\,\delta_{\ell}(u_{(\ell')}) f(u_{(\ell')}) = \frac{2}{|\ell\cdot\ell'|}\,f(0) \ .
\end{align}
Here again, our restriction to timelike separations $\ell\cdot\ell'>0$ will be convenient, enabling us to ignore this subtlety and proceed without writing absolute values. The basic 2-point star product \eqref{eq:2_pt_product} then reads:
\begin{align}
 \delta_\ell(Y)\star\delta_{\ell'}(Y) = \frac{2}{\ell\cdot\ell'}\,\exp\left(-\frac{iY\ell\ell' Y}{2\ell\cdot\ell'}\right) \ . \label{eq:Lorentzian_2_pt_product}
\end{align}
The 3-point product can again be deduced from this via eq. \eqref{eq:delta_Fourier_xi_first}, which leads to a Gaussian integral \eqref{eq:Gaussian_spinor} over $P(\ell'')$, with the quadratic form in the exponent inherited from that in \eqref{eq:Lorentzian_2_pt_product}:
\begin{align}
 A_{ab} = -\frac{i(\ell\ell' - \ell'\ell)_{ab}}{2\ell\cdot\ell'} \ .
\end{align} 
Similarly to the Euclidean case \eqref{eq:3_pt_product}, this Gaussian integral evaluates to:
\begin{align}
 \delta_\ell(Y)\star\delta_{\ell'}(Y)\star\delta_{\ell''}(Y) = \pm i \sqrt{\frac{\ell\cdot\ell''}{2(\ell\cdot\ell')(\ell'\cdot\ell'')}} \, \delta_\ell(Y)\star\delta_{\ell''}(Y) \ . \label{eq:Lorentzian_3_pt_product}
\end{align}
This time, however, the sign on the RHS is \emph{not} ambiguous! The value of an imaginary Gaussian integral $\int e^{ix^2} dx$ over the real line is $\sqrt{\pi}\,e^{\pi i/4}$, while that of $\int e^{-ix^2} dx$ is $\sqrt{\pi}\,e^{-\pi i/4}$. Therefore, the sign in \eqref{eq:Lorentzian_3_pt_product} is completely determined by the signature of the real quadratic form $\frac{1}{i}A_{ab}$ over $P(\ell'')$: it is $+$ for $(+,+)$ signature, and $-$ for $(-,-)$ signature (for timelike-separated points $(\ell,\ell',\ell'')$, the signature cannot be mixed, as one can see from calculating the determinant). Now, $A_{ab}$, and thus the sign in \eqref{eq:Lorentzian_3_pt_product}, is manifestly odd under the interchange $\ell\leftrightarrow\ell'$. The only possible Lorentz-invariant conclusion is that the sign depends on the time ordering of $(\ell,\ell',\ell'')$; more specifically, since time is circular, the sign must be distinguishing cyclic vs. anti-cyclic permutations of the ``default'' time ordering $\ell<\ell'<\ell''$. Which of the two leads to a positive sign, and which to a negative sign, depends on our arbitrary choice of orientation for the time circle. For concreteness, let's associate the ``default'' ordering $\ell<\ell'<\ell''$ with a positive sign. 

By recursion, the $n$-point trace \eqref{eq:trace_n_point_raw} for timelike-separated points on the Lorentzian boundary reads:
\begin{align}
 \tr_\star\big(\delta_{\ell_1}(Y)\star\ldots\star \delta_{\ell_n}(Y) \big) = \frac{4i^{n-2}(-1)^\chi}{\sqrt{\prod_{p=1}^n (-2\ell_p\cdot\ell_{p+1})}} 
  = 4i^{n-2}(4\pi)^n (-1)^\chi \prod_{p=1}^n G(\ell_p,\ell_{p+1}) \ , \label{eq:Lorentzian_trace_n_point}
\end{align}
where $G$ is the propagator \eqref{eq:Lorentzian_G}. The sign is again not arbitrary, but can be determined as follows. First, note that we can get from each point $\ell_p$ to the next one $\ell_{p+1}$ either along a future-pointing arc, or along its complementary past-pointing arc. The sign in \eqref{eq:Lorentzian_trace_n_point} is then $(-1)^\chi$, where $\chi+1$ is the winding number of the cyclic sequence $\ell_1\rightarrow\ell_2\rightarrow\dots\rightarrow\ell_n\rightarrow\ell_1$ around the time circle, choosing the future-pointing arc at each step. Alternatively, if we draw the first propagator $\ell_1\rightarrow\ell_2$ and the last propagator $\ell_n\rightarrow\ell_1$ as future-pointing, and for every intermediate propagator choose a future-pointing or past-pointing arc so as not to cross the time coordinate of $\ell_1$, then $\chi$ is the number of propagators that end up past-pointing.

Several conclusions follow. First, we see that to reproduce the reality properties of the correlators \eqref{eq:Lorentzian_correlators_scalar}, the twistor functions $\kappa^{(0)}(\ell;Y)$ that represent local spin-0 insertions must be chosen real $\kappa^{(0)}(\ell;Y)\sim \pm\delta_\ell(Y)$, unlike in the Euclidean. Second, the Lorentzian 3-point product \eqref{eq:Lorentzian_3_pt_product} and its $n$-point generalization \eqref{eq:Lorentzian_trace_n_point} \emph{respect} the discrete symmetries that were spontaneously broken by their Euclidean counterparts. This follows from the extra sign flips incurred upon reordering the star-product factors, due to the changed time ordering. In fact, the discrete symmetries \emph{must} be respected, since in the absence of sign ambiguities, there is no mechanism that might break them. By the same token, we don't expect any hidden non-linearities when integrating over the correlators' insertion points. 

On the other hand, the alternating signs in \eqref{eq:Lorentzian_trace_n_point} mean that, unlike in the Euclidean case \eqref{eq:HS_correlators_scalar}, the HS-algebraic traces \emph{can't} reproduce correctly the correlators \eqref{eq:Lorentzian_correlators_scalar} from the CFT path integral:
\begin{align}
 \left<J^{(0)}(\ell_1)\dots J^{(0)}(\ell_n)\right>_{\text{connected}} \ \nsim \ \tr_\star\!\big(\kappa^{(0)}(\ell_1;Y)\star\ldots\star\kappa^{(0)}(\ell_n;Y) \big) + \text{permutations} \ , \label{eq:Lorentzian_HS_correlators_scalar}
\end{align} 
In particular, for odd $n$, the HS-algebraic trace \eqref{eq:Lorentzian_trace_n_point} \emph{vanishes} when summed over permutations. This is in agreement with the discrete symmetries, but in obvious conflict with the CFT correlator. One could in principle introduce by hand some \emph{compensating} alternating signs into the sum over permutations in \eqref{eq:Lorentzian_HS_correlators_scalar}. However, then the RHS of \eqref{eq:Lorentzian_HS_correlators_scalar} can't be considered as the expansion of an HS-algebraic partition function, such as in \eqref{eq:F}-\eqref{eq:Z}. 

To summarize, the HS-algebraic ``Feynman diagrams'' $\tr_\star\!\big(\kappa^{(0)}(\ell_1;Y)\star\ldots\star\kappa^{(0)}(\ell_n;Y) \big)$ (for spin 0, at timelike-separated points) are no longer sign-ambiguous, no longer break discrete symmetries, and integrating over them does not appear to threaten with any non-linearities. However, their signs are different from those required to correctly reproduce the correlators from the CFT path integral.

\section{Explanation 2: On-shell vs. off-shell boundary particles} \label{sec:shell}

In the previous two sections, we ascribed the disagreement between the boundary CFT and HS algebra to certain sign issues. While perhaps satisfactory at the technical level, this explanation is rather anti-climactic. Is there some more conceptual reason for the disagreement? In this section, we will present such a reason: the CFT path integral deals with off-shell boundary particles, while HS algebra sees on-shell ones. We begin in section \ref{sec:shell:dimensions} with a degree-of-freedom counting argument, which applies to any spacetime signature. We will then focus on a Lorentzian boundary, where the issue can be framed more explicitly in terms of on-shell vs. off-shell (in Euclidean, on-shell particles exist only in a complexified sense). In section \ref{sec:shell:QM}, we will review the identification between HS algebra and the algebra of operators on the space of on-shell boundary particle states. We then apply it, in section \ref{sec:shell:correlators}, to understand the strange agreement-up-to-signs of the Lorentzian correlators from section \ref{sec:Lorentzian}. 

\subsection{Dimensionality of algebras} \label{sec:shell:dimensions}

Suppose that the CFT partition function \eqref{eq:Z_local_scalar} (with spin-0 sources, to simplify the discussion) in fact agreed with the HS-algebraic partition function \eqref{eq:Z}. The first is given in terms of traces $\tr(\sigma G)^n$ within the algebra of infinite-dimensional ``matrices'' $\Pi(\ell',\ell)$ over the space of boundary fields $\phi(\ell)$. The second is given in terms of traces $\tr_\star(F\star\ldots\star F)$ within HS algebra. The would-be equality of the partition functions is a statement of isomorphism between the two algebras, via the linear mapping \eqref{eq:F}, which for the spin-0 case reads simply:
\begin{align}
 F(Y) = \int d^3\ell\,\kappa^{(0)}(\ell;Y)\,\sigma(\ell) \ . \label{eq:F_scalar}
\end{align}
In more detail, the would-be isomorphism is between bilocal boundary functions $\Pi(\ell',\ell)$ and twistor functions $F(Y)$, such that:
\begin{enumerate}
 \item The function $\Pi(\ell',\ell) = \delta(\ell,\hat\ell)\delta(\ell',\hat\ell)$, which describes a local spin-0 source at a point $\hat\ell$, maps to the twistor function $F(Y) = \kappa^{(0)}(\hat\ell;Y)$.
 \item The matrix product $\Pi_1 G\Pi_2$, where $G(\ell,\ell')$ is the propagator \eqref{eq:G}, maps to the star product $F_1\star F_2$.
 \item The trace $\tr(\Pi G)$ maps to $-\frac{1}{4}\tr_\star\!F$.
\end{enumerate}
Note that the focus on spin 0 does not restrict the isomorphism's generality. Indeed, if we consider pairs of spin-0 insertions $\Pi_1(\ell',\ell) = \delta(\ell,\hat\ell_1)\delta(\ell',\hat\ell_1)$ and $\Pi_2(\ell',\ell) = \delta(\ell,\hat\ell_2)\delta(\ell',\hat\ell_2)$ at points $\hat\ell_1$ and $\hat\ell_2$, then their products $\Pi_1 G\Pi_2 = G(\hat\ell_1,\hat\ell_2)\delta(\ell,\hat\ell_1)\delta(\ell',\hat\ell_2)$ already span the entire algebra of bilocals $\Pi(\ell',\ell)$. Thus, an agreement of the partition functions \emph{even just for spin-0 sources} already calls for a complete isomorphism between the two algebras.

However, such an isomorphism is clearly impossible, by simple degree-of-freedom counting. The bilocal matrix algebra consists of functions $\Pi(\ell',\ell)$ of $3\times 2 = 6$ spacetime coordinates, while HS algebra consists of functions $F(Y)$ of 4 twistor components. This makes the disagreement between $Z_{\text{HS}}$ and $Z_{\text{local}}$ seem natural and unavoidable: the underlying algebras simply have different dimensions. 

Now, let's understand the origin of the dimensional mismatch itself. One way to do this is in terms of gauge redundancy. The twistor function $F(Y)$ contains only physical degrees of freedom. So does the local spin-0 source $\sigma(\ell)$. However, the sources $A^{(s)}_{\mu_1\dots\mu_s}(\ell)$ with $s>0$ are gauge-redundant, due to the conservation of the currents $J^{(s)}_{\mu_1\dots\mu_s}(\ell)$. The \emph{bilocal} sources $\Pi(\ell',\ell)$ are even \emph{more} gauge-redundant, due to the field equations satisfied by each ``leg'' of the bilocal operator $\bar\phi_I(\ell')\phi^I(\ell)$. And, as we've seen, when $Z_{\text{local}}$ is viewed algebraically, all the degrees of freedom in $\Pi(\ell',\ell)$ come into play, even if initially we were only interested in the spin-0 source $\sigma(\ell)$.

There is another, equivalent way of understanding the mismatch, which can be made particularly concrete in Lorentzian boundary signature. In the same way that bilocals $\Pi(\ell',\ell)$ are operators on the space of off-shell fields $\phi(\ell)$, the HS algebra elements $F(Y)$ can be viewed as operators on the space of \emph{on-shell} boundary particle states. Let us review this identification in detail.

\subsection{HS algebra as quantum mechanics} \label{sec:shell:QM}

Our definition of HS algebra in section \ref{sec:correlators:HS:algebra} was quite abstract. In this section, we will bring into play a more concrete point of view: HS algebra is just the \emph{operator algebra in the quantum mechanics} of a free massless particle in the 2+1d boundary spacetime. This identification has been made from various points of view in \cite{Eastwood:2002su,Segal:2002gd,Iazeolla:2008ix}. A simplified version, adapted to a 2+1d boundary and utilizing twistors, was presented in \cite{Neiman:2018ufb}; we will now summarize it here.

\subsubsection{Twistor space as phase space}

Consider a free massless particle in a conformally flat $d$-dimensional Lorentzian spacetime. What is the classical phase space of such a particle? A point in the phase space consists of a lightray, namely the particle's wordline, along with an affine ``magnitude'' to represent the particle's energy. Since the theory is conformal, we can represent the $d$-dimensional spacetime in an embedding-space formalism, as the projective lightcone in $\bbR^{2,d}$. Every point in the $d$-dimensional spacetime then becomes a lightray through the origin in $\bbR^{2,d}$, while a \emph{lightray} in $d$ dimensions becomes a \emph{totally null plane} through the origin in $\bbR^{2,d}$. Combining this with the ``magnitude'' that encoded the particle's energy, we conclude that the particle's phase space consists of totally null bivectors $L^{\mu\nu}$ in $\bbR^{2,d}$. In a flat conformal frame $\ell\cdot\ell_{\infty} = -\frac{1}{2}$ defined by a ``point at infinity'' $\ell_\infty^\mu \in \bbR^{2,d}$, the energy-momentum of a particle described by $L^{\mu\nu}$ reads:
\begin{align}
 p_\mu = 2L_{\mu\nu}\ell_{(\infty)}^\nu \ . \label{eq:p}
\end{align}
The Poisson brackets of $L^{\mu\nu}$ are fixed up to normalization by the $SO(2,d)$ conformal symmetry. The normalization can in turn be fixed \cite{Neiman:2018ufb} so as to match the identification of \eqref{eq:p} as a translation generator. The resulting Poisson brackets read:
\begin{align}
 \left\{L^{\mu\nu}, L_{\rho\sigma} \right\} = 4\delta^{[\mu}_{[\rho}\, L^{\nu]}{}_{\sigma]} \ ,
\end{align} 
implying that the $L^{\mu\nu}$ generate the conformal group. Now, in our case of interest $d=3$, a totally null bivector $L^{\mu\nu}$ is simply the square of a twistor $Y^a$, via:
\begin{align}
 L^{\mu\nu} = \frac{1}{8}Y\gamma^{\mu\nu}Y \ . \label{eq:L}
\end{align}
The Poisson bracket for this new phase space variable $Y^a$ is again fixed by $SO(2,3)$ symmetry, while the normalization can be fixed by matching to \eqref{eq:L}. The result reads:
\begin{align}
 \left\{Y^a, Y^b \right\} = 2I^{ab} \ , \label{eq:Poisson}
\end{align}
implying a symplectic form:
\begin{align}
 \Omega_{ab} = -\frac{1}{2}I_{ab} \ . \label{eq:Omega}
\end{align}
Now, consider the quantization of this particle mechanics, where the Poisson bracket \eqref{eq:Poisson} is upgraded into a commutator $[\hat Y^a,\hat Y^b] = 2iI^{ab}$. General quantum operators $\hat f$ can be represented as ordinary functions $f(Y)$ of the phase space coordinates $Y^a$, with the convention that $Y^{a_1}\dots Y^{a_n}$ represents the totally symmetrized product of the operators $(\hat Y^{a_1},\dots,\hat Y^{a_n})$. The operator product $\hat f\hat g$ is then represented by the Moyal star product $f(Y)\star g(Y)$, \emph{which is precisely the star product \eqref{eq:star}-\eqref{eq:star_int} of HS algebra}. Thus, HS algebra is just the algebra of operators in the quantum mechanics of the boundary particle. 

The trace operation $\tr_\star f(Y)$ is not quite identical to the quantum-mechanical trace $\tr\hat f$, but is closely related, via: 
\begin{align}
 \tr\hat f = \frac{1}{4}\int d^4Y f(Y) = \frac{1}{4} \tr_\star\!\left(f(Y)\star\delta(Y)\right) \ . \label{eq:tr_str}
\end{align}
Here, the factor of 4 is due to the ratio between our twistor measure $d^4Y$, constructed from $I_{ab}$, and the one constructed from the symplectic form \eqref{eq:Omega}. It is the same factor of 4 as in \eqref{eq:trace_matrix}, and can be thought of as an explanation for the factor of $\frac{1}{4}$ in the HS-algebraic partition function \eqref{eq:Z}. 

The twistor delta function $\delta(Y)$ represents an operator that flips the sign of $Y^a$, as can be evidenced by its adjoint action \eqref{eq:delta_flip}. From the point of view of the boundary particle, one might expect this sign flip to be invisible, since eq. \eqref{eq:L} defines $Y^a$ only up to sign. However, the issue is more subtle, and we'll return to it in section \ref{sec:discuss:spin_locality:antipodal}.

\subsubsection{Boundary spinor space as configuration space}

Having interpreted twistor space as a phase space, one may wonder if it can be decomposed into configuration variables and their conjugate momenta. This would necessitate splitting twistor space into two 2d subspaces that are Lagrangian, i.e. totally null under the symplectic form $\Omega_{ab}\sim I_{ab}$. Such a splitting is precisely given by a choice of two boundary points $\ell$ and $l$. The role of configuration and momentum variables is then played by the spinors $y_{(\ell)}\in P(\ell)$ and $y_{(l)}\in P(l)$ that make up the twistor $Y^a$. These spinors have a simple meaning within the boundary particle's mechanics. In particular, $y_{(l)}$ is just a square root of the on-shell energy-momentum \eqref{eq:p} in the conformal frame defined by $\ell$:
\begin{align}
 p_\mu = \frac{1}{4}\,y_{(l)}\gamma_\mu \ell\, y_{(l)} \ . \label{eq:p_spinors}
\end{align}
In other words, $y_{(l)}$ is a momentum spinor. Pure states of the boundary particle can be expressed as wavefunctions $\psi(y_{(l)})$. As we saw, the twistor delta function $\delta(Y)$ represents a sign-reversal operator on the phase space. Acting on pure states, this operator sends $\psi(y_{(l)})\rightarrow \psi(-y_{(l)})$. 

In secret, we already encountered this formalism, in section \ref{sec:partition:ISO3:momentum_spinors}. In particular, the transform \eqref{eq:boundary_uu'_basis} between $F(Y)$ and $\tilde C_{\ell,l}(u,u')$ is just the Wigner-Weyl transform between two representations of a quantum-mechanical operator $\hat F$: as a phase space function $F(Y)$, vs. as matrix elements $\tilde C_{\ell,l}(u,u')$ between the basis states $y_{(l)}=u$ and $y_{(l)}=u'$. Of course, section \ref{sec:partition:ISO3:momentum_spinors} was written in the context of a Euclidean boundary. However, the Euclidean signature changes little, except to complexify the phase-space coordinates $Y^a$ and configuration variables $y^a_{(l)}$. One difference is that, in Euclidean, the operator represented by $\delta(Y)$ sends $\psi(y_{(l)})$ not to $\psi(-y_{(l)})$, but to $-\psi(-y_{(l)})$, as we can see from the overall sign in eq. \eqref{eq:delta_matrix}. As a result, in Euclidean, there would be an overall minus sign in \eqref{eq:tr_str}. This extra sign arises from the minus in the decomposition \eqref{eq:twistor_boundary_decomposition} of the twistor metric, $d^4Y = -\frac{1}{4}d^2y_{(\ell)}d^2y_{(l)}$. On a Lorentzian boundary, this minus is suppressed, since the real integration contours impose positivity on all the measures and delta functions.

To sum up, we see that $Z_{\text{HS}}$ can be thought of as a determinant (or an exponentiated trace) over the boundary particle's Hilbert space, i.e. over the space of solutions to the boundary field equation $\Box\phi = 0$. This is particularly explicit in the $\tilde C_{\ell,l}(u,u')$ basis of section \ref{sec:partition:ISO3:momentum_spinors}, where the Hilbert space is realized as functions $\psi(u)$ of a boundary momentum-spinor $u\in P(l)$, and $\tilde C_{\ell,l}(u,u')$ is just a matrix element between the $\ket{u}$ state and the $\ket{u'}$ state.

\subsection{On-shell HS algebra vs. off-shell CFT correlators} \label{sec:shell:correlators}

The above understanding of HS algebra makes it even clearer that $Z_{\text{HS}}$ should disagree with the CFT path integral $Z_{\text{local}}$. The two are similar superficially, in that they both calculate determinants over boundary fields. However, HS algebra sees only \emph{on-shell} boundary fields, i.e. solutions to the source-free field equation $\Box\phi = 0$, which describe states of the free boundary particle. In contrast, the CFT path integral is over \emph{off-shell} fields, and is in fact composed of propagators $G=\Box^{-1}$ that solve the field equation $\Box G(\ell,\ell') = \delta(\ell,\ell')$ \emph{with source}. It is only natural that determinants over such different spaces will disagree! 

What begins to seem strange at this point is that HS algebra ever managed to reproduce the CFT's $n$-point correlators. Let us now explain this ``miracle'' from the point of view of boundary particle mechanics. We will work in Lorentzian signature, and explain both how the correlators match to the extent that they do, and how the sign disagreements come about. As in section \ref{sec:Lorentzian}, we restrict for simplicity to spin 0.

Consider the twistor function $\kappa^{(0}(\ell;Y)\sim\delta_\ell(Y)$ that describes a local spin-0 insertion at the point $\ell$. Recall that the 2d subspace $P(\ell)$ is totally null under the twistor metric, which makes it, from the point of view of boundary particle mechanics, a Lagrangian submanifold of the phase space. This implies that the phase space function $\delta_\ell(Y)$ describes (up to normalization) a \emph{projector} $\ket{0_\ell}\!\bra{0_\ell}$ onto a particular quantum state $\ket{0_\ell}$. Specifically, if we fix a second boundary point $l$ and decompose twistor space as $Y^a = y^a_{(\ell)} + y^a_{(l)}$, then $y_{(\ell)}$ and $y_{(l)}$ are canonical conjugates, and $\ket{0_\ell}$ is just the state with $y_{(l)}=0$. In other words, it is the state with vanishing energy-momentum \eqref{eq:p_spinors} in the flat conformal frame defined by $\ell$. 

Thus, the LHS of the HS-algebraic ``Feynman diagrams'' \eqref{eq:Lorentzian_trace_n_point} takes the form:
\begin{align}
 \tr_\star\big(\delta_{\ell_1}(Y)\star\ldots\star \delta_{\ell_n}(Y) \big) 
   \ \sim \ \braket{0_{\ell_1}|0_{\ell_2}}\braket{0_{\ell_2}|0_{\ell_3}}\dots\braket{0_{\ell_{n-1}}|0_{\ell_n}}\braket{0_{\ell_n}|0_{\ell_1}} \ , \label{eq:trace_n_point_QM}
\end{align}
where the inner product $\braket{0_{\ell'}|0_\ell}$ can be thought of as a wavefunction $\psi_\ell(\ell')$ of the state $\ket{0_\ell}$. What does this wavefunction look like? It must be a solution to $\Box\phi = 0$ that is symmetric around the chosen point $\ell$. If we restrict again for simplicity to timelike separations, then there's exactly one such solution in the neighborhood of $\ell$. The solution is:
\begin{align}
 \psi_\ell(\ell') \sim \pm G(\ell,\ell') \ , \label{eq:psi}
\end{align}
where $G$ is the propagator \eqref{eq:Lorentzian_G}, and the sign depends on whether $\ell'$ is to the future or to the past of $\ell$. Globally on the $S_1\times S_2$ boundary, the sign of $\psi_\ell(\ell')$ is ill-defined, since there's no globally consistent time ordering between $\ell$ and $\ell'$. This is in fact a general property of solutions to $\Box\phi = 0$ on the $S_1\times S_2$ topology. If we denote the frequency along the time circle by $\omega$, and the $S_2$ angular momentum by $j$, then the conformal Laplacian becomes:
\begin{align}
 \Box = \omega^2 - j(j+1) - \frac{1}{4} \ .
\end{align}
Thus, $\Box\phi = 0$ translates into $|\omega| = j+\frac{1}{2}$, making $\omega$ a half-integer. Therefore, not only $\psi_\ell(\ell')$, but \emph{any} on-shell solution will have a sign inconsistency upon traversing the time circle (of course, the inconsistency disappears if the time circle is decompactified). This sign issue won't bother us, because the HS-algebraic ``Feynman diagrams'' \eqref{eq:trace_n_point_QM} depend on the state $\ket{0_\ell}$ only in its squared form $\ket{0_\ell}\!\bra{0_\ell}$.

The agreement up to sign of the HS-algebraic ``Feynman diagrams'' with their CFT counterparts is now clear. The inner products $\braket{0_{\ell_p}|0_{\ell_{p+1}}}$ on the RHS of \eqref{eq:trace_n_point_QM} are given by the on-shell wavefunctions \eqref{eq:psi}, which reproduce the off-shell propagators $G(\ell_p,\ell_{p+1})$ up to time-ordering-dependent signs. These are exactly the time-ordering-dependent signs that we encountered on the RHS of \eqref{eq:Lorentzian_trace_n_point}.

\section{Aftermath: choosing $Z_{\text{HS}}$} \label{sec:discuss}

In this paper, we identified a disagreement between the partition function $Z_{\text{local}}$ of the boundary vector model and the HS-algebraic expression $Z_{\text{HS}}$, despite the matching of the corresponding $n$-point correlators (up to sign, in the $SO(2,3)$ case), and despite the absence of contact corrections to $Z_{\text{local}}$ in the spin-0 sector. We then analyzed the causes of this disagreement. After all is said and done, the disagreement must be accepted as a fact. We must either give up on $Z_{\text{HS}}$, and with it lose the power of manifest global HS symmetry, or give up on $Z_{\text{local}}$, and with it the connection to the local boundary field theory. We will now advocate for the second choice, and explore its implications. In section \ref{sec:discuss:spin_locality}, we will discuss replacing boundary locality with boundary spin-locality as a guiding principle. Then, in section \ref{sec:discuss:dS}, we will address our core motivations, and discuss how choosing $Z_{\text{HS}}$ can benefit the project of higher-spin dS/CFT.

\subsection{Boundary spin-locality} \label{sec:discuss:spin_locality}

\subsubsection{Leaving the local CFT behind}

From the point of view of AdS/CFT orthodoxy, choosing $Z_{\text{HS}}$ over $Z_{\text{local}}$ is madness. Isn't the CFT always right? Doesn't it provide the very definition of the bulk quantum gravity theory? Isn't disagreement with it synonymous with having made a mistake? Yet we can answer with counter-questions of our own. Is symmetry -- in this case, HS symmetry -- not the highest principle? Is there any actual evidence that the bulk HS gravity matches the boundary vector model at finite sources, as opposed to just the $n$-point correlators? 

Also, let us recall how boundary locality comes about in AdS/CFT, from the bulk point of view. The bulk quantum gravity theory is non-local, but its non-locality is usually \emph{confined to some length scale}, such as the string length. In the bulk$\rightarrow$boundary limit, this finite non-locality length becomes scaled down to zero, resulting in a local boundary theory. But, as we mentioned in the Introduction, recent results have shown that HS gravity remains non-local at all scales! In particular, the quartic-vertex results of \cite{Bekaert:2015tva,Sleight:2016dba,Sleight:2017pcz} reveal the bulk interactions to be as non-local as a massless propagator. Even under the infinite rescaling of the bulk$\rightarrow$boundary limit, a massless propagator does not become pointlike. The usual argument for boundary locality is no longer obvious at all!

Perhaps, then, choosing $Z_{\text{HS}}$ is not as crazy as one might think. But then we must face the consequences of leaving the CFT behind. Our formula \eqref{eq:Z} for $Z_{\text{HS}}$ was deduced from the HS-algebraic formula \eqref{eq:HS_correlators_spins} for the CFT correlators. But if we are content with a partition function that doesn't match the CFT one, then what's the point of matching the CFT correlators in the first place? The time has come to face this inconsistency. If we value HS symmetry higher than agreement with the CFT, and we cannot have both, then we must admit the most general partition function compatible with HS symmetry. This means replacing eq. \eqref{eq:Z} with:
\begin{align}
 Z_{\text{HS}}[F(Y)] = \exp\left(\sum_{n=1}^\infty c_n \tr_\star\Big(\underbrace{F(Y)\star\ldots\star F(Y)}_{n\text{ factors}}\Big)\right) \ , \label{eq:Z_general}
\end{align}
with an arbitrary coefficient $c_n$ at each order. However, this leads us straight into the impasse that was outlined in \cite{Sleight:2017pcz} for the bulk theory. Just like the bulk fields, the argument $F(Y)$ of the partition function can be subjected to non-linear redefinitions. As long as these are trivial at first order, they won't affect the basic interpretation of $F(Y)$ in terms of bulk perturbations, or in terms of boundary modes. While HS symmetry severely restricts the possible redefinitions, it still leaves us with the freedom of a single coefficient $a_n$ at each order:
\begin{align}
 F(Y) \ \longrightarrow \ F(Y) + \sum_{n=2}^\infty a_n \underbrace{F(Y)\star\ldots\star F(Y)}_{n\text{ factors}} \ . \label{eq:redefinitions}
\end{align}
Clearly, there is enough freedom in the redefinitions \eqref{eq:redefinitions} to change the coefficients $c_n$ in \eqref{eq:Z_general} into essentially anything! The theory has thus become empty: having cut our ties with the local CFT and placed our trust in HS symmetry alone, we are left with a partition function that carries \emph{no information}, apart from what can be reshuffled arbitrarily by field redefinitions. 

\subsubsection{A spin-local path integral} \label{sec:discuss:spin_locality:path_integral}

To rise from this despair, let us recall the emerging answer to the analogous situation in the bulk \cite{Gelfond:2018vmi,Didenko:2018fgx,Didenko:2019xzz,Gelfond:2019tac}. Let us rein in the freedom of redefinitions by imposing on the boundary not the standard spacetime locality of the CFT, but spin-locality. As we will see, this will allow us to reproduce the HS-algebraic partition function \eqref{eq:Z} from first principles, up to some choices of signs and reality conditions that we will need to fix by hand. We do not claim any strict equivalence between bulk spin-locality and the boundary spin-locality that we will now introduce. However, it should be clear that one is inspired by the other.

In the bulk, spin-locality means locality with respect to the spinor arguments of the master field $C(x;Y)\equiv C(x;y_{(-x)},y_{(x)})$. As we discussed in section \ref{sec:partition:ISO3:spinor_helicity}, as $x$ approaches the boundary point $\ell$ along the geodesic defined by a second point $l$, the limiting behavior of the master field $C(x;y_{(-x)},y_{(x)})$ is described by the function $\calC_{\ell,l}(u_R,u_L)$ of two boundary spinors $u_R,u_L\in P(l)$ (in the following, we will omit the subscripts on $\calC_{\ell,l}$, to reduce clutter). Therefore, the boundary version of spin-locality should be locality with respect to the spinor variables $u_R$ and $u_L$.

In the bulk, the project of \cite{Gelfond:2018vmi,Didenko:2018fgx,Didenko:2019xzz,Gelfond:2019tac} is to impose spin-locality as a requirement on the unfolded field equations, which are obtained from Vasiliev's equations by solving with respect to the auxiliary variables $Z^a$. On the boundary, we can take a much simpler route, constructed by analogy with the CFT path integral and its standard relation to bulk fields. We will demand that the partition function be expressed as the path integral of an action $S_{\text{spin-local}}[\calC,\varphi]$, over some set of dynamical variables $\varphi$, with $\calC(u_R,u_L)$ in the role of an external source that couples to a complete set of ``single-trace'' operators $\calO[\varphi]$, i.e. a set of operators from which all others can be constructed. The requirement of spin-locality is then simply that $S_{\text{spin-local}}[\calC,\varphi]$ should contain no more than one integral over $u_R,u_L$:
\begin{align}
 Z_{\text{spin-local}}[\calC(u_R,u_L)] &= \int \calD\varphi\, e^{-S_{\text{spin-local}}[\calC,\varphi]} \ ; \label{eq:Z_HS_path_integral} \\
 S_{\text{spin-local}}[\calC,\varphi] &= \int_{P(l)} d^2u_R\,d^2u_L\,L[\calC,\varphi] \ .
\end{align}
Here, $L[\calC,\varphi]$ is a spin-local ``Lagrangian'' that contains only quantities evaluated at the single point $(u_R,u_L)$. 

To this requirement of spin-locality we now add the requirement of global HS symmetry. For this purpose, it will be convenient to switch from the $(u_R,u_L)$ spinors to their linear combinations $(u,u')$ from section \ref{sec:partition:ISO3:momentum_spinors}, replacing $\calC(u_R,u_L)$ with $\tilde\calC(u,u')$. This doesn't affect the spin-locality condition: locality in $(u_R,u_L)$ is equivalent to locality in $(u,u')$. Now, as we recall from sections \ref{sec:partition:ISO3:momentum_spinors} and \ref{sec:shell:QM}, global HS symmetry means simply that we treat $\tilde\calC(u,u')$ as a matrix over the space of spinor functions $\psi(u)$, and restrict ourselves to matrix-algebra-like products that are invariant under linear transformations of $\psi(u)$. This restriction fixes the spin-local Lagrangian $L(\tilde\calC,\varphi)$ almost completely. Disregarding trivial terms that are either constant or linear in the dynamical variables $\varphi$, we find that only terms quadratic in $\varphi$ are possible:
\begin{align}
 L[\tilde\calC,\varphi_I,\varphi^I] &= \varphi_I(u)\left[\alpha^I{}_J\,\delta(u-u') + \beta^I{}_J\,\tilde\calC(u,u') \right]\varphi^J(u') \ ; \label{eq:Lagrangian} \\
 S_{\text{spin-local}}[\tilde\calC,\varphi_I,\varphi^I] &= \alpha^I{}_J\int d^2u\,\varphi_I(u)\,\varphi^J(u) + \beta^I{}_J\int d^2u\,d^2u'\,\varphi_I(u)\,\tilde\calC(u,u')\,\varphi^J(u') \ .
\end{align}
From the point of view of spin-locality alone, the dynamical variables $\varphi$ could have been functions of both $u$ and $u'$. However, the added requirement of HS symmetry implies that the only non-trivial (i.e. higher than linear) terms in the Lagrangian must have the form \eqref{eq:Lagrangian}, with some number $N$ of spinor functions $\varphi_I(u)$ of $u$ alone, and $N$ functions $\varphi^I(u')$ of $u'$ alone. So far, these variables can be either commuting or anti-commuting; as we will see, the correct choice will be the opposite of the usual one for the CFT fields. Instead of having $\varphi_I$ and $\varphi^I$ independent, we may linearly relate them as $\varphi_I(u) = g_{IJ}\varphi^J(u)$, which will restrict $\tilde\calC(u,u')$ to even spins. As with the local CFT, we will denote the range of $I$ in this case as $1\dots 2N$. We take the internal-space metric $g_{IJ}$ to be symmetric if the $\varphi^I$ commute, and anti-symmetric if they anti-commute.

Let's now address the constant coefficients $\alpha^I{}_J$ and $\beta^I{}_J$ in the Lagrangian \eqref{eq:Lagrangian}. The coefficient $\alpha^I{}_J$ of the $\tilde\calC$-independent term can be replaced by the identity $\delta^I_J$ by appropriate $u$-independent linear transformations of $\varphi^I$. This still leaves the coefficient $\beta^I{}_J$ of the second term arbitrary. To fix it, we invoke one of the principles we borrowed from AdS/CFT: our set of external sources $\tilde\calC(u,u')$ must be \emph{complete}, i.e. the set of operators to which they couple should generate all possible operators $\calO[\varphi]$ in the theory. This can only be true if we impose a symmetry on the internal-space indices: $U(N)$, $O(2N)$ or $Sp(2N)$, depending on whether $\varphi_I$ and $\varphi^I$ are linearly related, and on whether they commute or anti-commute. The coefficient $\beta^I{}_J$ must then be proportional to the identity, and can be normalized to the identity by rescaling $\tilde\calC$. This leaves us with a spin-local action quite analogous to the (bilocal form of) the free vector model \eqref{eq:S_bilocal}:
\begin{align}
 S_{\text{spin-local}}[\tilde\calC,\varphi_I,\varphi^I] &= \int d^2u\,\varphi_I(u)\,\varphi^I(u) + \int d^2u\,d^2u'\,\varphi_I(u)\,\tilde\calC(u,u')\,\varphi^I(u') \ . \label{eq:S_spin_local}
\end{align}
As an aside, we note that the set of sources $\tilde\calC(u,u')$ is now ``even more complete'' than is usual in AdS/CFT. Normally, the theory would also include ``multi-trace'' operators, constructed by multiplying together the ``single-trace'' operators $\varphi_I(u)\varphi^I(u')$ to which $\tilde\calC(u,u')$ couples. However, such ``multi-trace'' operators cannot be coupled to a source in a way that would be both spin-local and HS-invariant. In this sense, $\tilde\calC(u,u')$ exhausts all possible external sources in the theory.

We are now ready to perform the path integral \eqref{eq:Z_HS_path_integral} over the action \eqref{eq:S_spin_local}. The integral is Gaussian, leading to:
\begin{align}
 Z_{\text{spin-local}}[\tilde\calC(u,u')] = \left(\det[1 + \tilde\calC]\right)^{\mp N} \ , \label{eq:Z_spin_local_det}
\end{align}
where the sign in the exponent is $-$ for commuting variables $\varphi^I$, and $+$ for anti-commuting. Expanding the determinant in \eqref{eq:Z_spin_local_det}, we get:
\begin{align}
 \begin{split}
   &Z_{\text{spin-local}}[\tilde\calC(u,u')] = \exp\left(\mp N\tr\ln[1 + \tilde\calC]\right) = \exp\left(\pm N\sum_{n=1}^\infty \frac{(-1)^n}{n} \tr\tilde\calC^n \right) \\
   &\quad = \exp\left(\pm N\sum_{n=1}^\infty \frac{(-1)^n}{n}  
     \int_{P(l)} d^2u_1\dots d^2u_n\,\tilde\calC(u_1,u_2)\,\tilde\calC(u_2,u_3)\dots\tilde\calC(u_{n-1},u_n)\,\tilde\calC(u_n,u_1)\right) \ .
 \end{split} \label{eq:Z_spin_local_raw}
\end{align}
This is almost identical to the HS-algebraic effective action $Z_{\text{HS}}$ as given in \eqref{eq:Z_tilde_C}. A subtle difference still remains, which we will now address.

\subsubsection{Integer spins, antipodal symmetry and reality conditions} \label{sec:discuss:spin_locality:antipodal}

The difference between $Z_{\text{spin-local}}$ in \eqref{eq:Z_spin_local_raw} and $Z_{\text{HS}}$ in \eqref{eq:Z_tilde_C} can be bridged by recalling the double-cover relationships \eqref{eq:L},\eqref{eq:p_spinors} between the twistor/spinor variables and the state of the boundary particle. Since the boundary particle has spin 0, its wavefunction $\psi(u)$ should be even in the momentum-spinor $u$. Similarly, the matrix elements $\braket{u|\hat f|u'}$ of any operator $\hat f$ should be even in both $u$ and $u'$. By the same logic, we should require our spin-local ``fields'' $\varphi_I(u)$ and $\varphi^I(u')$ to be even in their spinor argument, making the sources $\tilde\calC(u,u')$ even in both $u$ and $u'$. We can then flip the sign of the last spinor in eq. \eqref{eq:Z_spin_local_raw}, bringing it into the form:
\begin{align}
 \begin{split}
   Z_{\text{spin-local}}[\tilde\calC(u,u')] = \exp\Big(\pm N&\sum_{n=1}^\infty \frac{(-1)^n}{n} 
       \int_{P(l)} d^2u_1\dots d^2u_n \\
   &{}\times\tilde\calC(u_1,u_2)\,\tilde\calC(u_2,u_3)\dots\tilde\calC(u_{n-1},u_n)\,\tilde\calC(u_n,-u_1)\Big) \ .    
 \end{split} \label{eq:Z_spin_local}
\end{align}
This agrees with the HS-algebraic partition function $Z_{\text{HS}}$ from \eqref{eq:Z_tilde_C}, as was derived originally from the CFT correlators, provided that we choose the lower sign in \eqref{eq:Z_spin_local_det}-\eqref{eq:Z_spin_local}, i.e. provided that the variables $\varphi$ in the spin-local path integral are \emph{anti-commuting}. This last requirement is a bit strange, but not obviously problematic. Since it originates from the overall sign in $\ln Z_{\text{HS}}$, this is a good time to recall again with humility that we had that sign wrong in our previous works \cite{Neiman:2017mel,Neiman:2018ufb}.

Another discrete choice that must be made by hand is the reality condition on $\tilde\calC(u,u')$, or, equivalently, on the twistor function $F(Y)$ that corresponds to it via \eqref{eq:boundary_uu'_basis}. As we saw in sections \ref{sec:correlators}-\ref{sec:partition}, the correlators for real CFT sources are reproduced by \emph{imaginary} twistor functions $\bar F(Y) = -F(Y)$, which translates into:
\begin{align}
 \overline{\tilde\calC(u,u')} = -\tilde\calC(\bar u',\bar u) \ .
\end{align}
Again, in the original versions of \cite{Neiman:2017mel,Neiman:2018ufb}, we falsely assumed the opposite, which in turn led us to the wrong overall sign in $\ln Z_{\text{HS}}$ discussed above.

One final awkwardness remains. When working with $\bbR^3$ modes on the boundary, one can indeed restrict to functions that are even in the spinors $u$ and $u'$, as we've done to go from \eqref{eq:Z_spin_local_raw} to \eqref{eq:Z_spin_local}; such was also our approach in \cite{Neiman:2018ufb}. However, with the global $S_3$ modes, this becomes impossible. As we saw in \eqref{eq:delta_matrix}, having $\tilde\calC(u,u')$ even in $u$ and $u'$ is equivalent to $F(Y)$ being \emph{odd} under star-multiplication by $\delta(Y)$. This is indeed the case for the local boundary insertions \eqref{eq:kappa_0},\eqref{eq:kappa_s}, as one can see from applying the identity \eqref{eq:spinor_twistor_delta} to a boundary point $\ell$:
\begin{align}
 \delta(Y)\star\delta_\ell(Y) = \delta_\ell(Y)\star\delta(Y) = -\delta_\ell(Y) \ .
\end{align}
However, this is \emph{not} the case for the twistor functions \eqref{eq:k_S3} describing $S_3$ modes, since star-multiplication by $\delta(Y)$ turns the right-handed $\delta_x(Y)$ into the left-handed $\delta_{-x}(Y)$. This mismatch can be seen equivalently from the point of view of bulk fields \cite{Neiman:2017mel}. Indeed, recall from \eqref{eq:C_antipodal} that star-multiplication by $\delta(Y)$ corresponds in the bulk to the antipodal map $x\rightarrow -x$. The boundary-to-bulk propagators, e.g. $1/(\ell\cdot x)$ for spin 0, are odd under this map. However, their superpositions into $S_3$ spherical harmonics, e.g. the zeroth harmonic $1/(x\cdot x_0 - 1)$, are neither even nor odd. In particular, they are regular on one branch of the bulk $EAdS_4$, but when analytically continued into the antipodal branch, they have a pole at $x = -x_0$. This is yet another spontaneous breaking of a discrete symmetry, which is quite general in AdS/CFT with massless bulk fields (for massive fields, the boundary-to-bulk propagators aren't antipodally symmetric in the first place). 

To summarize, we managed to recover the HS-algebraic partition function $Z_{\text{HS}}$ from first principles, using spin-locality and global HS symmetry. We were forced to appeal to the CFT correlators only for the choice between commuting and anti-commuting variables, and for the reality condition on the sources. The weak point in our argument is that the step from \eqref{eq:Z_spin_local_raw} to \eqref{eq:Z_spin_local} assumed a definite antipodal symmetry (which makes sense from the point of view of boundary particle mechanics, and is satisfied by local boundary insertions), whereas in our main application, i.e. in the $S_3$ harmonics, this antipodal symmetry is sacrificed in favor of regularity in the Euclidean bulk.

\subsection{Implications for dS/CFT} \label{sec:discuss:dS}

\subsubsection{A global maximum for the Hartle-Hawking wavefunction}

Finally, we come to our own motivation for the study of higher-spin theory. One of the great open problems in theoretical physics is quantum gravity with positive cosmological constant. Higher-spin gravity turns out to be especially relevant to this problem. This is due to a remarkable observation made in \cite{Anninos:2011ui}: the AdS/CFT duality between type-A higher-spin gravity in $AdS_4$ and the $O(2N)$ vector model \emph{does not obviously break} upon changing the sign of the cosmological constant. This provides a rare working model of dS/CFT \cite{Strominger:2001pn,Anninos:2012qw}, and thus a crucial theoretical laboratory for quantum gravity in de Sitter space (though, of course, HS gravity is not the realistic gravity of GR). 

Following \cite{Maldacena:2002vr}, the authors of \cite{Anninos:2011ui} interpret the partition function of the CFT on the future (or past) conformal boundary of $dS_4$ as the Hartle-Hawking wavefunction \cite{Hartle:1983ai} of the $dS_4$ quantum HS gravity:
\begin{align}
 \Psi_{\text{Hartle-Hawking}}[A^{(s)}_{\mu_1\dots\mu_s(\ell)}] = Z_{\text{local}}[A^{(s)}_{\mu_1\dots\mu_s(\ell)}] \ . \label{eq:HH_general}
\end{align}
For this interpretation to make sense, the partition function had better be peaked on empty de Sitter space, i.e. at vanishing sources. A problem immediately arises: the partition function \eqref{eq:Z_local} of the boundary vector model has a local \emph{minimum} at zero, as can be seen from the sign of its 2-point function. The solution proposed in \cite{Anninos:2011ui} was to change the vector model's fundamental fields $\phi^I(\ell)$ from commuting to anti-commuting, and accordingly change the internal symmetry group from $O(2N)$ to $Sp(2N)$. This serves to flip the sign of the 2-point function, and indeed of the entire effective action, giving the partition function the desired local maximum. The restriction to even spins is essential here: upon continuation from $AdS_4$ to $dS_4$, the signs of even-spin and odd-spin 2-point functions end up being opposite, so it's impossible to give the correct sign to both at once.

So far, then, we can arrange a partition function that is locally peaked on empty $dS_4$. But is this maximum a global one? This question was investigated in \cite{Anninos:2012ft}, and the answer turned out to be negative. In particular, the authors of \cite{Anninos:2012ft} evaluated the effective action for a constant spin-0 source $\sigma$ on $S_3$, as we reviewed in section \ref{sec:partition:local}. Their answer was the one in eq. \eqref{eq:Z_local_sigma}, but with an overall sign flip in the exponent, as discussed above:
\begin{align}
 \begin{split}
   \Psi_{\text{Hartle-Hawking}}[\sigma] &= Z_{\text{local}}[\sigma] = \exp\left(\frac{N\pi}{8}\int_1^{\sqrt{1+4\sigma}} t^2 \cot\frac{\pi t}{2}\,dt\right) \\
     &= \exp\left(-\frac{N\pi^2}{8}\left(\sigma^2 - \frac{2}{3}\sigma^3 + \left(\frac{\pi^2}{6} - 1\right)\sigma^4 + \dots \right)\right) \ . 
 \end{split} \label{eq:PsiHH}
\end{align}
As plotted in \cite{Anninos:2012ft}, the local maximum of this partition function at $\sigma=0$ is overshadowed by a series of higher maxima at $\sigma$ values of order 1. This is a disaster that can't be brushed aside by flipping some overall sign. A new foundation is apparently required for higher-spin dS/CFT, one that would produce a Hartle-Hawking state different from \eqref{eq:PsiHH}. 

Such a new foundation has been proposed in \cite{Anninos:2017eib}, and studied further in \cite{Anninos:2019nib}. The authors of \cite{Anninos:2017eib} argued that it's not enough to have a Hartle-Hawking wavefunction: the Hilbert space in which this Hartle-Hawking state is meant to live must itself be carefully defined, with HS symmetry taken into account. For this purpose, they replace the boundary CFT and its anti-commuting fields $\phi^I(\ell)$ by an alternative construction based on \emph{commuting} fields $Q^I(\ell)$. Instead of the Hartle-Hawking wavefunction \eqref{eq:HH_general} given by the CFT path integral over $\phi^I(\ell)$, they postulate a straightforward wavefunction $\Psi[Q^I(\ell)] = e^{\int d^3\ell\, Q\Box Q}$, which would normally be the path integral's \emph{integrand}. Functional integrals over $Q^I(\ell)$ do appear, but with a different interpretation: they arise when computing expectation values for operators, which correspond (via a dictionary involving the shadow transform) to the CFT's \emph{sources}. 

If one accepts these changes in the rules of AdS/CFT (in particular, in the role of the boundary path integral), then the construction of \cite{Anninos:2017eib} gets many things right. In particular, when computing propabilities for the constant spin-0 mode on $S_3$, the CFT result \eqref{eq:PsiHH} is reproduced for small enough $\sigma$, but the large-$\sigma$ region with the unwanted additional maxima is now excluded. This is because the construction effectively restricts the operator $\Box + \sigma$ (or, more generally for bilocal sources, $\Box + \Pi$) to be negative-definite. In fact, for better or worse, just like the bilocal approach to the CFT, the construction of \cite{Anninos:2017eib} is crucially concerned with matrix algebra on the space of off-shell boundary fields. In particular, the ``HS symmetry'' promoted in \cite{Anninos:2017eib} as a guiding principle is the group of either \emph{all} linear transformations of the off-shell boundary fields $\phi(\ell)$, or those that preserve the conformal Laplacian $\Box$. In contrast, actual HS algebra, as we reviewed in section \ref{sec:shell:QM}, is the much smaller algebra of linear transformations on the space of \emph{on-shell solutions} to $\Box\phi = 0$. 

We contend that the present paper provides an alternative resolution to the problem of the Hartle-Hawking wavefunction's maxima. In the preceding sections, we've explored an alternative to the CFT partition function $Z_{\text{local}}$, namely the HS-algebraic partition function $Z_{\text{HS}}$. We've analyzed how $Z_{\text{HS}}$ differs from $Z_{\text{local}}$ despite being derived from the same correlators. We've shown how it can be constructed from first principles, through the requirements of HS symmetry and spin-locality, bypassing altogether the spacetime-local CFT. Just as with $Z_{\text{local}}$, the requirement for $F(Y)=0$ to be a local maximum necessitates flipping the sign of $\ln Z_{\text{HS}}$ in eq. \eqref{eq:Z}: 
\begin{align}
 Z_{\text{HS}}[F(Y)] = \frac{1}{\left(\textstyle\det_\star[1+F(Y)]\right)^{N/4}} \ .
\end{align}
In the spin-local construction of section \ref{sec:discuss:spin_locality}, this requires us to change the dynamical variables $\varphi^I(u)$ from \emph{anti-commuting} to \emph{commuting}, and the internal symmetry group from $Sp(2N)$ to $O(2N)$. Let's now evaluate $Z_{\text{HS}}$ on a constant spin-0 source, and check if this produces a better-behaved Hartle-Hawking wavefunction than the one in \eqref{eq:PsiHH}. We already performed this calculation for $EAdS_4$ signature, in eq. \eqref{eq:Z_HS_sigma}. All that remains is to flip the sign in the exponent, which gives:
\begin{align}
\Psi_{\text{Hartle-Hawking}}[\sigma] = Z_{\text{HS}}[\sigma] = \frac{1}{(1 + \pi^2\sigma^2)^{N/8}} \ . \label{eq:PsiHH_HS}
\end{align}
Aside from being obviously simpler than \eqref{eq:PsiHH}, this wavefunction is globally peaked at $\sigma=0$, as desired. 

The similarity to the construction of \cite{Anninos:2017eib} is worth remarking on: in both cases, the boundary CFT with anticommuting fields is replaced by a more symmetry-driven construction with new, commuting dynamical variables on the boundary. However, there are also differences; our proposal to use $Z_{\text{HS}}$ is more conservative in some ways, and more radical in others. We give up boundary locality, but gain manifest invariance under the true, on-shell, HS symmetry. In addition, we retain the standard AdS/CFT relationship between (linearized) bulk fields and external sources in a boundary path integral, at the cost of making that path integral spin-local rather than local. The tradeoff seems worthwhile. We thus propose the HS-algebraic partition function $Z_{\text{HS}}$ as an improved foundation for higher-spin dS/CFT.

\subsubsection{A road towards physics on cosmological horizons}

As with our interest in HS theory itself, our interest in higher-spin dS/CFT has a particular agenda. There is something conceptually unsatisfactory about the focus of \cite{Maldacena:2002vr,Anninos:2011ui,Anninos:2017eib} on the wavefunction of the $dS_4$ universe at its future conformal boundary. Aside from any other possible concerns, the future conformal boundary of de Sitter space is unobservable! Thus, the approach of \cite{Maldacena:2002vr,Anninos:2011ui,Anninos:2017eib} implies an inflation-type scenario, in which the de Sitter phase is merely temporary, and its would-be conformal future boundary becomes contained in the past lightcone of post-inflationary observers. 

In our own work -- see e.g. \cite{Neiman:2017zdr,Neiman:2018ufb} -- we try to take seriously the causal structure of pure $dS_4$. We still take as our starting point the CFT on the unobservable Euclidean boundary, and its interpretation as a Hartle-Hawking wavefunction. Therefore, e.g. the global maximum of \eqref{eq:PsiHH_HS}, and its absence in \eqref{eq:PsiHH}, remain important. However, we then aim to take a next step, and extract the Lorentzian physics in an observable patch of the bulk $dS_4$, delineated by a pair of cosmological horizons (the so-called ``static patch''). The daunting difficulty of this task makes manifest HS symmetry essential. 

In this context, the switch from boundary locality to spin-locality becomes especially natural. On one hand, since the boundary is not observable, locality on it is quite beside the point. Similarly, the $n$-point correlators at separated boundary points are no longer of any interest: only the full partition function matters. On the other hand, the observable patch does intersect the boundary at a pair of points -- the boundary endpoints of the observer's horizons (or, equivalently, of her worldline). Let us denote these by $\ell$ and $l$. It is thus natural to use variables that live at one (or both) of these endpoints. An especially natural candidate would be the limit of the bulk master field $C(x;Y)$ as $x$ approaches the endpoint $\ell$ from the direction of the other endpoint $l$. But, as we've seen in sections \ref{sec:partition:ISO3} and \ref{sec:discuss:spin_locality}, these are precisely the variables $\calC_{\ell,l}(u_R,u_L)$ or $\tilde\calC_{\ell,l}(u,u')$ of the spin-local formalism! Moreover, as discussed in \cite{David:2019mos}, the same variables have an especially nice interpretation in terms of field modes on the $dS_4$ horizon.

To conclude, the HS-algebraic, spin-local partition function $Z_{\text{HS}}$ provides a better-behaved Hartle-Hawking wavefunction at the conformal boundary of de Sitter space. Furthermore, the variables in which spin-locality and HS symmetry are manifest are also the most natural variables for taking higher-spin dS/CFT from the unobservable Euclidean boundary into an observable Lorentzian bulk patch. Let's get to work.

\section*{Acknowledgements}

We are grateful to Eugene Skvortsov, Mikhail Vasiliev, Per Sundell, Mirian Tsulaia and Sudip Ghosh for discussions, and to Dionysios Anninos and Charlotte Sleight for email exchanges. This work was supported by the Quantum Gravity and Mathematical \& Theoretical Physics Units of the Okinawa Institute of Science and Technology Graduate University (OIST). YN's thinking was substantially informed by talks and discussions at the workshop ``Higher spin gravity -- chaotic, conformal and algebraic aspects'' at APCTP in Pohang.

\end{document}